%% file: main.tex
\documentclass{lmcs}
\pdfoutput=1
\usepackage[utf8]{inputenc}

\usepackage{lastpage}
\lmcsdoi{20}{3}{10}
\lmcsheading{}{\pageref{LastPage}}{}{}%
{Feb.~03,~2023}{Jul.~29,~2024}{}


\usepackage[T1]{fontenc}
\usepackage{amssymb}
\usepackage{mathpartir}
\usepackage{listings}
\usepackage[normalem]{ulem}
\usepackage{tikz}
\usetikzlibrary{calc,cd,decorations.pathmorphing}
\usepackage{multirow}
\usepackage{xparse}
\usepackage{textgreek}

\newlist{cas}{description}{3}
\newlength\caslength
\setlist[cas]{leftmargin=\caslength}
\newcommand\casaux[2]{\item[#1 \textnormal{#2}]}
\newcommand\case{\casaux{Case}}
\newcommand\subcase{\casaux{Subcase}}
\newcommand\subsubcase{\casaux{Subsubcase}}

\input{macros}

\begin{document}

\title[Quantitative and Faithful Generalized Applications]{A Faithful and Quantitative Notion of\texorpdfstring{\\}{ } Distant Reduction for
  the Lambda-calculus with Generalized Applications\titlecomment{Long version
of \cite{espiritosanto22}}}

\author[J.~Espírito Santo]{José Espírito
  Santo\lmcsorcid{0000-0002-6348-5653}}[a] \author[D.~Kesner]{Delia
  Kesner\lmcsorcid{0000-0003-4254-3129}}[b] \author[L.~Peyrot]{Loïc
  Peyrot\lmcsorcid{0000-0002-1398-7460}}[b]

\address{Universidade do Minho, Braga, Portugal}
\email{jes@math.uminho.pt}

\address{Université Paris Cité, CNRS, IRIF, Paris, France}
\email{kesner@irif.fr, lpeyrot@irif.fr}

\input{abstract}

\maketitle

\input{introduction}
\input{djn}
\input{strong-properties}
\input{isn}
\input{types_strong}
\input{faithfulness}
\input{equivalences}
\input{conclusion}

\section*{Acknowledgments}
\noindent The first author was financed by Portuguese Funds through FCT (Fundação para a
Ciência e Tecnologia) within the projects UIDB/00013/2020 and UIDP/00013/2020.

\bibliographystyle{alphaurl}
\bibliography{biblio}

\end{document}

%% file: macros.tex
\definecolor{mongreen}{RGB}{0,179,0}

\newcommand \termsName[1] {\mathtt{T}_{#1}}
\newcommand \redName[1] {\mathrm{#1}}
\newcommand \ctxName[1] {\mathtt{#1}}
\newcommand \ruleName[1] {(\textsc{#1})}

\newcommand \lterms {\termsName \Lambda}
\newcommand \esterms {\termsName{ES}}

\newcommand \jterms {\termsName J}

\newcommand \rewrel {\mathcal R}

\newcommand \calcwithrules[2] {#1[#2]}

\newcommand \alphaeq {=_{\alpha}}

\newcommand \lcalc {\lambda}
\newcommand \lmcalc {\lambda\mu}

\newcommand \escalc {\lambda ES}

\newcommand \brule {\redName{B}}
\newcommand \db {\redName{dB}}
\newcommand \sub {\redName{sub}}

\newcommand{\jcalc}{\Lambda J}
\newcommand{\jmcalc}{\Lambda J_m}
\newcommand{\ptwo}{\redName{p2}}

\newcommand{\dbpi}{\redName{d}\beta\pi}
\newcommand{\jn}{\redName{\beta,\pi}} 

\newcommand{\djncalc}{\lambda J_n}
\newcommand{\dbeta}{\redName{d}\beta}

\newcommand{\djn}{\dbeta} 

\newcommand{\whdj}{\redName{lr}}
\newcommand{\paraxjn}{\textsc{(VAR)}}
\newcommand{\parabsjn}{\textsc{(ABS)}}
\newcommand{\parappjn}{\textsc{(APP)}}
\newcommand{\pardbjn}{\textsc{(DB)}}

\newcommand{\jvcalc}{\Lambda J_v}

\newcommand \nf[1] {\operatorname{NF}_{#1}}
\newcommand \nef[1] {\operatorname{NE}_{#1}}

\newcommand \nans {\mathtt a}
\newcommand \nl {\mathtt n} 

\newcommand \sn[1] {\operatorname{SN}(#1)}
\newcommand \isn[1] {\operatorname{ISN}(#1)}
\DeclareMathOperator \isnj {ISNj}
\newcommand \isnvarrule {\ruleName{snvar}}
\newcommand \isnabsrule {\ruleName{snabs}}
\newcommand \isnapprule {\ruleName{snapp}}
\newcommand \isnbetarule {\ruleName{snbeta}}
\newcommand \isnredexa {\ruleName{snredex1}} 
\newcommand \isnredexb {\ruleName{snredex2}} 
\newcommand \isnbpvarrule {\ruleName{var}}
\newcommand \isnbphvarrule {\ruleName{hvar}}
\newcommand \isnbpabsrule {\ruleName{lambda}}
\newcommand \isnbppirule {\ruleName{pi}}
\newcommand \isnbpbetarule {\ruleName{beta}}
\newcommand \isnbpone {\ruleName{I}}
\newcommand \isnbptwo {\ruleName{II}}

\newcommand \ec {\diamond} 
\newcommand \fc {\ctxName C} 
\newcommand \lc {\ctxName L} 
\newcommand \dc {\ctxName D} 
\newcommand \dcnl {\ctxName {D_n}} 

\newcommand \whc {\ctxName W} 
\newcommand \lrc {\ctxName R} 

\newcommand \id {\mathtt I}

\newcommand \MM {\mathcal M}
\newcommand \MN {\mathcal N}

\newcommand \TypeVariable {BTV}
\newcommand \btvone {a}
\newcommand \btvtwo {b}
\newcommand \btvthree {c}

\newcommand \sysFullES {\cap ES}

\newcommand \sysFullJ {\cap J}

\newcommand \sysSTJ {ST}

\newcommand \many {\ruleName{many}}

\newcommand \ruleAxES{\ruleName{ax}}
\newcommand \ruleAbsES{\ruleName{$\to_i$}}

\newcommand \ruleAppES{\ruleName{$\to_e$}}
\newcommand \ruleES{\ruleName{es}}

\newcommand \ruleAxJ {\ruleName{var}}
\newcommand \ruleAbsJ {\ruleName{abs}}
\newcommand \ruleAppJ {\ruleName{app}}

\newcommand \lx {\lambda x}
\newcommand \ly {\lambda y}
\newcommand \lz {\lambda z}

\newcommand \ap[3] {(#1,#2.#3)}
\newcommand \gap[4] {#1\ap{#2}{#3}{#4}}

\newcommand{\letb}[3]{\text{\lstinline!let!}\ #1\ \text{\lstinline!=!}\ #2\ \text{\lstinline!in!}\ #3}

\newcommand \esub[2] {[#1 / #2]}

\newcommand \rsub[2] {\{#1/#2\}}

\newcommand \inter {\uplus}
\newcommand \emult {\mult{\, }}
\newcommand \mult[1] {[#1]}
\newcommand \multunion {\sqcup}

\newcommand{\seq}[3]{#1 \vdash #2:#3}
\newcommand{\der}[3]{#1 \Vdash #2:#3}
\newcommand{\dern}[4]{#1 \Vdash^{#4} #2:#3}
\newcommand{\derp}[4]{#1 \Vdash_{#4} #2:#3}
\newcommand{\dernp}[5]{#1 \Vdash^{#4}_{#5} #2:#3}

\newcommand{\demp}[5]{#1 \triangleright \derp{#2}{#3}{#4}{#5}}
\newcommand{\demnp}[6]{#1 \triangleright \dernp{#2}{#3}{#4}{#5}{#6}}

\newcommand \rrule[1] {\mapsto_{#1}}

\newcommand \rew[1] {\rightarrow_{#1}}
\newcommand \rewp[1] {\rew{#1}^+}
\newcommand \rewn[1] {\rew{#1}^*}

\newcommand \parred[1] {\rightrightarrows_{#1}}

\newcommand \obseq \equiv

\newcommand \fv[1] {\operatorname{fv}(#1)} 
\newcommand \ctx[2] {#1 \langle  #2 \rangle} 

\newcommand \tmsz[1] {|#1|} 
\newcommand \devdjn[1] {(#1)^{\dbeta}} 

\newcommand \maxred[2][] {||#2||_{#1}} 
\newcommand \rbmaxredbp[1] {\maxred[\jn]{#1}^{\beta}}

\newcommand{\rbmaxp}[1]{||#1||_{\pi}}

\newcommand \dom[1] {\operatorname{dom}(#1)}
\newcommand \choice[1] {\#(#1)} 
\newcommand \cmin[2] {#1 \setminus #2}
\newcommand \msetsz[1] {|#1|} 

\newcommand \mapesl[1] {#1^\downarrow}

\newcommand{\mapjl}[1]{#1^{\#}} 
\newcommand{\mapjes}[1]{{#1}^*} 
\newcommand{\mapjesn}[1]{#1^{\star}} 
\newcommand{\mapesj}[1]{{#1}^{\circ}}
\newcommand{\mapesjj}[1]{{#1}^{\bullet}} 
\newcommand{\mapjesj}[1]{{#1}^{\star\circ}}
\newcommand{\mapjesl}[1]{#1^{\square}}
\newcommand{\mapjj}[1]{#1^{\dagger}} 
\newcommand{\lvar}[1]{#1^\text{l}}
\newcommand{\rvar}[1]{#1^\text{r}}
\newcommand{\mapjessubs}[1]{\rsub{\lvar{#1}\rvar{#1}}{#1}}
\newcommand{\mapjesapp}[4]{\esub{\mapjesn{#1}}{\lvar{#3}}
\esub{\mapjesn{#2}}{\rvar{#3}} \mapjessubs{#3} \mapjesn{#4}}
\newcommand{\mapjesappempty}[4]{\esub{\mapjesn{#1}}{\lvar{#3}}
\esub{\mapjesn{#2}}{\rvar{#3}} \mapjesn{#4}}

\newcommand \eqdef \coloneqq 
\newcommand \tuple[1] {\langle #1 \rangle}
\newcommand \pair[2] {\tuple{#1, #2}}

\DeclareMathOperator \subproj P

\newcommand \deft[1] {\textbf{#1}}
\newcommand \gramTitle[1] {\mbox{\textbf{(#1)}}}
\definecolor{verdelight}{RGB}{171,232,194}
\definecolor{naranjalight}{RGB}{255,204,153}
\definecolor{celeste}{RGB}{178,255,255}

\newcommand{\ih}{\textit{i.h.}}
\newcommand{\ie}{\textit{i.e.}}
\newcommand{\eg}{\textit{e.g.}}
\newcommand{\cf}{\textit{c.f.}}

\newcommand \iI {i \in I}
\newcommand \jJ {j \in J}

\newcommand \multii[1] {\mult{#1}_{\iI}}
\newcommand \multjj[1] {\mult{#1}_{\jJ}}
\newcommand \chmultii[1] {\choice{\multunion_{\iI} {#1}_i}}
\newcommand \chmultjj[1] {\choice{\multunion_{\jJ} {#1}_j}}
\newcommand \interii {\inter_{\iI}}

\renewcommand{\Autoref}[1]{%
  \begingroup%
  \def\pageautorefname{Page}%
  \def\lemmaautorefname{Lemma}%
  \def\sectionautorefname{Section}%
  \def\subsectionautorefname{Subsection}%
  \def\theoremautorefname{Subsection}%
  \autoref{#1}%
  \endgroup%
}

\newcommand{\lhs}{l.~h.~s.}
\newcommand{\rhs}{r.~h.~s.}

%% file: abstract.tex
\begin{abstract}
  We introduce a call-by-name \textlambda-calculus $\djncalc$ with generalized
  applications which is equipped with {\it distant reduction}.
  This allows to unblock $\beta$-redexes without resorting to the standard
  permutative conversions of generalized applications used in the original
  $\jcalc$-calculus with generalized applications of Joachimski and Matthes.
  We show strong normalization of simply-typed terms, and we then fully
  characterize strong normalization by means of a quantitative (\ie\
  non-idempotent intersection) typing system.
  This characterization uses a non-trivial inductive definition of strong
  normalization --related to others in the literature--, which is based on a
  weak-head normalizing strategy.
  We also show that our calculus $\djncalc$ relates to explicit substitution
  calculi by means of a faithful translation, in the sense that it preserves
  strong normalization.
  Moreover, our calculus $\djncalc$ and the original $\jcalc$-calculus
  determine equivalent notions of strong normalization.
  As a consequence, $\jcalc$ inherits a faithful translation into explicit
  substitutions, and its strong normalization can also be characterized by the
  quantitative typing system designed for $\djncalc$, despite the fact that
  quantitative subject reduction fails for permutative conversions.
\end{abstract}

%% file: introduction.tex
\section{Introduction}
\label{sec:introduction}

In the original calculus with generalized applications $\jcalc$ of
Joachimski and Matthes~\cite{joachimski03,joachimski00}, the
standard syntax of the \textlambda-calculus is modified by generalizing
the application constructor $tu$ into a new shape $\gap t u y r$,
capturing a notion of sharing for applications: a term $\gap t u y r$
can intuitively be understood as a let-binding of the form $\letb y
{tu} r$.

This new constructor can be better understood in a typed framework.
Indeed, the simply-typed $\jcalc$-calculus is an interpretation of the
implicative fragment of von Plato's system of natural deduction with generalized
elimination rules~\cite{plato01} under the Curry-Howard correspondence.
For example,  generalized elimination of implication in von Plato's system is realised by the
  following rule with three premises.
  \[
    \inferrule*{
      \Gamma \vdash A \to B
      \and \Gamma \vdash A
      \and \Gamma, B \vdash C
    }{
      \Gamma \vdash C
    }
  \]
The resulting typing rules for the simply-typed $\jcalc$-calculus are given in
\autoref{sec:some-properties}.
Besides the logical reading, the syntax with generalized applications
constitutes also a minimal framework for studying the call-by-name (CbN) and
call-by-value (CbV) functional paradigms, as well as various kinds of
permutative conversions beyond the \textlambda-calculus.

The operational semantics of $\jcalc$ is given by a
call-by-name\footnote{A recent call-by-value variant, proposed
in~\cite{espiritosanto20}, is out of the scope of this paper.}
$\beta$-rule generalizing the one of the \textlambda-calculus, as well
as a permutative $\pi$-rule on terms. The two rules are as follows:
\[ \begin{array}{lll}
  \gap{(\lambda x.t)}{u}{y}{r} &  \rrule{\beta} & \rsub{\rsub{u}{x}{t}}{y}{r} \\
    \gap{\gap t u y {r}}{u'}{z}{r'} &  \rrule{\pi} & \gap{t}{u}{y}{\gap{r}{u'}{z}{r'}} \\
  \end{array} \]
In a typed setting, the reduction of terms in $\jcalc$ corresponds to
normalization in natural deduction with generalized elimination rules.
The $\pi$-rule corresponds to a permutation
(commutative
conversion) caused by the fact that a
same formula is a premiss and a conclusion of an elimination in the
associated logical system.
Indeed, in the redex above, the type of $\gap t u y r$ is the same as
the type of $r$.
A normalization process effectively transforms proofs into what are known as fully normal forms.
Fully-normal forms enjoy the subformula property and are in one-to-one
correspondence with the cut-free derivations of the sequent
calculus~\cite{plato01}.

Both reduction rules $\beta$ and $\pi$ make perfect sense in the untyped setting as
well.  The $\beta$-rule executes the call of the function $\lambda
x.t$ with argument $u$: this eliminates the shared application, and
the result $\rsub{u}{x}{t}$ has to be \emph{unshared} within the
continuation $r$. The $\pi$-rule permutes terms, eventually unblocking $\beta$-redexes.
For example, in the reduction sequence
$\gap{\gap t u y {\lambda x.r}}{u'}{z}{r'} \rew\pi
\gap{t}{u}{y}{\gap{(\lambda x.r)}{u'}{z}{r'}} \rew\beta
\gap{t}{u}{y}{\rsub{\rsub{u'}{x}{r}}{z}{r'}}$, the first step reveals
a $\beta$-redex, previously hidden.
Indeed, some $\beta$-redexes are obstructed by the syntax of terms,
and rearrangement of terms through
$\pi$-reduction steps is often necessary to obtain meaningful semantics.

Because of the interaction between computation, 
specified by $\beta$, and permutation, specified by $\pi$,
characterizing strong normalization in $\jcalc$ is not evident.
This has been done through the notion of typability in an (idempotent) intersection typing
system by~\cite{matthes00}: a term is typable if and only if
it is strongly normalizing.  However, this characterization is just
\emph{qualitative}.  A different flavor of intersection types, called
\emph{non-idempotent}~\cite{gardner94}, offers a more powerful
\emph{quantitative} characterization of strong normalization. Indeed, the length of the longest reduction sequence to normal form
starting at a typed term, as well as the size of its normal
form, is bounded by the size of its type derivation.  Non-idempotent
intersection (\textit{a.k.a.} \emph{quantitative}) type systems
can be seen as an inductive
representation of the \emph{relational model} of linear
logic~\cite{decarvalho07}.  However, quantitative types were never used
in the framework of generalized applications, and it is our purpose to
propose and study one such typing system.

Quantitative types allow for simple combinatorial proofs of strong
normalization, without any need to use reducibility or computability
arguments.  More remarkably, they also provide a refined tool to
understand permutations.
As we will observe, in the original $\jcalc$-calculus, rule $\pi$ is not
quantitatively sound (\ie\ $\pi$ does not enjoy quantitative subject reduction),
although $\pi$ becomes valid in a qualitative  framework (with idempotent
types).
This means that it is not possible to obtain a non-idempotent type system
for the original formulation of $\jcalc$. 
How can we then unblock redexes to reach normal forms in a quantitative model of
computation based on generalized applications?

Our solution relies on a different permutation rule
$\gap t u y {\lx.r} \rrule{\ptwo} \lx.\gap t u y r$.\footnote{Rule $\ptwo$ is
  used in~\cite{espiritosanto03,espiritosanto11} along with two other
  permutation rules $\redName{p1}$ and $\redName{p3}$ to reduce terms with
generalized applications to a form corresponding to ordinary $\lambda$-terms.}
More precisely, instead of considering rule $\ptwo$ independently from $\beta$,
we adopt the paradigm of \emph{distant reduction}~\cite{accattoli10,AK12}, which
extends the key concept of $\beta$-redex, so that it is possible to find a
$\lambda$-abstraction hidden under a certain context, in our case under a
sequence of nested generalized applications.
To do so, we directly integrate the $\ptwo$-permutations that are necessary to
unblock reduction together with $\beta$, creating a distant~$\beta$ rule, called
$\dbeta$.
This choice does not affect (strong) normalization, which is our focus, and
highlights the computational behavior of the calculus: every step has a
computational content given by the underlying $\beta$.

The syntax of the $\jcalc$-calculus will thus be equipped with an operational
call-by-name semantics given by the  distant rule $\dbeta$, but without $\pi$.
The resulting calculus is called $\djncalc$.
As a major contribution, we prove a characterization of strong normalization in
terms of typability in our quantitative type system.
In such proof, the soundness result (typability implies strong normalization) is
obtained by simple combinatorial arguments, with the size of typing derivations
decreasing at each $\dbeta$-step.
For the completeness result (strong normalization implies typability) we need an
inductive characterization of the terms that are strongly normalizing for
$\dbeta$: this is a non-trivial technical contribution of the paper.

Our new calculus $\djncalc$ is then compatible with a quantitative typing
system.
However, this type system designed for $\djncalc$ only partially captures strong
normalization for $\jcalc$ on a quantitative level, because the bound for
reduction lengths
given by the size of type derivations only holds for
$\beta$ and $\dbeta$, and not for $\pi$.
Nevertheless, using this partial bound, we can prove that the type system
designed for $\djncalc$ is also sound for strong normalization in the original
calculus $\jcalc$, in the sense that any typable term is strongly normalizing.
It immediately follows that if a term $t$ is strong normalizing in $\djncalc$, then it is strongly
normalizing in $\jcalc$.

Actually, we go further and prove that this implication is an equivalence.
The central role in the proof is again played by intersection type systems,
together with a new encoding of generalized applications into explicit
substitutions (ES).
More precisely, we consider a calculus with explicit substitutions, where a new
constructor $\esub u x t$, akin to a let-binding $\letb x u t$, is added to the
grammar of the \textlambda-calculus.
The reading given above of $\gap t u y r$ as a let-binding expressing the
sharing of the application $tu$ is similar to the intuitive and
known~\cite{espiritosanto07} translation of $\gap t u y r$ into the explicit
substitution $\esub {tu} y r$.
This translation, however, does not suit our goals, because it does not preserve
strong normalization: a non-terminating computation generated by the interaction
of $t$ with $u$ in $\gap t u y r$ will always have to be substituted for $y$ in
$r$, and thus may vanish if $y$ does not occur free in $r$ (a detailed example
will be given later).

We instead propose a new, type-preserving encoding of terms with
generalized applications into ES. Thanks to it,
we show the dynamic behavior of our calculus $\djncalc$ to be
\emph{faithful} to explicit substitutions: a term is strongly
normalizing in $\djncalc$ if and only if its new enconding into ES is
also strongly normalizing.  The proof of faithfulness essentially
relies on an analysis of typability in the type system designed for
$\djncalc$.
Thanks to the properties of the
  calculus with explicit substitutions, preservation of strong
  normalization from $\djncalc$ (and thus $\jcalc$) to and from the
\textlambda-calculus can be finally guaranteed.

\paragraph{Plan of the paper.}
\Autoref{sec:calculus} presents and motivates our calculus $\djncalc$ with
distant $\beta$.
\Autoref{sec:isn} provides an inductive characterization of strongly normalizing
terms in $\djncalc$.
\Autoref{sec:types} presents the non-idempotent intersection type system for
$\djncalc$, proves the characterization of strong normalization in $\djncalc$ as
typability in that system, and discusses why $\pi$ is not quantitative.
\Autoref{sec:faithfulness} defines the new translation into ES and proves it to
be faithful, in the sense of preserving and reflecting strong normalization.
\Autoref{sec:equivalences} contains comparisons with other calculi, obtained by
equipping the terms of $\djncalc$ with $\beta$,
$(\beta,\ptwo)$, and $(\beta,\pi)$.
The main focus there is to prove the respective notions of strong normalization
equivalent, but we also collect the results $\jcalc$ inherits from our study of
$\djncalc$. \Autoref{sec:conclusion} summarizes our contributions and discusses
future and related work.

%% file: djn.tex
\section{A calculus with generalized applications}
\label{sec:calculus}

In this section we define our calculus with generalized applications,
denoted $\djncalc$. Starting from the issue of stuck redexes, we discuss different
possibilities for the operational semantics. Next we prove some
introductory properties of the calculus we propose.

\subsection{Syntax}

We start with some general notations. 
  We consider an \deft{abstract reduction system} to be
  a set of objects together with a binary relation
  $\rew\rewrel$ defined  on this set, understood as a reduction relation.
    Given a reduction relation $\rew\rewrel$,
we write $\rewn\rewrel$ (resp. $\rewp\rewrel$) for the
reflexive-transitive (resp. transitive) closure of $\rew\rewrel$.  A
term $t$ is said to be in \deft{$\rewrel$-normal form} (written
$\rewrel$-nf) iff there is no $t'$ such that $t \rew{\rewrel} t'$.  A
term $t$ is said to be \deft{$\rewrel$-strongly normalizing} (written
$t \in \sn\rewrel$) iff there is no infinite $\rewrel$-reduction
sequence starting at $t$.  $\rewrel$ is strongly normalizing iff every
term is $\rewrel$-strongly normalizing.  When $\rew\rewrel$ is
finitely branching, $\maxred[\rewrel]{t}$ denotes the maximal length
of an $\rewrel$-reduction sequence to $\rewrel$-nf starting at $t$,
for every $t \in \sn\rewrel$.

\medskip

We now introduce the concrete syntax of our calculi. 
The set of terms is generated by the following grammar and is denoted by $\jterms$.
\[
  \gramTitle{Terms}\ t,u,r,s \Coloneqq x \mid \lx.t \mid \gap t u x r
\]
We use $\id$ to denote the identity
function $\lz.z$ and $\delta$ to denote the term $ \lx.\gap x x z z$.
The term $\gap{t}{u}{x}{r}$ is called a generalized application, and
the part $x.r$ is sometimes referred as the \emph{continuation} of the
generalized application.  \deft{Free variables} of terms are defined as
usual, notably $\fv{\gap{t}{u}{x}{r}} := \fv{t} \cup \fv{u} \cup
\fv{r} \setminus \{x\}$.  We  work modulo
$\alpha$-conversion, denoted $\alphaeq$, so that bound variables can
be systematically renamed.
We will follow Barendregt convention and assume that the free variables of
a given term are distinct from the bound ones.
The \deft{substitution} operation is capture-avoiding
and defined as usual, in particular $\rsub{u}{x}{(\gap{t}{s}{y}{r})}
\eqdef \gap{(\rsub{u}{x}t)}{\rsub{u}{x}s}{y}{\rsub{u}{x}r}$, where $y\notin \fv{u}$.

Contexts (terms with one occurrence of the hole $\ec$) and distant contexts are given by the following grammars:
\[ \begin{array}{rlll}
  \gramTitle{Contexts} & \fc & \Coloneqq & \ec \mid \lx.\fc \mid
  \gap \fc {u}{x}{r} \mid \gap{t}{\fc}{x}{r} \mid \gap{t}{u}{x}{\fc} \\
  \gramTitle{Distant contexts} & \dc & \Coloneqq & \ec \mid \gap{t}{u}{x}{\dc}
\end{array} \]
The term $\ctx{\fc}{t}$ denotes $\fc$ where $\ec$ is replaced by $t$, so that
capture of variables may eventually occur. We say that $t$ has an \deft{abstraction shape} iff $t = \ctx{\dc}{\lx.u}$.

Given a reduction rule $\rrule\rewrel \subseteq
\jterms \times \jterms$,  $\rew{\rewrel}$ denotes the reduction
relation generated by the closure of $\rrule\rewrel
$ under all contexts. The syntax of $\jterms$ can be equipped with
different rewriting rules. We use the generic notation $\calcwithrules
\jterms \rewrel$ to denote the calculus given by the syntax $\jterms$
equipped with the reduction relation $\rew\rewrel$. In particular, the
\deft{$\jcalc$-calculus}~\cite{joachimski00} is given by $\calcwithrules
\jterms {\beta,\pi}$, where we recall $\beta$ and $\pi$,
defined in \autoref{sec:introduction}~\footnote{For instance, we consider the reduction rule $\rrule{\beta} $ to be the set of all pairs having the displayed form. In this way we avoid the use of formal syntax from Higher-Order Rewriting~\cite{Terese03}. }:
\[ \begin{array}{lll}
  \gap{(\lambda x.t)}{u}{z}{r} &  \rrule{\beta} & \rsub{\rsub{u}{x}{t}}{z}{r} \\
  \gap{\gap t u y {r}}{u'}{z}{r'} &  \rrule{\pi} & \gap{t}{u}{y}{\gap{r}{u'}{z}{r'}} \\
\end{array} \]
Like in the \textlambda-calculus, $\alpha$-conversion is needed (and used
implicitely), if for instance $y \in \fv{u'} \cup \fv{r'}$ in rule $\pi$.

\subsection{Towards a Call-by-Name Operational Semantics}
\label{sec:semantics-djn}

Permutation $\pi$ in the $\jcalc$-calculus is not only relevant from a
  syntactical point of view, but also semantically.
  For example, consider the term
  $t_\Omega \eqdef \gap{\gap {x_1} y {x_2} \delta} \delta z z$,
  where $\delta = \lx.\gap x x z z$.
  The particular syntax of this term hides a redex $\gap \delta \delta z z$,
  being the source of  a non-terminating reduction sequence.
  Indeed, the interaction between the first and the second $\delta$ in
$t_\Omega$ is \emph{stuck} by a piece of syntax in between them.

As this example hints at, the set of strongly normalizing terms in
  $\calcwithrules\jterms\beta$
  is strictly smaller than the set of normalizing terms in
  $\calcwithrules\jterms{\beta,\pi}$.
  In this sense, solely considering $\beta$ reduction creates \emph{premature}
normal forms.
This is where the permutative rule $\pi$ plays the role of an unblocker for
$\beta$-redexes. Indeed,
  \[
  t_{\Omega} \rew\pi \gap {x_1} y {x_2} {\gap \delta \delta z z}
  \rew\beta \gap {x_1} y {x_2} {\gap \delta \delta z z} \rew\beta \ldots
  \]
After the permutation step reveals the redex, the reduct of $t_\Omega$ reduces
indefinitely to itself.

More generally, given $t \eqdef \gap{\ctx\dc{\lx.t'}}{u}{y}{r}$ with $\dc \neq
\ec$, a sequence of $\pi$-steps reduces this term $t$ to
$\ctx\dc{\gap {(\lx.t')} u y r}$.
A further $\beta$-step produces $\ctx\dc{\rsub{\rsub u x {t'}} y r}$.
So, the original $\jcalc$-calculus, which is exactly $\jterms[\beta,\pi]$,
has an associated derived notion of distant $\beta$ rule, \emph{based on $\pi$}.
This rule $\dbpi$ is specified as follows.
\begin{align}
  \label{dpi-rule}%
  \gap{\ctx\dc{\lx.t}} u y r \rrule\dbpi \ctx\dc{\rsub{\rsub u x t} y r}
\end{align}
Coming back to our example $t_\Omega$, we consider $\dc = \gap {x_1} y {x_2}
  \ec$ so that
\[
  t_\Omega = \gap{\ctx\dc\delta} \delta z z
  \rew\dbpi \ctx\dc{\gap \delta \delta z z}
  \rew\dbpi \ctx\dc{\gap \delta \delta z z} \rew\dbpi \ldots
\]

Still, the operational semantics that we propose in this
paper will not reduce as in \eqref{dpi-rule}, because such rule, as
well as $\pi$ itself, does not admit a quantitative semantics (see
\autoref{s:pi-not-quantitative}).
We then choose to unblock $\beta$-redexes with rule $\ptwo$ instead:
\[
  \gap t u y {\lx.r} \rrule\ptwo \lx.\gap t u y {r}
\]
On our example, this gives:
\[
  t_\Omega = \gap{\gap {x_1} y {x_2} {\lx.\gap x x z z}} \delta z z
  \rew\ptwo \gap{(\lx.\gap {x_1} y {x_2} {\gap x x z z})} \delta z z
  \rew\beta \gap {x_1} y {x_2} {\gap \delta \delta z z}
\]

More generally, the left-hand side term in \eqref{dpi-rule} is
  reduced as follows:
  \[
    \gap{\ctx\dc{\lx.t}} u y r
    \rewn\ptwo \gap{(\lx.\ctx\dc{t})}u y r
    \rew\beta \rsub{\rsub u x {\ctx\dc{t}}} y r
    = \rsub{\ctx\dc{\rsub u x t}} y r \enspace.
  \]

We turn this reduction into a new reduction rule $\dbeta$, called \emph{distant} $\beta$:
  \begin{equation}
    \gap {\ctx\dc{\lx.t}} u y r
    \rrule\dbeta \rsub{\ctx\dc{\rsub u x t}} y r \enspace.
\end{equation}
Coming back again to our example,
  and by taking the same $\dc = \gap {x_1} y {x_2} \ec$ as before,
  we get the correct $\dbeta$-reduction from $t_\Omega$ to
$\gap {x_1} y {x_2} {\gap \delta \delta z z}$.
\begin{defi}
  The distant calculus with generalized applications is given by
  $\djncalc \eqdef \calcwithrules \jterms \dbeta$.
\end{defi}

Reducing the term $t_\Omega$ with $\dbpi$ or $\dbeta$ gives exactly the
  same result.
  This is however not always the case.
  In the right-hand side of rule~$\dbpi$, a
    unique copy of the distant context $\dc$ lies
  outside of the two substitutions, regardless of the number of occurrences of $y$
  in $r$.  On the contrary, in rule~$\dbeta$, the distant
  context may be erased or duplicated according to the number of
  occurrence of the variable $y$ in the term $r$.  Take for instance
  $\id = \lx.x$, the context $\dc = \gap {x} y {y'} \ec$ and the term
  $t = \gap{\gap {x} y {y'} \id}{\id}{z}{z'} =
  \gap{\ctx\dc\id}{\id}{z}{z'}$.  We have:
\begin{align*}
  t &\rew\dbpi \ctx\dc{\rsub{\rsub \id x x}{z}{z'}}
  = \ctx\dc{z'} = \gap {x} y {y'}{z'}  \\
  t &\rew\dbeta \rsub{\ctx\dc{\rsub \id x x}} {z}{z'} = z'
\end{align*}

The first behavior has a CbV flavor, neither erasing nor duplicating
  applications, while the second  behavior has a CbN flavor.
  Notice how, if we were to substitute a \textlambda-abstraction for $x$
      in the term $t$, the number of reduction steps to the same normal form $z'$
    would differ.
  This should give a first intuition on why neither $\pi$ nor $\dbpi$ are
quantitatively correct in a CbN calculus.

In summary, $\calcwithrules \jterms \dbpi$ does not provide
a sound semantics for a resource-aware model, such as the one
given by a quantitative type system.
More precisely, quantitative subject reduction does not hold for (a rule relying
on) $\pi$, as is shown in \autoref{s:pi-not-quantitative}.
We adopt $\calcwithrules \jterms \dbeta$ instead.

%% file: strong-properties.tex
\subsection{Some Properties of \texorpdfstring{$\djncalc$}{lambda Jn}}%
\label{sec:some-properties}

In this section we discuss some untyped properties of the new calculus:
  $\dbeta$-normal forms and confluence; as well as some properties of the simply
typed $\djncalc$: the subformula property and $\dbeta$-strong normalization.

\paragraph{Characterization of normal forms}
We describe the set of normal forms by means of the following context free
grammar.

\begin{lem}
  \label{l:nf-cbn}
  The grammar $\nf\djn$ characterizes $\djn$-normal forms.
  \begin{align*}
    \nf\djn &\Coloneqq x \mid \lx.\nf\djn \mid \gap {\nef\djn} {\nf\djn} x {\nf\djn}\\
    \nef\djn &\Coloneqq x \mid \gap {\nef\djn} {\nf\djn} x
    {\nef\djn}
  \end{align*}
\end{lem}

  \begin{proof}
  We need to show the following two properties:
  \begin{enumerate}
    \item $t\in\nef\djn \iff t$ is in $\djn$-nf and does not have an abstraction shape;
    \item $t\in\nf\djn \iff t$ is in $\djn$-nf.
  \end{enumerate}
  Soundness is by simultaneous induction on $t \in \nef\djn$ and $t \in
  \nf\djn$,
  while completeness is by induction on $t$.
  Both inductions are straightforward.
\end{proof}

We already saw that, once $\beta$ is generalized to $\dbeta$, $\pi$ is not
needed anymore to unblock $\beta$-redexes; the next lemma says that $\pi$
preserves $\djn$-nfs, so it does not bring anything new to $\djn$-nfs either.
\begin{lem}
  If $t$ is a $\djn$-nf, and $t
  \rew{\pi} t'$, then $t'$ is a $\djn$-nf.
\end{lem}
\begin{proof}
  Given \autoref{l:nf-cbn}, the proof proceeds by simultaneous induction on
  $\nf\djn$ and $\nef\djn$,
  where we generalize the induction hypothesis for $\nef\djn$ by stating that a term
  in $\nef\djn$ does not have an abstraction shape. Besides that, the proof is
straightforward.
\end{proof}

\paragraph{Confluence}%
\label{sec:confluence-djn}

We now prove confluence of the calculus.
For this, we adapt the proof of \cite{takahashi95}.
The same proof method is used for $\jcalc$ by \cite{joachimski00} and
by \cite{espiritosanto20} for $\jvcalc$.
We begin by defining the following \deft{parallel reduction} $\parred\djn$:
\begin{mathpar}
  \inferrule*[right=\paraxjn]{ }{x \parred\djn x}
  \and \inferrule*[right=\parabsjn]{
    t \parred\djn t'}{
  \lx.t \parred\djn \lx.t'}
  \and \inferrule*[right=\parappjn]{
    t \parred\djn t' \and u \parred\djn u' \and r \parred\djn r'}{
  \gap t u x r \parred\djn \gap{t'}{u'}{x}{r'}}
  \and \inferrule*[right=\pardbjn]{
    \ctx\dc t \parred\djn t' \and u \parred\djn u' \and r
    \parred\djn r'}{
    \gap{\ctx\dc{\lx.t}} u y r \parred\djn
  \rsub{\rsub{u'} x {t'}} y {r'}}
\end{mathpar}

The particularity of our proof is the following lemma which deals with
distance.
\begin{lem}
  \label{l:djn-parred-auxabs}%
  Let $t_1 = \ctx\dc t \parred\djn t_2$.
  Then there are $\dc', t'$ such that $t_2 = \ctx{\dc'}{t'}$
  and $\ctx\dc{\lx.t} \parred\djn \ctx{\dc'}{\lx.t'}$.
\end{lem}
\begin{proof}
  By induction on $\dc$.
  \begin{cas}
  \item[Case] $\dc = \ec$.
      We take $\dc' = \ec$, $t' = t_2$.
      We have $\lx.t_1 \parred\djn \lx.t_2$ by rule \parabsjn.
    \item[Case] $\dc = \gap s u y {\dc_0}$ and $t_1 = \gap s u y {\ctx{\dc_0} t}
      \parred\djn \gap {s'}{u'} y r = t_2$ by rule $\parappjn$.
      By hypothesis, we have $s \parred\djn s'$, $u \parred\djn u'$ and
      $\ctx{\dc_0} t \parred\djn r$.
      By \ih\ $r = \ctx{\dc_1}{t'}$ and $\ctx{\dc_0}{\lx.t} \parred\djn
      \ctx{\dc_1}{\lx.t'}$.
      We conclude by taking $\dc' = \gap {s'}{u'} y {\dc_1}$.
    \item[Case] $\dc = \gap {\ctx{\dc_0}{\lz.s}} u y {\dc_1}$ and
      $t_1 = \gap {\ctx{\dc_0}{\lz.s}} u y {\ctx{\dc_1} t}
      \parred\djn \rsub {\rsub {u'} x {s'}} y r = t_2$
      by $\pardbjn$.
      By hypothesis, we have $\ctx{\dc_0}{\lz.s} \parred\djn s'$,
      $u \parred\djn u'$ and $\ctx{\dc_1} t \parred\djn r$.
      By \ih\ $r = \ctx{\dc_2}{t''}$ and $\ctx{\dc_1}{\lx.t} \parred\djn
      \ctx{\dc_2}{\lx.t''}$.
      We can assume by $\alpha$-equivalence that the free variables of $u'$ and
      $s'$ are not bound by $\dc_2$.
      We take $\dc' = \rsub {\rsub {u'} z {s'}} y {\dc_2}$ and
      $t' = \rsub {\rsub {u'} z {s'}} y {t''}$.
      Thus, we have $\ctx{\dc'}{\lx.t'} = \rsub {\rsub {u'} z {s'}} y
      {\ctx{\dc_2}{\lx.t''}}$ and we can conclude $\ctx\dc{\lx.t} =
      \gap {\ctx{\dc_0}{\lz.s}} u y {\ctx{\dc_1}{\lx.t}} \parred\djn
      \ctx{\dc'}{\lx.t'}$ by \ih\ and rule~$\parabsjn$.
      \qedhere
  \end{cas}
\end{proof}

\begin{lem}
  \label{l:perm-rsub}
  Let $y \notin \fv u$. Then
  $\rsub u x {\rsub r y t} =
  \rsub {\rsub u x r} y {\rsub u x t}$.
\end{lem}
\begin{proof}
By   straightforward induction on $t$.
\end{proof}

\begin{lem}
  \label{l:djn-parred}%
  Let $t_1, t_2, u_1, u_2 \in \jterms$. Then:
  \begin{enumerate}
    \item If $t_1 \rew\djn t_2$, then $t_1 \parred\djn t_2$.
    \item If $t_1 \parred\djn t_2$, then $t_1 \rewn\djn t_2$.
    \item If $t_1 \parred\djn t_2$ and $u_1 \parred\djn u_2$,
      then $\rsub{u_1} z {t_1} \parred\djn \rsub{u_2} z {t_2}$.
  \end{enumerate}
\end{lem}
\begin{proof}
  The proof of the first statement is by induction on $t_1 \rew\djn t_2$.
  In the base case $t_1 = \gap{\ctx\dc{\lx.t}} u y r
  \rew\dbeta \rsub{\rsub u x {\ctx\dc t}} y r = t_2$,
  we use rule~$\pardbjn$ with premises $\ctx\dc t \parred\djn \ctx\dc t$,
  $u \parred\djn u$ and $r \parred\djn r$.
  The other cases are straightforward by \ih\ and rules $\parabsjn$ or
  $\parappjn$.

  The proof of the second statement is by induction on $t_1 \parred\djn t_2$.
  The base case $\paraxjn$ is by an empty reduction $t_1 = x = t_2$.
  The cases $\parabsjn$ and $\parappjn$ are direct by \ih\
  The case left is $\pardbjn$,
  with $t_1 = \gap{\ctx\dc{\lx.t}} u y r
  \parred\djn \rsub{\rsub {u'} x {t'}} y {r'} = t_2$
  with hypothesis $\ctx\dc t \parred\djn t'$, $\ctx\dc u \parred\djn u'$
  and $\ctx\dc r \parred\djn r'$.
  By \autoref{l:djn-parred-auxabs}, there are $\dc', t''$ such that
  $\ctx\dc{\lx.t} \parred\djn \ctx{\dc'}{\lx.t''}$ and $t' = \ctx{\dc'}{t''}$.
  By \ih\ we have $\ctx\dc{\lx.t} \rewn\djn \ctx{\dc'}{\lx.t''}$,
  $u \rewn\djn u'$ and $r \rewn\djn r'$.
  We have the following reduction:
  \[ t_1 \rewn\djn \gap{\ctx{\dc'}{\lx.t''}}{u'} y {r'}
    \rew\djn \rsub{\rsub{u'} x {\ctx{\dc'}{t''}}} y {r'}
  = t_2. \]

  The proof of the third statement is also by induction on $t_1 \parred\djn t_2$.
  \begin{cas}
    \case{\paraxjn}
      Then $t_1$ is a variable.
      If $t_1 = z$, we have $\rsub{u_1} z {t_1} = u_1$,
      $\rsub{u_2} z {t_2} = u_2$ and this is direct by the second hypothesis.
      If $t_1 = y \neq z$, we have $\rsub{u_1} z {t_1} = y = \rsub{u_2} z
      {t_2}$, this is direct by \paraxjn.
      \case{\parabsjn}
      Then $t_1 = \lx.t \parred\djn \lx.t' = t_2$,
      where w.l.o.g. $x \neq z$ and $x \notin \fv{u_1} \cup \fv{u_2}$
      and such that $t \parred\djn t'$.
      By \ih\ we have $\rsub{u_1} z {t_1} = \lx.\rsub{u_1} z t
      \parred\djn \lx.\rsub{u_2} z {t'} = \rsub{u_2} z {t_2}$.
      \case{\parappjn}
      Then $t_1 = \gap t u x r \parred\djn \gap{t'}{u'} x {r'} = t_2$,
      where w.l.o.g. $x \neq z$ and $x \notin \fv{u_1} \cup \fv{u_2}$
      and such that $t \parred\djn t'$, $u \parred\djn u'$ and $r \parred\djn
      r'$.
      By \ih\ we have $\rsub{u_1} z {t_1}
      = \gap{\rsub{u_1} z t}{\rsub{u_1} z u} x {\rsub{u_1} z r}
      \parred\djn \gap{\rsub{u_1} z {t'}}{\rsub{u_1} z {u'}} x {\rsub{u_1} z
        {r'}}
      = \rsub{u_1} z {t_2}$.
      \case{\pardbjn}
      Then $t_1 = \gap{\ctx\dc{\lx.t}} u y r \parred\djn
      \rsub{\rsub{u'} x {t'}} y r = t_2$
      where w.l.o.g $x,y \neq z$ and $x,y \notin \fv{u_1} \cup \fv{u_2}$,
      $\dc$ does not capture free variables of $u_1, u_2$,
      and such that $\ctx\dc t \parred\djn t'$,
      $u \parred\djn u'$ and $r \parred\djn r'$.
      By \ih\ we have $\rsub{u_1} z {\ctx\dc t} \parred\djn \rsub{u_2} z
      {t'}$,
      $\rsub{u_1} z u \parred\djn \rsub{u_2} z {u'}$
      and $\rsub{u_1} z r \parred\djn \rsub{u_2} z {r'}$.
      Let $\rsub{u_1} z {\ctx\dc t} =
      \ctx{\dc_{\rsub{u_1} z {}}}{t_{\rsub{u_1} z {}}}$.
      By rule \pardbjn, we infer
      \begin{align*}
        \rsub{u_1} z t_1
        &= \gap{\ctx{\dc_{\rsub{u_1} z {}}}{\lx.t_{\rsub{u_1} z {}}}}{
        \rsub{u_1} z u} y {\rsub{u_1} z r}\\
        &\parred\djn \rsub{\rsub{\rsub{u_2} z {u'}} x {\rsub{u_2} z {t'}}} y
        {\rsub{u_2} z r}\\
        &= \rsub{u_2} z {t_2} &\text{(by \autoref{l:perm-rsub} twice)}
        &\qedhere
      \end{align*}
  \end{cas}
\end{proof}

Statements~(1) and~(2) of the previous lemma imply that $\rewn\djn$ is the
transitive and reflexive closure of $\parred\djn$.
We now only need to prove the \emph{diamond} property for $\parred\djn$
to conclude.
The difference between Takahashi's method and the more usual Tait and
Martin-Löfs's method \cite[\S3.2]{barendregt84} is to replace the proof of
diamond for the parallel reduction by a proof of the \emph{triangle property}.

\begin{defi}[Triangle property]
  Let $\rew\rewrel$ be a reduction relation on $\jterms$ and $f$ a function.
  $(\rew\rewrel, f)$ satisfies the triangle property if,
  for any $t \in \jterms$, $t \rew\rewrel t'$ implies $t' \rew\rewrel f(t)$.
\end{defi}

\begin{defi}[Developments]
The $\dbeta$-development $\devdjn t$  of a $\jterms$-term $t$ is defined as follows.
  \[ \begin{array}{ll}
    \devdjn x = x & \multirow{2}{*}{$
      \devdjn{\gap t u y r} =
      \begin{cases}
        \rsub{\rsub{\devdjn u} x {\devdjn{\ctx\dc{t'}}}} y {\devdjn r}, &\text{if }
        t = \ctx\dc{\lx.t'}\\
        \gap{\devdjn t}{\devdjn u} x {\devdjn r}, &\text{otherwise}
    \end{cases}$}\\
    \devdjn{\lx.t} = \lx.\devdjn t &
  \end{array} \]
\end{defi}

\begin{lem}[Triangle property of $(\parred\djn, \devdjn\cdot)$]
  Let $t_1 \parred\djn t_2$.
  Then $t_2 \parred\djn \devdjn{t_1}$.
\end{lem}
\begin{proof}
  By induction on $t_1$.
  \begin{cas}
    \case{$t_1 = x$}
      Then $t_1 = t_2 = \devdjn{t_1}$ and we conclude with rule \paraxjn.
      \case{$t_1 = \lx.t$}
      Then $t_1 \parred\djn t_2 = \lx.t'$ by rule \parabsjn.
      We have $\devdjn{t_1} = \lx.\devdjn t$.
      By \ih\ $t' \parred\djn \devdjn t$.
      By \parabsjn, $\lx.t' \parred\djn \lx.\devdjn t$.
      \case{$t_1 = \gap t u y r$, where $t \neq \ctx\dc{\lx.t'}$}
      Then $t_1 \parred\djn t_2 = \gap{t'}{u'} y {r'}$ by rule \parappjn.
      We have $\devdjn{t_1} = \gap{\devdjn t}{\devdjn u} y {\devdjn r}$.
      By \ih\ $t' \parred\djn \devdjn t$, $u' \parred\djn \devdjn u$
      and $r' \parred\djn \devdjn r$.
      By \parappjn, $\gap{t'}{u'} y {r'} \parred\djn \devdjn{t_1}$.
      \case{$t_1 = \gap {\ctx\dc{\lx.t}} u y r$}
      Then $\devdjn{t_1} =
        \rsub{\rsub{\devdjn u} x {\devdjn{\ctx\dc t}}} y {\devdjn r}$.
      There are two subcases. In both cases we have $u \parred\djn u'$ and $r
      \parred\djn r'$ and thus by \ih\ $u' \parred\djn \devdjn u$ and $r
      \parred\djn \devdjn r$.
      \begin{cas}
        \subcase{\parappjn}
          Then $\ctx\dc{\lx.t} \parred\djn t'$.
          By a reasoning similar to \autoref{l:djn-parred-auxabs},
          we can show that $t' = \ctx{\dc'}{\lx.t''}$
          and that $\ctx\dc t \parred\djn \ctx{\dc'}{t''}$.
          Thus $t_2 = \gap{\ctx{\dc'}{\lx.t''}}{u'} y {r'}$
          and by \ih\ $\ctx{\dc'}{t''} \parred\djn \devdjn{\ctx\dc t}$.
          We use rule \pardbjn\ with the three \ih\ as premises to derive
          $t_2 \parred\djn \rsub{\rsub{\devdjn u} x {\devdjn{\ctx\dc t}}} y {\devdjn
          r} = \devdjn{t_1}$.
        \subcase{\pardbjn}
          Then $t_2 = \rsub{\rsub{u'} x {t'}} y {r'}$
          and $\ctx\dc t \parred\djn t'$.
          By \ih\ $t' \parred\djn \devdjn{\ctx\dc t}$.
          By \ih\ and two applications of \autoref{l:djn-parred}(3),
          we have:
          \[
            \rsub{\rsub {u'} x {t'}} y {r'}
            \parred\djn \rsub{\rsub{\devdjn u} x {\devdjn{\ctx\dc t}}} y {\devdjn r}
            = \devdjn{t_1}
          \qedhere
          \]
      \end{cas}
  \end{cas}
\end{proof}

\begin{prop}
The reduction relation  $\rew\djn$ is confluent.
\end{prop}

\begin{proof}
  The triangle property of $(\parred\djn, \devdjn\cdot)$ implies that
  $\parred\djn$ is diamond, since for any $t_2$ such that $t_1 \parred\djn t_2$,
  $t_2 \parred\djn \devdjn{t_1}$.
  This implies in turn that $\parred\djn = \rewn\djn$ is diamond
  and thus that $\rew\djn$ is confluent.
\end{proof}

\paragraph{Properties of simply typed terms}
Let us now briefly discuss two properties related to \deft{simple typability}
for generalized applications, using the original type system of
\cite{joachimski00}, which is called here $\sysSTJ$.
Recall the following typing rules, where
$A, B, C \Coloneqq \btvone \mid A \to B$,
and $\btvone$ belongs to a set of constants.
Symbol~$\Gamma$ denotes a \deft{type environment} mapping distinct
variables to simple types.
\begin{mathpar}
  \inferrule{ }{\seq{\Gamma;x:A}{x}{A}}
  \and \inferrule{\seq{\Gamma;x:A}{t}{B}}{\seq{\Gamma}{\lx.t}{A \to B}}
  \and \inferrule{
  \seq{\Gamma}{t}{A \to B} \and \seq{\Gamma}{u}{A} \and
\seq{\Gamma;y:B}{r}{C}}{
\seq\Gamma{\gap{t}{u}{y}{r}}{C}}
\end{mathpar}
We denote the existence of a type derivation for $t$ ending in the
  sequent $\seq \Gamma t A$ in system~$\sysSTJ$ by writing
  $\derp{\Gamma}{t}{A}{\sysSTJ}$.
  We write $\demp{\Phi}{\Gamma}{t}{A}{\sysSTJ}$ to give the name $\Phi$ to such
a derivation.

\paragraph{Subformula property}

The subformula property for normal forms is an important property of proof
systems, being useful notably for proof search.
It holds for von Plato's generalized natural deduction, and therefore also for
the original calculus $\jcalc$.
Despite the minimal amount of permutations used, which does not provide full normal forms,
this property is still true in our system.
\begin{lem}[Subformula property]
  If $\demp{\Phi}{\Gamma}{\nf\djn}{A}{\sysSTJ}$ then every formula  in the
  derivation $\Phi$ is a subformula of $A$ or a subformula of some formula in
  $\Gamma$.
\end{lem}
\begin{proof}
  The lemma is proved together with another statement:
  if $\demp\Psi {\Gamma}{\nef\djn}{A}{\sysSTJ}$ then every formula in $\Psi$ is
  a subformula of some formula in $\Gamma$.
  The proof is by simultaneous induction on $\Phi$ and $\Psi$.
\end{proof}
The subformula property confirms that executing only needed permutations still
gives rise to a reasonable notion of normal form.

\paragraph{Strong normalization}

The second property for typed terms we show
states that they are
$\djncalc$-strongly normalizable.
The proof is achieved by mapping $\djncalc$ into the
$\lcalc$-calculus equipped with the following
$\sigma$-rule~\cite{regnier94}:
	\[ (\lx.M)NN'\rrule{\sigma_1}(\lx.MN')N\]
The map into the $\lcalc$-calculus, based
  on~\cite{espiritosanto07}, is given by $\mapjl{x} :=x$,
  $\mapjl{(\lx.t)}:=\lx.\mapjl{t}$, and $\mapjl{\gap t u x
    r}:=(\lx.\mapjl r)(\mapjl t \mapjl u)$.

  \begin{lem}
    \label{lem:map-into-lambda}
    (1) $\mapjl{(\rsub u x t)}=\rsub{\mapjl u}x{\mapjl t}$, and
    (2) $\mapjl{(\ctx\dc{\lx.t})}\mapjl{u}
    \rew{\beta\sigma_1}\mapjl{(\ctx\dc{\rsub u x t})}$.
\end{lem}
\begin{proof} (1) is proved by a routine induction on $t$.
  (2) is proved by induction on $\dc$.
  More precisely, in the base case, one performs a $\beta$-reduction step and uses statement (1).
  In the inductive case, one performs a $\sigma_1$-step and applies the induction hypothesis.
\end{proof}

\begin{thm}[Strong normalization]
  \label{l:sn-with-simple-types}%
  If $t$ is simply typable, \ie\ $\derp{\Gamma}{t}{\sigma}{\sysSTJ}$, then $t \in \sn{\djn}$.
\end{thm}
\begin{proof}
  Map $\mapjl{(\cdot)}$ produces the following simulation: if $t_1\rew{\djn}t_2$ then
  $\mapjl{t_1}\rew{\beta\sigma_1}^+\mapjl{t_2}$. The proof of the simulation
  result is by induction on $t_1\rew{\djn}t_2$. 
  We just show the base case, the inductive cases being easy.
\[
\begin{array}{rcll}
\mapjl{(\gap{\ctx\dc{\lx.t} }u y r)} & = & (\ly.\mapjl r)(\mapjl{(\ctx\dc{\lx.t})}\mapjl{u})&\text{(by def.~of $\mapjl{(\cdot)}$)}\\
&\rew{\beta\sigma_1}^+ & (\ly.\mapjl r)\mapjl{(\ctx\dc{\rsub u x t})} & \text{(by Lemma \ref{lem:map-into-lambda})}\\
& \rew\beta & \rsub{\mapjl{(\ctx\dc{\rsub u x t})}}y{\mapjl r}&\\
&=& \mapjl{(\rsub{\ctx\dc{\rsub u x t}}y r)} & \text{(by Lemma \ref{lem:map-into-lambda})}
\end{array}
\]

Now, given a simply typable term $t \in \jterms$, the $\lambda$-term $\mapjl
t$ is also simply typable in the \textlambda-calculus. Hence, $\mapjl t
\in\sn{\beta}$. It is well known that this is equivalent
\cite{regnier94} to $\mapjl t \in\sn{\beta,\sigma_1}$. By the
simulation result, $t\in\sn{\djn}$ follows.
\end{proof}

%% file: isn.tex
\section{Inductive Characterization of Strong Normalization}%
\label{sec:isn}

In this section we give an inductive characterization of strong
normalization (ISN) for $\djncalc$, written $\isn \dbeta$, and prove it correct.
This characterization will be useful to show completeness of the type system
that we are going to present in \autoref{s:types}, as well as to compare strong
normalization of $\djncalc$ to the ones of $\calcwithrules\jterms{\beta,\ptwo}$
and $\jcalc$ in \autoref{sec:equivalences}.

\subsection{ISN in the \texorpdfstring{$\lambda$}{lambda}-calculus with Weak-Head Contexts}%
\label{subsec:isn-lambda}

We write $\isn\rewrel$ the set of strongly normalizing terms under
$\rewrel$ given by an inductive definition.  As an
introduction, we first look at the case of ISN for the
\textlambda-calculus (written $\isn\beta$), on which our forthcoming definition
of $\isn\djn$ elaborates.  A usual way to define $\isn\beta$ is by the
following rules~\cite{raamsdonk96}, where the general notation $M \vec
P$ abbreviates $(\ldots (M P_1) \dots) P_n$ for some $n \geq 0$.
\begin{mathpar}
  \inferrule{P_1, \ldots, P_n \in \isn\beta }{x \vec P \in \isn\beta}
  \and \inferrule{M \in \isn\beta}{\lx.M \in \isn\beta}
  \and \inferrule{\rsub N x M \vec P, N \in \isn\beta }{
  (\lx.M) N \vec P \in \isn\beta}
\end{mathpar}
One then shows that $M \in  \sn\beta$ if and only if $M \in \isn \beta$.

Notice that this definition is deterministic (up to the order of the
independent evaluations of the arguments $P_1, ..., P_n$).
As such, a reduction strategy emerges from this definition: it
is a \emph{strong strategy} based on a preliminary \emph{weak-head strategy}.
The strategy is the following:
first reduce a term to a \emph{weak-head normal form} $\lx.M$ or $x \vec P$,
and then iterate reduction under abstractions and  inside arguments (in any order),
without any need to come back to the head of the term.
Formally, \deft{weak-head normal forms}, which are those produced by the first level of the strategy,  are of two kinds:
\begin{align*}
  \gramTitle{Neutral terms} &\quad \nl \Coloneqq x \mid \nl M\\
  \gramTitle{Answers} &\quad \nans \Coloneqq \lx.M
\end{align*}
Neutral terms cannot produce any head $\beta$-redex.
They are the terms of the shape $x \vec P$.
On the contrary, answers can create a $\beta$-redex when given at least one argument.
In the case of the \textlambda-calculus, these are only abstractions.
If the term is not a weak-head normal form, a redex can be located inside a
\[ \gramTitle{Weak-head context} \quad \whc\,\Coloneqq\, \ec \mid \whc t. \]

These concepts give rise to a different definition of $\isn\beta$: 
\begin{mathpar}
  \inferrule{ }{x \in \isn\beta}
  \and \inferrule{\nl, M \in \isn\beta}{\nl M \in \isn\beta}
  \and \inferrule{M \in \isn\beta}{\lx.M \in \isn\beta}
  \and \inferrule{\ctx\whc{\rsub N x M}, N \in \isn\beta}{
  \ctx\whc {(\lx.M) N} \in \isn\beta}
\end{mathpar}
Weak-head contexts are an alternative to the meta-syntactic notation
$\vec{P}$ of vectors of arguments used in the first definition of $\isn\beta$. Notice in the alternative definition that there is one rule for
each kind of neutral term, one rule for answers and one rule for terms
which are not weak-head normal forms.

\subsection{ISN for \texorpdfstring{$\dbeta$}{dbeta}}%
\label{subsec:isn-dbeta}

We now define $\isn\djn$ with the same tools used in the last subsection.
Hence, we first have to define neutral terms, answers and a special notion of context.
We call the contexts left-right contexts ($\lrc$), and the underlying strategy
the left-right strategy.
This approach gives the counterpart of the weak-head strategy for the \textlambda-calculus.
\begin{defi}
  \label{def:left-right}%
  We consider the following grammars:
  \[ \begin{array}{rllll}
    \gramTitle{Neutral terms} & \nl & \Coloneqq & x \mid \gap \nl u x \nl \\
    \gramTitle{Answers} & \nans & \Coloneqq & \lx.t \mid \gap \nl u x \nans \\
    \gramTitle{Left-right contexts} & \lrc & \Coloneqq &\ec \mid \gap \lrc u x r \mid \gap \nl u x \lrc
  \end{array} \]
\end{defi}
Notice that $\nl$ and $\nans$ are disjoint and stable by $\djn$-reduction.
Notice also that this time, answers are not only abstractions, but also
abstractions under a special distant context.
Moreover $\gap \nl u x r$ is never a $\dbeta$-redex, whereas $\gap \nans u x r$
is always a $\dbeta$-redex. 
The terminology ``left-right'' suggests that the hole
$\ec$ may appear in the left (viz $\gap \lrc u x r$) or right (viz $\gap \nl u x
\lrc$) component of a generalized application.
If this last form of $\lrc$ was forbidden, then we would define the contexts by
$\whc \Coloneqq \ec \mid \gap \whc u x r$, a generalized form of weak-head
contexts from the \textlambda-calculus, actually implicitly used in
\cite{matthes00} for $\jcalc$ (see also Fig.~\ref{fig:SN} in
\autoref{sec:equivalences}).
However, these contexts $\whc$ are not convenient for defining an inductive
predicate of strong normalization based on  the distant rule $\dbeta$, as
we argue below in \autoref{rem:weak-head}.

To achieve a characterization of $\isn \dbeta$,  we still need to obtain a
deterministic decomposition of terms, that we explain by means of an
example.

\begin{exa}[Decomposition]
  \label{e:one}
  Let $t = \gap{\gap{x_1}{x_2}{y_1}{\gap \id \id  z \id }} {x_3} y
  {\gap \id \id z z}$
  Then, there are two decompositions of $t$ in terms of a $\dbeta$-redex $r$ and
  a left-right context $\lrc$, \ie\ there are two ways to write $t$ as $\ctx\lrc
  r$: either with $\lrc=\ec$ and $r=t=\gap{\ctx\dc{\id}} {x_3} y {\gap
  \id \id z z}$, for $\dc = \gap{x_1}{x_2}{y_1}{\gap \id \id
  z \ec}$;
  or $\lrc = \gap{\gap{x_1}{x_2}{y_1}\ec} {x_3} y {\gap \id \id z z}$
  and $r=\gap \id \id z \id$.
  Notice how in the second case all the three rules in the grammar of left-right
  contexts are needed to generate $\lrc$.
\end{exa}

In the previous example, we will rule out the first decomposition  by defining next a restriction of the $\dbeta$-rule, securing uniqueness of such kind of decomposition in all cases. For that, we introduce a restricted notion of distant context:
\[ \begin{array}{rllll} \gramTitle{Neutral distant contexts} & \dcnl & \Coloneqq & \ec \mid \gap{\nl}{u}{x}{\dcnl}\end{array} \]
Notice that $\dcnl \subsetneq \lrc$;  moreover, $\ctx\dcnl{\lx.t}$ is an answer $\nans$, and conversely every answer has that form.

In the same spirit as weak-head reduction for the
\textlambda-calculus, the reduction relation underlying our definition of $\isn \dbeta$ is the \deft{left-right reduction} $\rew\whdj$,
defined as the closure under $\lrc$ of the following \deft{restricted} $\dbeta$-rule:
\[ \gap{\ctx\dcnl{\lx.t}} u y r \rrule{} \rsub{\ctx\dcnl{\rsub u y t}} y r. \]
Coherently with the \textlambda-calculus, left-right normal
forms are either neutral terms or answers.
We write $\nf\whdj$ to denote the set of $\jterms$-terms that are in $\whdj$-normal form. 

\begin{lem}
  \label{l:nfwh}%
  Let $t \in \jterms$. Then  $t$ is in $\whdj$-normal form \textit{iff} $t \in \nl \cup \nans$.
\end{lem}
\begin{proof}
  First, we show that $t$ $\whdj$-normal implies $t \in \nl \cup \nans$,
  by induction on $t$.
  If $t = x$, then $t \in \nl$. If $t = \lx.s$, then $t \in \nans$.
  Let $t = \gap s u x r$ where $s$ and $r$ are $\whdj$-normal.
  Then $s \notin \nans$, otherwise the term would $\whdj$-reduce at root.
  Thus by the \ih\ $s \in \nl$.
  By the \ih\ again $r \in \nl \cup \nans$ so that $t \in \nl \cup \nans$.

  Second, we show that $t \in \nl \cup \nans$ implies $t$ is $\whdj$-normal,
  by  simultaneous induction on $\nl$ and $\nans$.
  The cases $t = x$ (\ie\ $t \in \nl$) and $t = \lx.s$ (\ie\ $t \in \nans$)
  are straightforward.
  Let $t = \gap s u x r$ where $s \in \nl$ and $r \in \nl \cup \nans$.
  Since $r, s \in \nl \cup \nans$, by the \ih\ $t$ does not $\whdj$-reduce in $r$ or $s$.
  Since $s \in \nl$, $t$ does not $\whdj$-reduce at root either. Then, $t$ is $\whdj$-normal.
\end{proof}

The restriction of $\dc$ to a neutral distant context $\dcnl$ is what
allows determinism of our reduction relation $\rew\whdj$ (\autoref{l:wh-deterministic})  and correctness of our forthcoming definition of $\isn \dbeta$ (\autoref{d:isn}).

\begin{exa}[Decomposition]
  Going back to Example \ref{e:one}, did we obtain a decomposition $\ctx\lrc r$
  for $t$, with $r$ a restricted $\dbeta$-redex?
  The first option fails because $\dc = \gap{x_1}{x_2}{y_1}{\gap{\id}{ \id}{ z} \ec}$ is not a neutral distant context  due to the inner redex;
  and the second option succeeds because $r = \gap \id \id z \id$ is of course a
  restricted redex.
\end{exa}

\begin{lem}
  \label{l:wh-deterministic}%
  The reduction $\rew\whdj$ is deterministic.
\end{lem}
\begin{proof}
  Let $t$ be a $\whdj$-reducible term. We reason by induction on $t$.  If $t$ is a variable or an
  abstraction, then $t$ does not $\whdj$-reduce so that $t$ is
  necessarily an application $\gap{t'} u y r$. By \autoref{l:nfwh} we have three possible cases for $t'$.
  \begin{cas}
    \case{$t = \gap{t'} u y r$ with $t' \in \nans$} Then
    $t = \gap{\ctx\dcnl{\lx.s}} u y r$, so $t$ reduces at the root.
    Since $t' \in \nans$, then we know by \autoref{l:nfwh} that  (1) $t' \in \nf\whdj$, (2) $t' \notin \nl$, so that $t$ does not $\whdj$-reduce in $t'$ or $r$.
    \case{$t = \gap{t'} u y r$ with $t' \in \nl$} Then
    $t$ does not $\whdj$-reduce at the root.
    By \autoref{l:nfwh}, we know that  $t' \in \nf\whdj$ and thus $t$ necessarily reduces in $r$. By the \ih\ this reduction is deterministic.
    \case{$t = \gap{t'} u y r$ with $t' \notin \nf\whdj$} Then
    in particular  by \autoref{l:nfwh} we know that (1) $t'$ does not have an
    abstraction shape so that $t$ does not reduce at the root,
    and (2) $t' \notin \nl$ so that $t$ does not reduce in $r$.
    Thus $t$ $\whdj$-reduces only in $t'$. By the \ih\ this reduction is deterministic.
    \qedhere
  \end{cas}
\end{proof}

\begin{rem} \label{rem:weak-head} Consider again the term
  $t = \gap{\gap{x_1}{x_2}{y_1}{\gap {\gap \id \id z z} z \id}}
  {x_3} y {\gap \id \id z z}$ in \autoref{e:one}.
  As we explained before,  if the form $\gap \nl u x \lrc$ of the grammar of
  $\lrc$ was disallowed, then it would not be possible to decompose $t$ as
  $\ctx\lrc r$, with $r$ a restricted $\dbeta$-redex. Moreover, the reduction
  strategy associated with the intended definition of $\isn\djn$ would consider
  $t$ as a left-right normal form, and start reducing the subterms of $t$,
  including $\gap \id \id z \id $.
  Now, this latter (\emph{internal}) subterm would eventually reach $\id$ and
  suddenly the whole term $t' = \gap{\gap{x_1}{x_2}{y_1}{\id}} {x_3} y {r'}$
  would become an \emph{external} left-right redex: the typical separation
  between an initial external reduction phase followed by an internal reduction
  phase ---like in the \textlambda-calculus--- would
  be lost in our framework.
  This point, due to the \emph{distant} character of rule $\dbeta$, explains the
  subtlety of the upcoming \autoref{d:isn}.
\end{rem}

Our inductive definition of strong normalization follows.
\begin{defi}[Inductive strong normalization]
  \label{d:isn}
  We consider the following inductive predicate:  \begin{mathpar}
    \inferrule{ }{x \in \isn\djn} \isnvarrule
    \and \inferrule{
    \nl, u, r \in \isn\djn \\ r \in \nf\whdj}{
    \gap \nl u x r \in \isn\djn}  \isnapprule
    \and \inferrule{t \in \isn\djn}{\lambda x.t \in \isn\djn} \isnabsrule
    \and \inferrule{\ctx\lrc{\rsub {\ctx\dcnl {\rsub u x t}} y r}, \ctx\dcnl t, u \in \isn\djn}{
    \ctx\lrc{\gap{\ctx\dcnl{\lambda x.t}} u y r} \in \isn\djn} \isnbetarule
  \end{mathpar}
\end{defi}
Notice that every term can be written according to the conclusions of the
previous rules, so that the following grammar also defines the syntax $\jterms$.
\begin{equation}
  \label{eq:alt-syntax-dj}%
  t,u,r \Coloneqq x \mid \lambda x.t \mid \gap \nl u x {\nf\whdj}
  \mid \ctx\lrc{\gap{\ctx\dcnl{\lx.t}} u y r}
\end{equation}
Hence, at most one rule in the previous definition applies to each term, \ie\
the rules are deterministic.
An equivalent, but non-deterministic
definition of $\isn{\dbeta}$, can be given by removing the side
condition ``$r \in \nf\whdj$'' in rule~$\isnapprule$.  Indeed, this (weaker) rule would
overlap with rule~$\isnbetarule$ for terms in which the left-right
context lies in the last continuation, as for instance in $\gap {\gap x u y
y}{u'}{y'}{\gap \id \id z z}$.  Notice  the difference with
the \textlambda-calculus: due to the definition of left-right contexts $\lrc$, the head of a term with generalized
applications can be either on the left of the term (as in the
\textlambda-calculus), or recursively on the left in a continuation.

To show that our definition corresponds to strong normalization
(\autoref{l:cbn-sn-isn}),
we need a few intermediate statements (\autoref{l:stability-rsub} to
\autoref{l:sndbn}).

\begin{lem}
  \label{l:stability-rsub}%
  If $t_0 \rew\djn t_1$, then
  \begin{itemize}
    \item $\rsub u x {t_0} \rew\djn \rsub u x {t_1}$, and
    \item $\rsub {t_0} x u \rewn\djn \rsub {t_1} x u$.
  \end{itemize}
\end{lem}
\begin{proof}
  In the base cases, we have
  $t_0 = \gap{\ctx{\dc}{\lz.t}}{s}{y}{r} \rrule{\dbeta}
  \rsub{\rsub{s}{z}{\ctx \dc t}}{y}r = t_1$.
  By $\alpha$-equivalence we can suppose that $y,z \notin \fv u$ and $x \neq y$, $x \neq z$.
  The inductive cases and the base case for item~(2) are straightforward.
  We detail the base case of item~(1).
  \begin{align*}
    \rsub u x {t_0}
    &= \gap{\rsub u x {\ctx\dc{\lz.t}}}{\rsub u x s} y {\rsub u x r} \\
    &\rew\djn \rsub{
      \rsub{\rsub u x s} z {\rsub u x {\ctx\dc t}}
    } y {\rsub u x r} \\
    & =_{\autoref{l:perm-rsub}} \rsub{
    \rsub u x {(\rsub s z {\ctx\dc t})}}
    y {\rsub u x r} \\
    &=_{\autoref{l:perm-rsub}}
    \rsub u x {\rsub{\rsub{s}{z}{\ctx \dc t}}{y}r} \\
    &= \rsub u x {t_1}
    \qedhere
    \end{align*}
  \end{proof}

  \begin{rem}
    \label{r:snans}%
    By definition of $\dc$ contexts, for any $\jterms$-term $\ctx\dc{\lx.t} \in
    \sn\djn \iff \ctx\dc t \in \sn\djn$.
  \end{rem}

  \begin{lem}
    \label{l:sndbn}%
    Let $t_0 = \ctx\lrc{\rsub{\rsub{u}{x}{\ctx\dc t}}{y} r}, \ctx\dc t, u \in \sn\djn$.
    Then $t'_0 = \ctx\lrc{\gap{\ctx\dc{\lx.t}}{u}{y}{r}} \in \sn\djn$.
  \end{lem}
  \begin{proof}
    In this proof we use a notion of reduction of contexts which is the expected one: $\fc \rew{} \fc' $ iff
    the hole in $\fc$ is outside the redex contracted in the reduction step.
    By hypothesis we also have $r \in \sn\djn$.  We use
    the lexicographic order to reason by induction on
    $\langle \maxred[\djn]{t_0}, \maxred[\djn]{\ctx\dc t}, \maxred[\djn]{u} \rangle$.
    To show $t'_0 \in \sn\djn$ it is sufficient to show that all its
    reducts are in $\sn\djn$. We analyze all possible cases.
    \begin{cas}
      \case{$t'_0 \rew\djn t_0$} We conclude by the hypothesis.
      \case{$t'_0 \rew{\djn} \ctx\lrc{\gap{\ctx\dc{\lx.t'}}{u}{y}{r}} =t'_1$,
      where $t \rew{\djn} t'$} Thus also $\ctx\dc{t} \rew{\djn} \ctx\dc{t'}$.
      We then have $\ctx\dc{t'}\in \sn\djn$ and $u \in \sn\djn$ and by item~(2)
      $t_0= \ctx\lrc{\rsub{\rsub{u}{x}{\ctx\dc{t}}}{y} r} \rewn{\djn} \ctx\lrc{\rsub{\rsub{u}{x}{\ctx\dc{t'}}}{y} r} = t_1$,
      so that also $t_1 \in \sn\djn$.
      We can conclude that $t'_1\in \sn\djn$ by the \ih\  since
      $\maxred[\djn]{t_1}  \leq \maxred[\djn]{t_0}$ and $\maxred[\djn]{\ctx\dc{t'}} < \maxred[\djn]{\ctx\dc t}$.
      \case{$t'_0 \rew{\djn} \ctx\lrc{\gap{\ctx\dc{\lx.t}}{u'}{y}{r}} = t'_1$,
      where $u \rew{\djn} u'$}
      We have $\ctx\dc t, u' \in \sn\djn$ and by
      item~(2)
      $t_0 = \ctx\lrc{\rsub{\rsub{u}{x}{\ctx\dc{t}}}{y} r} \rewn{\djn} \ctx\lrc{\rsub{\rsub{u'}{x}{\ctx\dc{t}}}{y} r}= t_1$,
      so that also $t_1\in \sn\djn$.
      We conclude $t'_1\in \sn\djn$ by the \ih\ since
      $\maxred[\djn]{t_1}\leq  \maxred[\djn]{t_0}$
      and $\maxred[\djn]{u'} < \maxred[\djn] u$.
      \case{$t'_0 \rew{\djn} \ctx\lrc{\gap{\ctx\dc{\lx.t}}{u}{y}{r'}} = t'_1$,
      where $r \rew{\djn} r'$}
      We have $\ctx\dc t, u \in \sn\djn$ and by
      item~(1)
      $t_0 = \ctx\lrc{\rsub{\rsub{u}{x}{\ctx\dc{t}}}{y} r} \rew{\djn} \ctx\lrc{\rsub{\rsub{u}{x}{\ctx\dc{t}}}{y}{r'}} = t_1$.
      We conclude $t'_1\in \sn\djn$ by the \ih\ since since
      $\maxred[\djn]{t_1} < \maxred[\djn]{t_0}$.
      \case{$t'_0 \rew{\djn} \ctx\lrc{\gap{\ctx{\dc'}{\lx.t}}{u}{y}{r}} = t'_1$,
      where $\dc  \rew{\djn} \dc'$}
      We have $\ctx{\dc'} t, u \in \sn\djn$ and by \autoref{l:stability-rsub}
      $t_0 = \ctx\lrc{\rsub{\rsub{u}{x}{\ctx{\dc} t}}{y} r} \rewn{\djn} \ctx\lrc{\rsub{\rsub{u}{x}{\ctx{\dc'} t}}{y} r} = t_1$,
      so that also $t_1 \in \sn\djn$.
      We conclude $t'_1\in \sn\djn$ by the \ih\ since
      $\maxred[\djn]{t_1} \leq \maxred[\djn]{t_0}$ and
      $\maxred[\djn]{\ctx{\dc'} t} < \maxred[\djn]{\ctx\dc t}$.
      \case{$t'_0 \rew{\djn} \ctx{\lrc'}{\gap{\ctx\dc{\lx. t}}{u}{y}{r}} = t'_1$,
      where $\lrc \rew\djn \lrc'$}\!\!\! Thus $t_0 =
      \ctx\lrc{\rsub{\rsub u x {\ctx\dc t}} y r} \rew{\djn} \ctx{\lrc'}{\rsub{\rsub{u}{x}{\ctx\dc t}}{y} r} = t_1$.
      We have $t_1, \ctx\dc t, u \in \sn\djn$.
      We conclude that $t'_1\in \sn\djn$ by the \ih\  since 
      $\maxred[\djn]{t_1} < \maxred[\djn]{t_0}$.
      \case{$\lrc = \ctx{\lrc'}{\gap\dcnl{u'}{y'}{r'}}$
      and $r = \ctx{\dc''}{\lx'.t'}$}
      This is the only case left.
      Indeed, there is no redex in $\ctx\dc{\lx.t}$ other than in $\dc$ or
      $\lx.t$.
      Then,
      \[
        t_0' = \ctx{\lrc'}{\gap{\ctx\dcnl{\gap{\ctx\dc{\lx.t}} u y {\ctx{\dc''}{\lx'.t'}}}}{u'}{y'}{r'}}
      \]
      Let $\dc' = \ctx\dcnl{\gap{\ctx\dc{\lx.t}} u y {\dc''}}$.
      The reduction we need to consider is:
      \begin{align*}
        t_0' &= \ctx{\lrc'}{\gap{\ctx{\dc'}{\lx'.t'}}{u'}{y'}{r'}}\\
             &\rew\djn
             \ctx{\lrc'}{\rsub{\rsub{u'}{x'}{\ctx{\dc'}{t'}}}{y'}{r'}}\\
             &= \ctx{\lrc'}{\rsub{\rsub{u'}{x'}
                 {\ctx\dcnl{\gap{\ctx\dc{\lx.t}} u y {\ctx{\dc''}{t'}}}}
             }{y'}{r'}} = t_1'
           \end{align*}
           We will show that $t_1' \in \sn\djn$.

           For this we show that $t_1 =
           \ctx{\lrc'}{\rsub{\rsub{u'}{x'} {\ctx\dcnl{
               \rsub{\rsub u x {\ctx\dc t}} y {\ctx{\dc''}{t'}}}}
           }{y'}{r'}}\in \sn\djn$, that $\ctx{\dc'}{t'}\in \sn\djn$ and
           that $u'\in \sn\djn$.  We have $t_0 \rewp\djn t_1$ so that $t_1 \in
           \sn\djn$ and $\maxred[\djn]{t_1} < \maxred[\djn]{t_0}$.  $u'$ is a
           subterm of $t_0$, which is in $\sn\djn$, so that $u' \in
           \sn\djn$.  To show that $\ctx{\dc'}{t'} \in \sn\djn$, we
           consider $t_2 = \ctx{\dcnl}{\rsub {\rsub u x {\ctx\dc t}}
           y {\ctx{\dc''}{\lx'.t'}}}$.  We have $t_0 =
           \ctx{\lrc'}{\gap{t_2}{u'}{y'}{r'}}$.  We can show that
           $\maxred[\djn]{t_2} < \maxred[\djn]{t_0}$ (so that $t_2 \in \sn\djn$).
           Indeed, $\maxred[\djn]{\ctx{\lrc'}{\gap{t_2}{u'}{y'}{r'}}}
           \geq \maxred[\djn]{\gap{t_2}{u'}{y'}{r'}} \geq \maxred[\djn]{t_2} + 1 >
           \maxred[\djn]{t_2}$.  The second inequality holds since $t_2$ has an
           abstraction shape, and abstraction shapes are stable under
           substitution, and thus $\gap{t_2}{u'}{y'}{r'}$ is also a redex.
           We can then conclude that $t'_2=\ctx\dcnl{\gap{\ctx\dc{\lx.t}} u
           y {\ctx{\dc''}{\lx'.t'}}} = \ctx{\dc'}{\lx'.t'} \in
           \sn\djn$ by the \ih\ since $u, \ctx\dc t \in \sn\djn$.
           Thus, $\ctx{\dc'}{t'} \in \sn\djn$ by \autoref{r:snans}.

           We then have $t_1, \ctx{\dc'}{t'}, u' \in \sn\djn$
           and we can conclude $t'_1 \in \sn\djn$ since
           $\maxred[\djn]{t_1} < \maxred[\djn]{t_0}$. We conclude $t'_1 \in \sn\djn$ as required.
           \qedhere
         \end{cas}
       \end{proof}

       \begin{thm}
         \label{l:cbn-sn-isn}%
         $\sn\djn = \isn\djn$.
       \end{thm}
       \begin{proof}
         First, we show $\isn\djn \subseteq \sn\djn$.
         We proceed  by induction on $t \in \isn\djn$.
         \begin{cas}
           \case{$t = x$} Straightforward.
           \case{$t = \lx.s$, where $s \in \isn\djn$}
           By the \ih\ $s \in \sn\djn$, so that $t \in \sn\djn$
           trivially holds.
           \case{$t = \gap s u x r$ where $s, u, r \in \isn\djn$, $s \in \nl$
           and $r \in \nf\whdj$}
           Since $s \in \nl$,
           in particular $s$ is not an answer and can not
           $\djn$-reduce to one.
           Therefore any kind of reduction starting at $t$ only occurs in the subterms $s$, $u$ and $r$.
           We conclude since by the \ih\ we have $s, u, r \in \sn\djn$.
           \case{$t = \ctx\lrc{\gap{\ctx\dcnl{\lx.s}} u y r}$, where $\ctx\lrc{\rsub{\rsub u x {\ctx\dcnl s}} y r},
           \ctx\dcnl s, u \in \isn\djn$}
           The \ih\ gives $\ctx\lrc{\rsub{\rsub u x {\ctx\dcnl s}} y r} \in \sn\djn$,
           $\ctx\dcnl s \in \sn\djn$ and  $u \in \sn\djn$ so that by \autoref{l:sndbn}
           $t = \ctx\lrc{\gap{\ctx\dcnl{\lx.s}} u y r} \in \sn\djn$ holds, with $\dc = \dcnl$.
         \end{cas}
         Next, we show $\sn\djn \subseteq \isn\djn$.
         Let $t \in \sn\djn$. We reason by induction on $\tuple{\maxred[\djn] t, \tmsz t}$
         w.r.t. the lexicographic order.
         If $\tuple{\maxred[\djn] t, \tmsz t}$ is minimal, \ie\ $\tuple{0, 1}$,
         then $t$ is a variable and thus in $\isn\djn$ by rule~$\isnvarrule$. Otherwise we proceed by case analysis.
         \begin{cas}
           \case{$t = \lx.s$} Since $\maxred[\djn] s \leq  \maxred[\djn] t$
           and $\tmsz s < \tmsz t$, we conclude by the \ih\ and rule~$\isnabsrule$.
           \case{$t$ is an application}
           There are two cases.
           \begin{cas}
             \subcase{$t \in \nf\whdj$}
             Then $t = \gap s u x r \in \sn\djn$ implies  $s, u, r \in \sn\djn$. Moreover, $t \in \nf\whdj$ implies   $s \in \nl$
               and $r \in  \nf\whdj$.
             We have $\maxred[\djn] s \leq \maxred[\djn] t$,
             $\maxred[\djn] u \leq \maxred[\djn] t$,
             $\maxred[\djn] r \leq \maxred[\djn]  t$,
             $|s| < |t|$, $|u| < |t|$ and $|r| < |t|$.
             By the \ih\ $s, u,r \in \isn\djn$,
             and since $r \in \nf\whdj$ then we conclude
             $t \in \isn\djn$ by rule~$\isnapprule$.
             
             \subcase{$t \notin \nf\whdj$}
             By definition there is a context $\lrc$
             s.t.  $t = \ctx\lrc{\gap{\ctx\dcnl{\lx.s}} u y r}$.
             Moreover, $t \in \sn\djn$ implies in particular
             $\ctx\lrc{\rsub{\rsub u x {\ctx\dcnl s}} y r},
             u \in \sn\djn$,
             so that they are in $\isn\djn$ by the \ih\
             Moreover, $t \in \sn\djn$ also implies $\ctx\dcnl{\lx.s} \in \sn\djn$.
             Since the abstraction $\lx.s$ is never applied nor an argument,
             this is equivalent to $\ctx\dcnl s \in \sn\djn$, thus $\ctx\dcnl s \in \isn\djn$ by the \ih\ 
             We conclude by rule~$\isnbetarule$.
             \qedhere
           \end{cas}
         \end{cas}
       \end{proof}

%% file: types_strong.tex
\section{Quantitative Types Capture Strong Normalization}%
\label{sec:types}

We proved in \autoref{sec:some-properties} that simply typed terms
are strongly normalizing.
In this section we use non-idempotent intersection types to fully characterize
strong normalization, so that not only typable terms are strongly normalizing,
but  also strongly normalizing terms are typable.
First we introduce the typing system, next we prove the characterization, and
finally we study the quantitative behavior of the permutative rule $\pi$ by
giving in particular an example of failure of type preservation along $\pi$.
%
\subsection{The Typing System}
\label{s:types}

We  define the  quantitative type system $\sysFullJ$ for $\jterms$-terms and
we show that strong normalization in $\djncalc$ exactly corresponds to
$\sysFullJ$-typability.

Given a countable infinite set $\TypeVariable$ of constants $\btvone, \btvtwo,
\btvthree, \dots$, we define the following sets of types and multiset types:
\[ \begin{array}{rcll}
  \gramTitle{Types} & \sigma, \tau, \rho & \Coloneqq &  \btvone\mid \MM \to \sigma \\
  \gramTitle{Multiset types} & \MM, \MN  & \Coloneqq & \multii{\sigma_i}
  \text{ where $I$ is a finite set}
\end{array} \]
The empty multiset is denoted $\emult$.
We use $\msetsz\MM$ to denote the size of the multiset, thus if $\MM =
\multii{\sigma_i}$ then $\msetsz\MM = |I|$.
We introduce a \deft{choice operator} on multiset types: if $\MM \neq \emult$,
then $\choice\MM = \MM$, otherwise $\choice\emult = \mult\sigma$, where
$\sigma$ is an  arbitrary type.
This operator will be used to guarantee that there is always a typing witness for all
the subterms of typed terms.

\deft{Typing environments} (or just \deft{environments}), written $\Gamma,
\Delta, \Lambda$, are functions from variables to multiset types assigning the
empty multiset to all but a finite set of variables.
Typing environments will be written $x_1:\MM_1; \ldots; x_n:\MM_n$. 
For instance, $\Gamma_1 \eqdef x:\mult{\sigma_1,\sigma_2};
  y:\mult{\tau}$ is a typing environment, which can also
  can be written as  $x:\mult{\sigma_1,\sigma_2};y:\mult\tau; z:\emult$,
by explicitely indicating the empty multiset for some particular variable.
The domain of an environment $\Gamma$ is given by $\dom\Gamma \eqdef \{x
\mid \Gamma(x) \neq \emult\}$.
The \deft{union of environments}, written $\Gamma \inter \Delta$, is defined by
$(\Gamma \inter \Delta)(x) :=  \Gamma(x) \sqcup \Delta(x)$, where $\sqcup$
denotes multiset union.
For instance, if $\Gamma_2 \eqdef y:\mult\tau$, then
$\Gamma_1 \inter \Gamma_2 = x:\mult{\sigma_1,\sigma_2}; y:\mult{\tau,\tau}$.
This notion is extended to several environments as expected, so that
$\inter_{\iI} \Gamma_i$ denotes a finite union of environments
($\inter_{\iI}\Gamma_i$ is to be understood as the empty environment when $I =
\emptyset$).
We write $\cmin{\Gamma}{x}$ for the environment such that $(\cmin{\Gamma}{x})(y)
= \Gamma(y)$ if $y \neq x$ and $(\cmin\Gamma x)(x) = \emult$.
For instance, $\cmin{(\Gamma_1 \inter \Gamma_2)} x =
y:\mult{\tau,\tau}$.
We write $\Gamma; \Delta$ for $\Gamma \inter \Delta$ when
$\dom{\Gamma} \cap \dom{\Delta} = \emptyset$.
A \deft{sequent} has the form $\seq \Gamma t \sigma$ or $\seq \Gamma t \MM$, where $\Gamma$ is an
environment, $t$ is a term, $\sigma$ is a type and $\MM$ a multiset type.

The type system $\sysFullJ$ is given by the following typing rules.
\begin{mathpar}
  \inferrule*[right=\ruleAxJ]{ }{\seq{x:\mult\sigma} x \sigma}
  \and \inferrule*[right=\ruleAbsJ]{\seq{\Gamma; x:\MM} t \sigma}{
  \seq \Gamma {\lx.t}{\MM \to \sigma}}
  \and \inferrule*[right=\many]{\left( \seq{\Gamma_i}{t}{\sigma_i}\right)_{\iI} \and I \neq
  \emptyset}{
  \seq{\inter_{\iI} \Gamma_i}{t}{\multii{\sigma_i}}}
  \and \inferrule*[right=\ruleAppJ]{
    \seq{\Gamma}{t}{\choice{\mult{{\MM_i} \to {\tau_i}}_{\iI}}}
    \and \seq{\Delta}{u}{\choice{\multunion_{\iI} \MM_i}}
  \and \seq{\Lambda; x:\mult{\tau_i}_{\iI}}{r}{\sigma}}{
\seq{\Gamma \inter \Delta \inter \Lambda}{\gap t u x r}{\sigma}}
\end{mathpar}

We write $\derp \Gamma t \sigma \sysFullJ$ for the existence of
  a type derivation ending in the sequent $\seq \Gamma t \sigma$ in
  system~$\sysFullJ$.
  The $\sysFullJ$ annotation can be omitted when it is clear from the context.
  The \deft{size} of a type derivation is given by the number of its typing
  rules distinct from~$\many$.
  We use the notation $\demnp \Phi \Gamma t \sigma n \sysFullJ$ to call $\Phi$ a
derivation of size $n$, and  the annotation for the size $n$ is optional.

The typing system handles sequents assigning a type $\sigma$ or a  multiset
$\multii{\sigma_i}$, with $I\neq\emptyset$.
According to the rule $\many$, the latter kind of sequents should be understood
as a shorthand for a set of sequents of the former kind.
Still, the case $I=\emptyset$ is possible in rule $\ruleAppJ$, this is precisely
when the subtle use of the choice operator is required. 
Indeed, if $I$ is empty in $\ruleAppJ$, meaning in particular that
$x$ is assigned the empty multiset $\emult$ in the typing environment of the
third premise, then the multisets $\mult{{\MM_i} \to {\tau_i}}_{\iI}$ and
$\multunion_{\iI} \MM_i$ are also both empty.
Therefore, the choice operator must be used to type both terms $t$ and $u$,
which cannot be assigned the empty multiset type.
In this case, the resulting types $\choice{\mult{{\MM_i} \to {\tau_i}}_{\iI}}$
and $\choice{\multunion_{\iI} \MM_i}$ are non-empty multiset types, but they are
not necessarily related (\cf\ forthcoming example).
If $I$ is not empty, then the multiset typing $t$ is non-empty as well,
however, the choice operator
is needed  to type $u$ if $\multunion_{\iI} \MM_i$ is empty, \eg\ if $\mult{\emult \to
\sigma}$ types the term $t$.

\begin{exa}
  Let $\rho_i \eqdef \mult{{\mult {\sigma_i}} \to \sigma_i, \sigma_i} \to \sigma_i$
  and $\tau_i \eqdef \mult {\sigma_i} \to \sigma_i$, for $i=1,2$.
  The term $\gap \delta \delta x z$ can be typed with the following
  derivation, in the environment $z:\mult\tau$ (different $i$'s can be chosen to
  emphasize that $\sigma_1$ and $\sigma_2$ as well as $\rho_1$ and $\rho_2$ are
  unrelated):
  \[
    \inferrule*[right=\ruleAppJ]{
      \inferrule*[right=\many]{
      \seq{\emptyset} \delta { \rho_1}}{
    \seq{\emptyset} \delta { \mult{ \rho_1}}}
    \and \inferrule*[right=\many]{
    \seq{\emptyset} \delta { \rho_2}}{
  \seq{\emptyset} \delta { \mult{ \rho_2}}}
\and \inferrule*[Right=\ruleAxJ]{ }{\seq{z:\mult \tau; x: \emult}{z}{\tau}}}{
\seq{z:\mult \tau}{\gap \delta  \delta {x}{z}}{\tau}} \]
where the term $\delta$ is typed with $\rho_i$ as follows:
\[ \inferrule*[Right=\ruleAppJ]
  {\inferrule*[right=\many]
    {\inferrule*[Right=\ruleAxJ]{ }
    {\seq{y:\mult{\tau_i}}{y}{\tau_i}}}
    {\seq{y:\mult{\tau_i}}{y}{ \mult{\tau_i}}}
    \and \inferrule*[right=\many]
    {\inferrule*[Right=\ruleAxJ]{ }{\seq{y: \mult{\sigma_i}}{y}{ \sigma_i }}}
    {\seq{y: \mult{\sigma_i}}{y}{\mult{\sigma_i}}}
  \and \inferrule*[Right=\ruleAxJ]{ }{\seq{w:\mult {\sigma_i}}{w}{\sigma_i}}}
  {\inferrule*[Right=\ruleAbsJ]
    {\seq{y:\mult{{\mult {\sigma_i}} \to {\sigma_i}, \sigma_i}}{\gap{y}{y}{w}{w}}{\sigma_i}}
  {\seq{\emptyset}{\ly. \gap{y}{y}{w}{w}}{{\mult{{\mult{ \sigma_i}} \to {\sigma_i}, \sigma_i}} \to {\sigma_i}}}}
\]
Since $x$ does not appear in the subterm $z$ of $\gap \delta \delta x z$,
it is assigned the empty multiset $\emult$ in $\seq{z:\mult\tau; x:\emult} z
\tau$ on the premiss of rule~$\ruleAppJ$ (because the system has no weakening).
Thus, on the other two premiss of the application, we do not ask for a
derivation $\der \emptyset \delta \emult$, that is not valid in our system, but
we require the existence of a witness type for each: $\rho_1$ and $\rho_2$.
This choice operator, as well as the side-condition on $\many$ is what forces
every subterms to be typed.
\end{exa}

The two following technical lemmas will be useful for the forthcoming proofs.
First, system $\sysFullJ$ lacks weakening: it is \emph{relevant}.
\begin{lem}[Relevance]
  \label{l:relevance-sysFullJ}%
  If $\der{\Gamma}{t}{\sigma}$, then
  $\fv t = \dom \Gamma$.
\end{lem}
\begin{proof}
  By straightforward induction on $\der{\Gamma}{t}{\sigma}$. 
  \end{proof}

\begin{lem}[Split]
  \label{l:split-many-strong}\mbox{}
  \begin{itemize}
    \item If $\dern \Gamma {t}{\MM} n$, then for any decomposition
      $\MM = \sqcup_{\iI} \MM_i$ where $\MM_i \neq \emptyset$ for all $\iI$, then we have
      $\dern{\Gamma_i}{t}{\MM_i}{n_i}$ such that
      $\sum_{\iI} n_i = n$ and $\inter_{\iI}\Gamma_i = \Gamma $.
    \item If $\dern{\Gamma_i}{t}{\MM_i}{n_i}$ for all $\iI$
      and $I \neq \emptyset$,
      then $\dern \Gamma {t}{\MM} n$,
      where 
      $\MM = \sqcup_{\iI} \MM_i$, 
      $n= \sum_{\iI} n_i$ and $\Gamma= \inter_{\iI}\Gamma_i$.
  \end{itemize}
\end{lem}
\begin{proof}
  Straightforward by induction on the derivations.
\end{proof}

\subsection{Characterization of Strong
\texorpdfstring{$\djn$}{dbeta}-Normalization by Typing}

We start by proving soundness ($\dbeta$-normalization of typed terms).
  As it is usual with quantitative types, soundness relies on quantitative subject reduction, stating that
typing is preserved during reduction, but also that a step of reduction
decreases the size of the type derivation.
In general, this gives rises to simple combinatorial proofs of soundness.

However, by nature, subject reduction in quantitative type systems for
strong normalization does not hold.  Indeed, all subterms are typed,
even the ones that will be erased, but in most cases, these subterms
have free variables, that are typed in the corresponding environment.
Therefore,  when the term is erased, some parts of the
environment are lost, which means that typing is not preserved by
reduction steps (remember that every free variable is typed in
  the environment by relevance).
\begin{exa}
  \label{ex:erasing-reduction}%
  Let $t = \gap {(\lx.\id)} y z z \rew\dbeta \id$ where $\id = \lx.x$.
  The term $t$ can be typed with the derivation below, with the environment
  $y:\mult\sigma$.
  However, by relevance, the reduct $\id$ can only be typed with an empty
  environment, since $\id$ has no free variables, so that a proof
  $\der{y:\mult\sigma}{\id}{\mult\tau \to \tau}$ does not exist.
  \[
    \inferrule*{
      \inferrule*{
        \inferrule*
        {\inferrule*
          {\inferrule*{ }{\seq{x:\mult\tau} x \tau}}
          {\seq {} {\id} {\mult\tau \to \tau}}
      }{\seq {} {\lx.\id} {\emult \to \mult\tau \to \tau}}}
      {\seq {} {\lx.\id} {\mult{\emult \to \mult\tau \to \tau}}}
      \and \inferrule*
      {\inferrule*{ }{\seq{y:\mult\sigma} y \sigma}
      }{\seq{y:\mult\sigma} y \emult}
      \and \inferrule*{ }{\seq{z:\mult{\mult\tau \to \tau}} z {\mult\tau \to \tau}}
    }{\seq {y:\mult\sigma}{\gap {(\lx.\id)} y z z} {\mult\tau \to \tau}}
  \]
\end{exa}

We thus prove subject reduction only for \emph{non-erasing} steps
(\autoref{l:erasing-sr}).
This proof relies on a substitution lemma
(\ref{l:substitution-lemma-right-strong}) and an auxiliary lemma
(\ref{l:perml-typ}).
\begin{defi}[Erasing step]
  A reduction step $t_1 \rew\djn t_2$ is said to be \deft{erasing} iff the
  reduced $\dbeta$-redex in ${t_1}$ is of the form
  $\gap{\ctx\dc{\lx.t}} u y r$
  with $x \notin \fv t$ or $y \notin \fv r$.
\end{defi}

\begin{lem}[Substitution lemma]
  \label{l:substitution-lemma-right-strong}%
  Let $t, u \in \jterms$ with $x \in \fv t$.
  If both $\dern{\Gamma; x:\MM}{t}{\sigma}{n}$ and
  $\dern{\Delta}{u}{\MM}{m}$ hold,
  then $\dern{\Gamma \inter \Delta}{\rsub{u}{x}t}{\sigma}{k}$
  where $k = n + m - \msetsz\MM$.
\end{lem}
\begin{proof}
  By induction on the type derivation of $t$.
  We extend the statement to derivations ending with $\many$, for which the property is straightforward by the \ih
  \begin{cas}
    \case{$t = x$} Then $n = 1$ and by hypothesis $\Gamma = \emptyset$ and $\MM = \mult{\sigma}$
      (so that $\msetsz \MM = 1$). Moreover, $\dernp{\Delta}{u}{\MM}{m}{\sysFullJ}$
      necessarily comes from $\dernp{\Delta}{u}{\sigma}{m}{\sysFullJ}$
      by rule $\many$.
      Let $k=m$, then we conclude
      $\dern{\emptyset \inter \Delta}{u}{\sigma}{1+m-1}=
      \dern{\Gamma \inter \Delta}{\rsub{u}{x}x}{\sigma}{k}$.
      \case{$t = \ly.s$ where $y \neq x$ and $y \notin \fv{u}$}
      By definition we have $\sigma = {\MN} \to {\tau}$
      and $\dern{\Gamma;x:\MM;y:\MN}{s}{\tau}{n-1}$.

      By the \ih\ $\dern{(\Gamma; y:\MN) \inter \Delta}{\rsub u x
      s}{\tau}{k'}$ with $k' = n-1+m - \msetsz\MM$. By the relevance
      \autoref{l:relevance-sysFullJ} $y \notin \dom{\Delta}$
      so that $(\Gamma; y:\MN) \inter \Delta= \Gamma \inter \Delta; y:\MN$. By rule~$\ruleAbsJ$ we obtain $\dern{\Gamma \inter \Delta}{{\ly.
      \rsub u x s}}{{\MN} \to {\tau}}{k'+1}$.
      Let $k = k'+1$. We conclude
      because $\ly.\rsub u x s = \rsub u x {(\ly.s)}$ and $k = k'+1 = n+m -\msetsz \MM$.
      \case{$t = \gap s o y r$,  where $y \neq x$ and $y \notin \fv u$}
      We only detail the case where $x \in \fv{s} \cap \fv{o} \cap \fv{r}$, the other cases being similar. By definition we have
      $\dern{\Gamma_1; x:\MM_1}{s}{\choice{\mult{{\MN_i} \to {\tau_i}}_{\iI}}}{n_1}$,
      $\dern{\Gamma_2; x:\MM_2}{o}{ \choice{\multunion_{\iI} \MN_i}}{n_2}$ and
      $\dern{\Gamma_3; x:\MM_3; y:\mult{\tau_i}_{\iI}}{r}{\sigma}{n_3}$
      where
      $\Gamma = \Gamma_1 \inter \Gamma_2 \inter \Gamma_3$, $\MM = \MM_1 \sqcup \MM_2 \sqcup \MM_3$,
      and $n = 1+n_1+n_2+n_3$.
      Moreover, by \autoref{l:split-many-strong}
      we have $\dern{\Delta_1}{u}{\MM_1}{m_1}$, $\dern{\Delta_2}{u}{\MM_2}{m_2}$
      and $\dern{\Delta_3}{u}{\MM_3}{m_3}$
      where $\Delta = \Delta_1 \inter \Delta_2 \inter \Delta_3$ and $m=m_1+m_2+m_3$.
      The \ih\ gives
      $\dern{\Gamma_1 \inter \Delta_1}{\rsub u x s}{\choice{\mult{{\MN_i} \to {\tau_i}}_{\iI}}}{k_1}$,
      $\dern{\Gamma_2 \inter \Delta_2}{\rsub u x o}{ \choice{\multunion_{\iI} \MN_i}}{k_2}$ and
      $\dern{\Gamma_3 \inter \Delta_3;y:\mult{\tau_i}_{\iI}}{\rsub u x r}{\sigma}{k_3}$, where $k_i = n_i + m_i - \msetsz{\MM_i}$ for $i=1,2,3$.
      Then we have a derivation $\dern{\Gamma_1 \inter \Delta_1 \inter \Gamma_2
        \inter \Delta_2 \inter \Gamma_3 \inter \Delta_3}{\gap{\rsub u x s}{\rsub u
        x o} y {\rsub u x r}}{\sigma}{k}$
        where $k = 1 +_{i=1,2,3} k_i$.
        We conclude since $\Gamma \inter \Delta = \Gamma_1 \inter \Delta_1 \inter \Gamma_2 \inter \Delta_2 \inter \Gamma_3 \inter \Delta_3$,
        $\rsub u x {\gap s o y r} = \gap{\rsub u x s}{\rsub u x o} y {\rsub u x r}$ and
        $k= 1 +_{i = 1,2,3} k_i = 1 +_{i = 1,2,3} (n_i+m_i - \msetsz{\MM_i}) = n+m-\msetsz\MM$.
        \qedhere
    \end{cas}
  \end{proof}

  \begin{lem}
    \label{l:perml-typ}%
    Let $t\in\jterms$, and $\dc$ a list context.
    Then  $\dernp \Gamma {\ctx\dc{\lx.t}}{\sigma} n \sysFullJ$
    if and only if 
    $\dernp \Gamma {\lx.\ctx\dc t}{\sigma} n \sysFullJ$.
  \end{lem}
  \begin{proof}
    Both implications are proved by induction on $\dc$. The base case $\dc = \ec$ is trivial.
    Notice that we always have $\sigma = {\MN} \to {\rho}$.
    Let consider the inductive case $\dc = \gap s u y {\dc'}$.

    We first consider the left-to-right implication.
    So that let $\dern \Gamma {\ctx\dc{\lx.t}}{\sigma} n$.
    We have the following derivation, with 
    $n = k + l + m + 1$.
    \[ \inferrule*{
        \dern{\Pi}{s}{\choice{\mult{{\MM_i} \to {\tau_i}}_{\iI}}}{k}
        \and \dern{\Delta}{u}{\choice{\multunion_{\iI} \MM_i}}{l}
        \and \dern{\Lambda;y:\mult{\tau_i}_{\iI}}{\ctx{\dc'}{\lx. t}}{\sigma}{m}
      }{
        \seq{\Pi \inter \Delta \inter \Lambda}
        {\gap s u y {\ctx{\dc'}{\lx.t}}}{\sigma}
    } \]
    The \ih\ gives a derivation
    $\dern{\Lambda;y:\mult{\tau_i}_{\iI}}{\lx.\ctx{\dc'} t}{\sigma}{m}$
    and thus a derivation
    $\dern{\Lambda;y:\mult{\tau_i}_{\iI};x:\MN}{\ctx{\dc'} t}{\rho}{m-1}$.
    By $\alpha$-conversion, $y \notin \fv s \cup \fv u$,
    so that $y \notin \dom{\Pi \inter \Delta}$
    by \autoref{l:relevance-sysFullJ}.
    We can then build the following derivation of the same size:
    \[ \inferrule*{
        \dern{\Pi}{s}{\choice{\mult{{\MM_i} \to {\tau_i}}_{\iI}}}{k}
        \and \dern{\Delta}{u}{\choice{\multunion_{\iI}\MM_i}}{l}
      \and \dern{\Lambda;y:\mult{\tau_i}_{\iI};x:\MN}{\ctx{\dc'} t}{\rho}{m-1}}{
      \inferrule*{
        \seq{\Pi \inter \Delta \inter (\Lambda;x:\MN)}
      {\gap s u y {\ctx{\dc'} t}}{\rho}}{
      \seq{\Pi \inter \Delta \inter \Lambda}
{\lx. \gap s u y {\ctx{\dc'} t}}{\sigma}}} \]
For the right-to-left implication, we build the first derivations from the second similarly to the previous case.
\end{proof}

\begin{lem}[Non-erasing subject reduction]
  \label{l:qsr-strong}%
  Let $\dernp{\Gamma}{t_1}{\sigma}{n_1}{\sysFullJ}$. If $t_1 \rew\djn
  t_2$ is a non-erasing step, then $\dernp{\Gamma}{t_2}{\sigma}{n_2}{\sysFullJ}$ with $n_1 > n_2$.
\end{lem}
\begin{proof}
  By induction on $t_1 \rew{} t_2$.
  \begin{cas}
    \case{$t_1 = \gap{\ctx\dcnl{\lx.t}}{u}{y}{r} \rrule{\beta}
    \rsub{\ctx\dcnl{\rsub u x t}}{y}r=t_2$}
      Because the step is non-erasing, the types of $y$ and $x$ are not empty by
      \autoref{l:relevance-sysFullJ}, so that
      we have the following derivation, with $\Gamma = \inter_{\iI} \Sigma_i \inter_{\iI} \Delta_i \inter \Lambda$,
      $n_1 = \sum_{\iI} (n_t^i + 1 + n_u^i) + n_r + 1$ and $I \neq \emptyset$.
      \[ \inferrule*{
          \inferrule*{
          \left(\dern{\Sigma_i}{\ctx\dcnl{\lx.t}}{{\MM_i} \to {\tau_i}}{n_{\lambda}^i}\right)_{\iI}}{
        \seq{\inter_{\iI} \Sigma_i}{\ctx\dcnl{\lx.t}}{\mult{{\MM_i} \to
      {\tau_i}}_{\iI}}} \and\mkern-18mu
      \dern{\inter_{\iI} \Delta_i}{u}{\multunion_{\iI} \MM_i}{n_u} \and\mkern-18mu
      \dern{\Lambda; y:\mult{\tau_i}_{\iI}}{r}{\sigma}{n_r}}{
      \seq{\inter_{\iI} \Sigma_i \inter_{\iI} \Delta_i \inter
  \Lambda}{\gap{\ctx\dcnl{\lx.t}}{u}{y}{r}}{\sigma}} \]

  For each $\iI$, we use \autoref{l:split-many-strong} to retrieve
    derivations $\dern{\Delta_i}{u}{\MM_i}{n_u^i}$ such that
    $n_u = \sum_{\iI} n_u^i$.
  Furthermore,
  \autoref{l:perml-typ} gives a derivation
  $\dern{\Sigma_i}{\lx.\ctx\dcnl t }{{\MM_i} \to {\tau_i}}{n_{\lambda}^i}$
  and therefore we have a derivation
  $\dern{\Sigma_i;x:\MM_i}{\ctx\dcnl t }{\tau_i}{n_t^i}$ where $n_t^i = n_{\lambda}^i -1$.
  Moreover, the substitution \autoref{l:substitution-lemma-right-strong} gives
  $\dern{\Sigma_i \inter \Delta_i}{\rsub u x {\ctx\dcnl t}}{\tau_i}{k_i}$, where $k_i = n_t^i + n_u^i -\msetsz{\MM_i}$,
  so that we have a derivation
  $\dern{\inter_{\iI} \Sigma_i \inter_{\iI} \Delta_i}{\rsub u x
  {\ctx\dcnl t}}{\mult{\tau_i}_{\iI}}{+_{\iI} k_i}$.
  Applying the substitution \autoref{l:substitution-lemma-right-strong}
  again gives $\dern\Gamma{t_2 = \rsub{\rsub u x {\ctx\dcnl t}} y r}{\sigma}{n_2}$
  with $n_2 = n_r +_{\iI} k_i < n_1$.

  \case{$t_1 =\lx.t \rew{} \lx.t' = t_2$, where $t \rew{} t'$}
  By hypothesis, we have $\sigma = {\MM} \to {\tau}$
  and $\dern{\Gamma;x:\MM}{t}{\sigma}{n_1 -1}$.
  By the \ih\ we have $\dern{\Gamma;x:\MM}{t'}{\tau}{k}$
  for $n_1 -1 > k$.
  We can build a derivation of size $n_2=k+1$
  and we get $n_1 > n_2$. 

  \case{$t_1 = \gap t u x r$ and the reduction is internal}
  By hypothesis, we have the derivations
  $\dern{\Sigma}{t}{\choice{\mult{{\MM_i} \to {\tau_i}}_{\iI}}}{n_t}$,
  $\dern\Delta{u}{\choice{\multunion_{\iI} \MM_i}}{n_u}$ and
  $\dern{\Lambda; x:\mult{\tau_i}_{\iI}}{r}{\sigma}{n_r}$
  with $\Gamma = \Sigma \inter \Delta \inter \Lambda$ and
  $n_1 = 1 + n_t + n_u + n_r$.
  \begin{cas}
    \subcase{$t_1 \rew{} \gap {t'} u x r = t_2$, where $t \rew{} t'$}
      If $I \neq \emptyset$, we have $\Sigma = \inter_{\iI} \Sigma_i$, $n_t = \sum_{\iI} n_t^i$
      and derivations $\dern{\Sigma_i}{t}{{\MM_i} \to {\tau_i}}{n_t^i}$.
      If $I = \emptyset$, we have $\choice{\mult{{\MM_i} \to {\tau_i}}_{\iI}} = \mult{\tau}$
      and a derivation $\dern{\Sigma}{t}{\tau}{n_t}$.
      In both cases, we apply the \ih\ and derive
      $\dern{\Sigma}{t'}{\choice{\mult{{\MM_i} \to {\tau_i}}_{\iI}}}{k}$
      with $k < n_t$.
      We can build a derivation of size $n_2 = 1 + k + n_u + n_r$ and we get
      $n_1 > n_2$. 
      \subcase{$t_1 \rew{} \gap t {u'} x r = t_2$, where $u \rew{} u'$}
      Let $\choice{\multunion_{\iI} \MM_i} = \mult{\rho_j}_{\jJ}$.
      In particular, if $\multunion_{\iI} \MM_i = \emult$, then $J$ is a
      singleton.
      We have $\Delta = \inter_{\jJ} \Delta_j$, $n_u = \sum_{\jJ} n_u^j$
      and derivations $\dern{\Delta_j}{u}{\rho_j}{n_u^j}$.
      We apply the \ih\ and derive
      $\dern{\Delta}{u}{\choice{\multunion_{\iI} \MM_i}}{k}$
      with $k < n_u$.
      We can build a derivation of size $n_2 = 1 + n_t + k + n_r$ and we get
      $n_1 > n_2$.
      \subcase{$t_1 \rew{} \gap t u x {r'} = t_2$, where $r \rew{} r'$}
      By the \ih\ we have
      $\dern{\Lambda;x:\mult{\tau_i}_{\iI}}{r}{\sigma}{k}$
      with $k < n_r$.
      We can build a derivation of size $n_2 = 1 + n_t + n_u + k$ and we get
      $n_1 > n_2$.
      \qedhere
  \end{cas}
\end{cas}
\end{proof}

Although subject reduction does not always hold, the characterization
of normalizable terms as typable does.  To prove this, we need a
weaker form of subject reduction: the fact that the right-hand term of
an erasing reduction is still typed.  An important point is
  that the size of the type derivation will still decrease with
  erasing steps.  This is the goal of the following lemma.  Notice
that we do not consider general $(\jn)$-reductions, but only those
occurring inside a left-right context $\lrc$.  
 We will use the syntax
of terms given in \autoref{eq:alt-syntax-dj} to conclude the proof
(\autoref{l:bound}).
\begin{lem}[Erasing subject reduction]
  \label{l:erasing-sr}
  Let $t = \gap{\ctx\dcnl{\lx.s}} u y r$ and
  $t' = \rsub{\ctx\dcnl{\rsub u x s}} y r$ such that
  for some $\lrc$ there is $\dernp\Gamma{\ctx\lrc t}{\sigma}k \sysFullJ$.
  Then,
  \begin{enumerate}
    \item If $y \notin \fv{r}$, then there are typing derivations for
      $\ctx\lrc{t'} = \ctx\lrc{r}$, $\ctx\dcnl{s}$ and
      $u$ having measures $k_{\ctx\lrc{t'}}$,
      $k_{\ctx\dcnl{s}}$ and $k_{u}$ resp. such that
      $k > 1 + k_{\ctx\lrc{t'}} +
      k_{\ctx\dcnl{s}} + k_{u}$.
    \item If $y \in \fv{r}$ and $x \notin \fv{s}$, then there are typing
      derivations for
      $\ctx\lrc{t'} = \ctx\lrc{\rsub {\ctx\dcnl {s}} y
      r}$ and $u$ having measures
      $k_{\ctx\lrc{t'}}$ and $k_{u}$
      resp. such that
      $k > 1+ k_{\ctx\lrc{t'}} + k_{u}$.
  \end{enumerate}
\end{lem}
\begin{proof}
  We prove a stronger statement:
  the derivation for $\ctx\lrc{t'}$ is of the shape
  $\dernp{\Gamma'}{\ctx\lrc{t'}}{\sigma}{k_{\ctx\lrc{t'}}}\sysFullJ$
  with the same $\sigma$ but $\Gamma' \sqsubseteq \Gamma$.
  We proceed by induction on $\lrc$:
  \begin{cas}
    \case{$\lrc = \ec$}\hfill
    \begin{enumerate}
      \item The derivation of $t$ has premises
        $\dern{\Gamma_{\lambda}}{\ctx\dcnl{\lx.s}}{\tau}{k_{\lambda}}$,
        $\dern{\Delta}{u}{\rho}{k_u}$ and
        $\dern{\Lambda}{r}{\sigma}{k_{t'}}$, for some appropriate $\tau$ and
        $\rho$, such that
        $\Gamma = \Gamma_{\lambda} \inter \Delta \inter \Lambda$ and
        $k=k_{\lambda} + k_u + k_{t'} + 1$.
        By \autoref{l:perml-typ}, we have a derivation
        $\dern{\Gamma_{\lambda}}{\lx.\ctx\dcnl s}{\tau}{k_{\lambda}}$.
        Then, $\tau = {\MM} \to {\tau'}$ with $\MM$ potentially empty and we have a derivation
        $\dern{\Gamma_{\lambda};x:\MM}{\ctx\dcnl s}{\tau'}{k_{\lambda}-1}$.
        Let $k_{\ctx\dcnl s} = k_{\lambda}-1$.
        We have $k > 1 + k_{t'} + k_{\ctx\dcnl s} + k_u $ and 
        we let $\Gamma' = \Lambda$ since $t'=r$.
        We can conclude since $\Gamma' \sqsubseteq \Gamma$.
      \item The derivation of $t$ has premises
        $\dern{\Gamma_{\lambda}}{\ctx\dcnl{\lx.s}}{\mult{{\emult} \to {\tau_i}}_{\iI}}{k_{\lambda}}$, 
        and thus
        $(\dern{\Gamma_{\lambda}^i}{\ctx\dcnl{\lx.s}}{{\emult} \to {\tau_i}}{k_{\lambda}^i})_{\iI}$
        with $\Gamma_{\lambda} = \inter_{\iI} \Gamma_{\lambda}^i$ and $k_{\lambda} = +_{\iI} k_{\lambda}^i$, as well as 
        $\dern{\Delta}{u}{\rho}{k_u}$ and
        $\dern{\Lambda;\mult{\tau_i}_{\iI}}{r}{\sigma}{k_r}$, where 
        $\Gamma = \Gamma_{\lambda} \inter \Delta \inter \Lambda$ and
        $k = k_{\lambda} + k_u + k_r + 1 = k$
        and $I \neq \emptyset$.
        By \autoref{l:perml-typ}, we have derivations
        $(\dern{\Gamma_{\lambda}^i}{\lx.\ctx\dcnl s}{{\emult} \to {\tau_i}}{k_{\lambda}^i})_{\iI}$
        and thus derivations
        $(\dern{\Gamma_{\lambda}^i}{\ctx\dcnl s}{\tau_i}{k_{\lambda}^i-1})_{\iI}$.
        By rule~$\many$ we have a derivation
        $\dern{\Gamma_{\lambda}}{\ctx\dcnl s}{\mult{\tau_i}_{\iI}}{k_{\ctx\dcnl s}}$
        where $k_{\ctx\dcnl s} = +_{\iI} (k_{\lambda}^i - 1) = k_{\lambda} - |I|$.
        Using the substitution \autoref{l:substitution-lemma-right-strong} we construct a derivation
        $\dern{\Lambda \inter \Gamma_{\lambda}}{\rsub{\ctx\dcnl s} y r}{\sigma}{k_{t'}}$
        with $k_{t'} = k_r + k_{\ctx\dcnl s} - |I|$.
        We have $k = k_{\ctx\dcnl s} + |I| + k_u + k_r + 1 = 1 + k_{t'} + 2\times|I| + k_u > 1 + k_{t'} + k_u$. We let $\Gamma' = \Lambda \inter \Gamma_{\lambda} $.
        We can then conclude since $\Gamma' \sqsubseteq \Gamma$.
    \end{enumerate}

    \case{$\lrc = \gap{\lrc'}{u'} z {r'}$}
    The derivation of $\ctx\lrc t$ has three premises of the form:
    $\dern{\Gamma_1}{\ctx{\lrc'} t}{\choice{\mult{{\MM_i} \to {\tau_i}}_{\iI}}}{k_{\ctx{\lrc'} t}}$, 
    $\dern{\Delta}{u'}{\choice{\multunion_{\iI} \MM_i}}{k_{u'}}$ and
    $\dern{\Lambda;z:\mult{\tau_i}_{\iI}}{r'}{\sigma}{k_{r'}}$
    such that $k = 1 + k_{\ctx{\lrc'} t} + k_{u'} + k_{r'}$
    and $\Gamma = \Gamma_1 \inter \Delta \inter \Lambda$.
    By \ih\ we get from the first premise:
    \begin{enumerate}
      \item In cases~(1) and~(2) a derivation
        $\dern{\Gamma_2}{\ctx{\lrc'}{t'}}{\choice{\mult{{\MM_i} \to {\tau_i}}_{\iI}}}{k_{\ctx{\lrc'}{t'}}}$
        such that $\Gamma_2 \sqsubseteq \Gamma_1$
        and a typing derivation for $u$ of measure $k_u$.
      \item In case~(1) a typing derivation for $\ctx\dcnl s$ of measure $k_{\ctx\dcnl s}$
        and the fact that $k_{\ctx{\lrc'} t} > 1 + k_{\ctx{\lrc'}{t'}} + k_{\ctx\dcnl s} + k_u$.
      \item In case~(2) $1 + k_{\ctx{\lrc'}{t'}} + k_u$.
    \end{enumerate}
    Using the type derivations for $\ctx{\lrc'}{t'}$, $u'$ and $r'$ we can build a derivation
    $\dern{\Gamma_2 \inter \Delta \inter
    \Lambda}{\gap{\ctx{\lrc'}{t'}}{u'} z {r'}}{\sigma}{k_{\ctx\lrc{t'}}}$, where $k_{\ctx\lrc{t'}} = 1 + k_{k_{\ctx{\lrc'}{t'}}} + k_{u'} + k_{r'}$.
    We have $\Gamma_2 \inter \Delta \inter \Lambda \sqsubseteq \Gamma$.
    In case~(1) we can conclude because
    $k = 1 + k_{\ctx{\lrc'} t} + k_{u'} + k_{r'}
    >_{\ih} 1 + (1 + k_{\ctx{\lrc'}{t'}} + k_{\ctx\dcnl s} + k_u) + k_{u'} + k_{r'}
    = 1 + k_{\ctx\lrc{t'}} + k_{\ctx\dcnl s} + k_u$.
    In case~(2) in the same way, but without adding $k_{\ctx\dcnl s}$ in the sum.

    \case{$\lrc = \gap \nl {u'} z {\lrc'}$}
    The derivation of $\ctx\lrc t$ has premises:
    $\dern{\Gamma_{\nl}}{\nl}{\choice{\mult{{\MM_i} \to {\tau_i}}_{\iI}}}{k_{\nl}}$,
    $\dern{\Delta}{u'}{\choice{\multunion_{\iI} \MM_i}}{k_{u'}}$ and
    $\dern{\Lambda_1;z:\mult{\tau_i}_{\iI}}{\ctx{\lrc'} t}{\sigma}{k_{\ctx{\lrc'} t}}$.
    We have $\Gamma = \Gamma_{\nl} \inter \Delta \inter \Lambda_1$
    and $k = 1 + k_{\nl} + k_{u'} + k_{\ctx{\lrc'} t}$.
    By the \ih\ we get from the third premise:
    \begin{enumerate}
      \item In cases~(1) and~(2) a derivation
        $\dern{\Lambda_2;z:\mult{\tau_i}_{\iI'}}{\ctx{\lrc'}{t'}}{\sigma}{k_{\ctx{\lrc'}{t'}}}$
        such that $\Lambda_2 \sqsubseteq \Lambda_1$,
        and $I' \subseteq I$ ($I'$ possibly empty), 
        and a typing derivation for $u$ of measure $k_u$.
      \item In case~(1) a typing derivation for $\ctx\dcnl s$ of measure $k_{\ctx\dcnl s}$
        and the fact that $k_{\ctx{\lrc'} t} > 1 + k_{\ctx{\lrc'}{t'}} + k_{\ctx\dcnl s} + k_u$.
      \item In case~(2) $1 + k_{\ctx{\lrc'}{t'}} + k_u$.
    \end{enumerate}
    To build a derivation for $\ctx\lrc{t'}$, we need in particular derivations
    of type $\choice{\mult{{\MM_i} \to {\tau_i}}_{\iI'}}$ for $\nl$
    and $\choice{\multunion_{\iI'} \MM_i}$ for $u'$.
    \begin{cas}
      \subcase{$I' \neq \emptyset$} Then $\choice{\mult{{\MM_i} \to
      {\tau_i}}_{\iI'}} = \mult{{\MM_i} \to {\tau_i}}_{\iI'}$
      and by \autoref{l:split-many-strong} it is possible to construct a derivation
      $\dern{\Gamma'_{\nl}}{\nl}{\mult{{\MM_i} \to {\tau_i}}_{\iI'}}{k'_{\nl}}$
      from the original one for $\nl$ verifying $\Gamma'_{\nl} \sqsubseteq \Gamma_{\nl}$ and $k'_{\nl} \leq k_{\nl}$. 
      For $u'$ we build a derivation $\dern{\Delta'}{u'}{\choice{\multunion_{\iI'} \MM_i}}{k'_{u'}}$ verifying $\Delta' \sqsubseteq \Delta$ and $k'_{u'} \leq k_{u'}$.
      There are three cases:
      \begin{cas}
      \item[Subsubcase] $(\MM_i)_{\iI}$ are all empty, and therefore $(\MM_i)_{\iI'}$
        are all empty.
        Then we set $\choice{\multunion_{\iI'} \MM_i} = \choice{\multunion_{\iI} \MM_i}$.
        We take the original derivation
        so that $\Delta' = \Delta$, $k'_{u'} = k_{u'}$.
      \item[Subsubcase] $(\MM_i)_{\iI'}$ are all empty but
        $(\MM_i)_{\iI}$ are not all empty.
        As a consequence, $\multunion_{\iI} \MM_i \neq \emptyset$ and we
        take an arbitrary type $\rho$ of $\multunion_{\iI}
        \MM_i$ as a witness for $u'$, so that,
        $\dern{\Delta_{\rho}}{u'}{\rho}{k_{\rho}}$
        holds by \autoref{l:split-many-strong}. We have the expected
        derivation with rule~$\many$ taking $\Delta' =
        \Delta_{\rho}$, $\choice{\multunion_{\iI'} \MM_i} =
        \mult\rho$ and $k'_{u'} = k_{\rho}$.
      \item[Subsubcase] $\choice{\multunion_{\iI'} \MM_i} =
        \multunion_{\iI'} \MM_i$.
        By \autoref{l:split-many-strong} it is possible to construct the expected derivation from the original ones for $u'$.
      \end{cas}
      Finally, we conclude by the following derivation
      for $\ctx{\lrc}{t'}$:
      \[ \inferrule*{
          \dern{\Gamma'_{\nl}}{\nl}{\mult{{\MM_i} \to {\tau_i}}_{\iI'}}{k'_{\nl}} \and
          \dern{\Delta'}{u'}{\choice{\multunion_{\iI'} \MM_i}}{k'_{u'}} \and
        \Phi}{
      \seq{\Gamma'}{\gap{\nl}{u'}{y}{\ctx{\lrc'}{t'}}}{\sigma }} \]
      where $\Phi = \dern{\Lambda_2;
      z:\mult{\tau_i}_{\iI'}}{\ctx{\lrc'}{t'}}{\sigma}{k_{\ctx{\lrc'}{t'}}}$,
      where $\Gamma' = \Gamma'_{\nl} \inter\Delta' \inter \Lambda_2$, and the total measure of the derivation is $k_{\ctx{\lrc}{t'}} = 1+ k'_{\nl}+k'_{u'}+k_{\ctx{\lrc'}{t'}}$. 
      We have $k > 1 + k'_{\nl} + k'_{u'} + k_{\ctx{\lrc'} t} >_{\ih} 1 +
      k'_{\nl} + k'_{u'} + 1 + k_{\ctx{\lrc'}{t'}} + k_{\ctx\dcnl s} + k_u > 1
      + k_{\ctx{\lrc}{t'}} + k_{\ctx\dcnl s} + k_u$ in case~(1). Similarly but without $k_{\ctx\dcnl s} $ in
      case~(2).
      We can conclude since $\Gamma' \sqsubseteq \Gamma$. 
      \case{$I = I' = \emptyset$} We are done by taking the original derivations.
      \case{$I \neq \emptyset = I'$} Let us take an
      arbitrary $j \in I$: the type $\mult{{\MM_j} \to {\tau_j}}$ is set as a witness for $\nl$,
      whose derivation $\dern{\Gamma'}{\nl'}{\mult{{\MM_j} \to
      {\tau_j}}}{k_{\nl'}}$ is obtained from the derivation
      $\dern{\Gamma_{\nl}}{\nl}{\choice{\mult{{\MM_i} \to {\tau_i}}_{\iI}}}{k_{\nl}}$ by the split \autoref{l:split-many-strong}.
      For $u'$, we take as a witness an arbitrary $\rho \in \choice{\multunion_{\iI} \MM_i}$
      and we set $\choice{\multunion_{\iI'} \MM_i} = \mult\rho$.
      If $\multunion_{\iI} \MM_i = \emult$, then $\rho$ is the original witness.
      Otherwise $\rho$ is a type of one of the $\MM_i$'s.
      In both cases we use the split \autoref{l:split-many-strong} to get a derivation
      $\dern{\Delta'}{u'}{\mult\rho}{k'_{u'}}$
      where $\Delta' \sqsubseteq \Delta$ and $k'_{u'} \leq k_{u'}$.
      Using the type derivation given by the \ih\ for
      $\ctx{\lrc'}{t'}$, we conclude by the following
      derivation for $\ctx\lrc{t'}$:
      \[ \inferrule*{
          \dern{\Gamma_{\nl}'}{\nl'}{\mult{{\MM_j} \to {\tau_j}}}{k_{n'}}
          \and\mkern-9mu \dern{\Delta'}{u'}{\mult\rho}{k'_{u'}}
        \and\mkern-9mu \dern{\Lambda_2;z:\mult{\tau_i}_{\iI'}}{\ctx{\lrc'}{t'}}{\sigma}{k_{\ctx{\lrc'}{t'}}}}{
    \seq{\Gamma'}{\gap \nl {u'} y {\ctx{\lrc'}{t'}}}{\sigma}} \]
    where $\Gamma_{\nl}' \sqsubseteq \Gamma_{\nl}$, $\Delta' \sqsubseteq \Delta$, $k'_{\nl} \leq k_{\nl}$, $k'_{u'} \leq k_{u'}$.
    We have $\Gamma' = \Gamma'_{\nl} \inter \Delta' \inter \Lambda_2 \sqsubseteq \Gamma$.

    In case~(1) we can conclude because
    $k = 1 + k_{\nl} + k_{u'} + k_{\ctx{\lrc'} t}
    > 1 + k_{\nl}' + k'_{u'} + (1 + k_{\ctx{\lrc'}{t'}} + k_{\ctx\dcnl s} + k_u)
    = 1 + k_{\ctx\lrc{t'}} + k_{\ctx\dcnl s} + k_u$.
    Similarly but without $k_{\ctx \dcnl s}$ in case~(2).
    \qedhere
  \end{cas}
\end{cas}
\end{proof}

\begin{exa} Take again the erasing reduction step
    $t = \gap {(\lx.\id)} y z z \rew\dbeta \id = t'$ from
    \autoref{ex:erasing-reduction}.
    The previous lemma applies by taking $\lrc = \ec$.
    More precisely, we are in the second case where $z \in
    \fv{z}$, but $x \notin \fv \id$.  The typing derivation given in
    \autoref{ex:erasing-reduction} has size $6$.
    Although there is no derivation for $t' = \id$ under the same
    typing environment $y:\mult\sigma$, it is easy to see that
    there is a derivation for $t'$ of size $2$ under the empty
    environment.  There is also a trivial
    derivation for $\seq {y:\mult\sigma} y \sigma$ of size 1.  It is
    then verified that $6 > 1 + 2 + 1$.
\end{exa}

We now finish the proof of soundness by proving that every typable term has a finite maximal reduction length, bounded by the size of any typing derivation for the term.
The maximal reduction length of a term $t$ is written $\maxred[\djn] t$ for a term $t\in \sn\djn$.
\begin{lem} 	The function
  $\maxred[\djn]{\cdot}:\sn{\djn}\to\mathbb{N}_{0}$  verifies the following equalities\footnote{These equalities can be seen as giving an alternative, recursive definition of function $\maxred[\djn]{\cdot}$, based on the inductive definition of $\sn{\djn}$ given in \autoref{d:isn}.}:   
  \[ \begin{array}{llll}
    \maxred[\djn]{x} & = & 0 \\
    \maxred[\djn]{\lx.t} & = & \maxred[\djn]{t}\\
    \maxred[\djn]{\gap \nl u x r} & = & \maxred[\djn]{\nl} + \maxred[\djn]{u} + \maxred[\djn]{r} \\
    \maxred[\djn]{\ctx\lrc{\gap{\ctx\dcnl{\lx.s}} u y r}} & = &
    \begin{cases}
      1 + \maxred[\djn]{\ctx\lrc{r}} + \maxred[\djn]{\ctx\dcnl{s}} +
      \maxred[\djn]{u} & \text{ if $y \notin \fv{r}$; else:}\\
      1 + \maxred[\djn]{\ctx\lrc{\rsub {\ctx\dcnl {s}} y r}} + \maxred[\djn]{u} & \text{ if } x \notin \fv{s};\\
      1 + \maxred[\djn]{\ctx\lrc{\rsub {\ctx\dcnl {\rsub u x s}} y r}} & \text{ if } x \in \fv{s}
    \end{cases}
  \end{array} \]
\end{lem}

  \begin{proof}
  The first three equalities are obvious.
  For the fourth one, we trivially have
  that the \rhs\  is smaller or equal than
  the \lhs, because there is a reduction sequence starting at $\ctx\lrc{\gap{\ctx\dcnl{\lx.s}} u y r}$ whose length is the \rhs. We now show the opposite direction, that is, the length of an arbitrary reduction sequence
    to normal form starting at
  	$\ctx\lrc{\gap{\ctx\dcnl{\lx.s}} u y r}$ is bound above by the \rhs.

        Take any reduction sequence
  to normal form starting at
  $t=\ctx\lrc{\gap{\ctx\dcnl{\lx.s}} u y r}$. Then it has
  necessarily the following form
  \[ t= \ctx\lrc{\gap{\ctx\dcnl{\lx.s}} u y r} \rewn\dbeta
    \ctx{\lrc'}{\gap{\ctx{\dcnl'}{\lx.s'}} {u'} y {r'}} \rew\dbeta
   \ctx{\lrc'}{\rsub{\ctx{\dcnl'}{\rsub{u'}{x}{s'}}}{y}{r'}} =t' \rew\dbeta \ldots \] 
   where $\lrc \rewn\dbeta\lrc'$ has length $l_{\lrc}$,
   $\dcnl\rewn\dbeta\dcnl'$ has length $l_{\dcnl}$,
   $s \rewn\dbeta s'$ has length $l_{s}$,
   $u \rewn\dbeta u'$ has length $l_{u}$, and
     $r \rewn\dbeta r'$ has length $l_{r}$, the reduction from $t'$ has length $l_{t'}$ so that
   the previous sequence has length $l_{\lrc}+ l_{\dcnl} + l_{s} + l_{u} + l_{r} + 1+l_{t'}=:l_1$
   \begin{cas}
   \case{$y \in \fv{r}$ and $x \in \fv{s}$}
     We want to show that $l_1\leq 1 + \maxred[\djn]{\ctx\lrc{\rsub {\ctx\dcnl {\rsub u x s}} y r}}$.
     Take the following reduction sequence to normal form 
   \[ \begin{array}{l}
	\ctx{\lrc}{\rsub{\ctx{\dcnl}{\rsub{u}{x}{s}}}{y}{r}} \rewn\dbeta
	\ctx{\lrc'}{\rsub{\ctx{\dcnl'}{\rsub{u'}{x}{s'}}}{y}{r'}} =t' \rew\dbeta \ldots
\end{array} \]  
continuing from $t'$ as before, with length $l_{\lrc}+ (l_{u} \times |s|_x+l_s+l_{\dcnl}) \times |r|_y + l_r+l_{t'}:=l_2$. Since this reduction sequence starts at $	\ctx{\lrc}{\rsub{\ctx{\dcnl}{\rsub{u}{x}{s}}}{y}{r}}$, we conclude $1+l_2\leq 1 + \maxred[\djn]{\ctx\lrc{\rsub {\ctx\dcnl {\rsub u x s}} y r}}$. To finish the argument, 
we just need $l_1\leq 1+l_2$. This is immediate, since $1 \leq |r|_y$ and $1 \leq |s|_x$.


 \case{$y \in \fv{r}$ and $x \notin \fv{s}$} 
  Then also $x \notin \fv{s'}$ so that 
  $t' = \ctx{\lrc'}{\rsub{\ctx{\dcnl'}{s'}}{y}{r'}}$. In this case, we want to show that $l_1\leq 1 + \maxred[\djn]{\ctx\lrc{\rsub {\ctx\dcnl {s}} y r}} + \maxred[\djn]{u}$. Since $l_{u}\leq \maxred[\djn]{u}$, we just need $l_{\lrc}+ l_{\dcnl} + l_{s}  + l_{r} +l_{t'} \leq \maxred[\djn]{\ctx\lrc{\rsub {\ctx\dcnl {s}} y r}}$. Consider the reduction

\[
\ctx\lrc{\rsub {\ctx\dcnl {s}} y r} \rewn\dbeta \ctx{\lrc'}{\rsub {\ctx{\dcnl'} {s'}} y r'} = t'\rewn\dbeta\cdots
\]  
continuing from $t'$ as before, with length $l_{\lrc}+ (l_{\dcnl} + l_{s})
\times |r|_y + l_{r} +l_{t'}:=l_2$. Since this reduction sequence starts at $\ctx\lrc{\rsub {\ctx\dcnl {s}} y r}$, then $l_2\leq \maxred[\djn]{\ctx\lrc{\rsub {\ctx\dcnl {s}} y r}}$. We are done, since since $1 \leq |r|_y$.

%
%

 \case{$y \notin \fv{r}$} 
   Then also $y \notin \fv{r'}$ so that 
  $t' = \ctx{\lrc'}{r'}$. In this case, we want to show $l_1\leq 1 + \maxred[\djn]{\ctx\lrc{r}} + \maxred[\djn]{\ctx\dcnl{s}} +
  	\maxred[\djn]{u}$. The reduction 
  	\[
  	\ctx{\lrc}{r}\rewn\dbeta\ctx{\lrc'}{r'}=t'\rewn\dbeta\cdots
  	\]
  continuing from $t'$ as before, has length $l_{\lrc}+ l_{r} +l_{t'}\leq \maxred[\djn]{\ctx\lrc{r}}$. Also the reduction $\ctx\dcnl{s}\rewn\dbeta\ctx{\dcnl'}{s'}$ has length $l_{\dcnl} + l_{s}\leq\maxred[\djn]{\ctx\dcnl{s}}$. Since $l_u\leq \maxred[\djn]{u}$, we are done.
  \qedhere

%
     \end{cas}
 \end{proof}

  \begin{lem}[Soundness]
  \label{l:bound}

  If $\dernp{\Gamma}{t}{\sigma}{k}{\sysFullJ}$, then $t \in \sn\djn$
  and $\maxred[\djn]{t} \leq k$.

  \end{lem}

\begin{proof} We proceed by induction on $k$ and we reason by case analysis on
  $t$ according to the alternative grammar (\autoref{eq:alt-syntax-dj} on
  Page~\pageref{eq:alt-syntax-dj}).
  \begin{cas}
    \case{$t=x$} The type
    derivation is just an axiom so that $t=x$ and $k=1$.
    We trivially  get $x \in  \sn\djn$.
    We also have 
    $\maxred[\djn]{x}=0 < 1=k$.
    \case{$t = \lx.u$} There is a typing derivation for $u$ of size $k-1 < k$. The \ih\ gives $u\in  \sn\djn$ and  $\maxred[\djn]{u} \leq k-1$, so that
    we trivially get $\lx.u \in \sn\djn$  and  $\maxred[\djn]{t} = \maxred[\djn]{u} \leq k$.
    \case{$t= \gap \nl u x r$}
    Where $r \in \nf \whdj$ according to the
    alternative grammar.  There are typings of $\nl$, $u$ and
    $r$ with measures $k_{\nl}$, $k_u$ and $k_r$ resp. such that $k =
    1+ k_{\nl}+ k_u+ k_r$. By the \ih\ we get
      $\nl, u, r\in \sn\djn$, 
      $\maxred[\djn]{\nl} \leq  k_{\nl}$,
      $\maxred[\djn]{u} \leq k_u$ and
      $\maxred[\djn]{r} \leq k_r$.
    We then get $t \in \sn \djn$ by \autoref{d:isn} \isnapprule\
      and \autoref{l:cbn-sn-isn}.
    We also  get $\maxred[\djn]{t}= \maxred[\djn]{\nl} + \maxred[\djn]{u} +
    \maxred[\djn]{r} \leq_{\ih} k_{\nl}+ k_u+ k_r \leq k$.
    \case{$t= \ctx\lrc{\gap{\ctx\dcnl{\lx.s}} u y r}$}
    There are three possible cases:
    \begin{cas}
      \subcase{$x \in \fv{s}$ and $y \in \fv{r}$} Then $t \rew{\djn} \ctx\lrc{\rsub {\ctx\dcnl {\rsub u x s}} y r} = t_0$.
      Moreover, the subject reduction \autoref{l:qsr-strong} gives
      $\dernp{\Gamma}{t_0}{\sigma}{k'}{\sysFullJ}$ with $k' < k$. By the
      \ih\ we have $t_0 \in \sn\djn$ and  $\maxred[\djn]{t_0} \leq k'$. Moreover, $t_0 \in \sn\djn$ implies
        in particular $\ctx\dcnl{s}\in \sn\djn$ and $u \in  \sn\djn$. By \autoref{d:isn} \isnbetarule\  and \autoref{l:cbn-sn-isn} we get $t \in \sn\djn$. 
      We also conclude $\maxred[\djn]{t} = 1+ \maxred[\djn]{t_0} \leq 1+ k' \leq k$.

      \subcase{$y \notin \fv{r}$} Then $t \rew{\djn} \ctx\lrc{r} = t_0$ 
      By subject reduction for erasing steps (\autoref{l:erasing-sr})
      there are typings of
      $t_0$, $\ctx\dcnl{s}$ and $u$ having measures
      $k_{t_0}$, $k_{\ctx\dcnl{s}}$ and $k_{u}$ resp. such that
      $k > 1 + k_{t_0} + k_{\ctx\dcnl{s}} + k_{u}$.
      By the \ih\ we get $ t_0, \ctx\dcnl{s}, u  \in \sn\djn$,
        $\maxred[\djn]{t_0} \leq k_{t_0}$,
        $\maxred[\djn]{\ctx\dcnl{s}} \leq k_{\ctx\dcnl{s}}$, and
        $\maxred[\djn]{u} \leq k_{u}$. By \autoref{d:isn} \isnbetarule\
        and \autoref{l:cbn-sn-isn} we get
        $t \in \sn\djn$. 
        We also conclude $\maxred[\djn]{t} = 1 + \maxred[\djn]{t_0} + \maxred[\djn]{\ctx\dcnl{s}} + \maxred[\djn]{u}
      \leq_{\ih} 1 + k_{t_0} + k_{\ctx\dcnl{s}} + k_{u} < k$.
      
      \subcase{$x \notin \fv{s}$ and $y \in \fv{r}$} Then $t \rew{\djn} \ctx\lrc{\rsub {\ctx\dcnl {s}} y r} = t_0$.
By subject reduction for erasing steps (\autoref{l:erasing-sr})
      there are typings of
      $t_0$ and $u$ having measures
      $k_{t_0}$ and $k_{u}$ resp. such that
      $k > 1+ k_{t_0} + k_{u}$.
By the \ih\ we get $ t_0,  u  \in \sn\djn$,
        $\maxred[\djn]{t_0} \leq k_{t_0}$, and
  $\maxred[\djn]{u} \leq k_{u}$. By \autoref{d:isn} \isnbetarule\
  and \autoref{l:cbn-sn-isn} we get
        $t \in \sn\djn$. 
      Thus
      we conclude $\maxred[\djn]{t} = 1 + \maxred[\djn]{t_0} + \maxred[\djn]{u}
      \leq_{\ih} 1+ k_{t_0} + k_{u} < k$.

      and $\maxred[\djn]{t}=1 + \maxred[\djn]{t_0} + \maxred[\djn]{u}$.      
      \qedhere
    \end{cas}
  \end{cas}
\end{proof}

Completeness (forthcoming \autoref{l:completeness})
  relies on two key points: firstly, the fact that all normal forms
  are typable (\autoref{l:strong-normal-forms-typable}), and secondly,
  the subject expansion property (\autoref{l:qse-strong}), which is
  the dual of subject reduction.  Once again, this last property only
  holds for non-erasing steps.  To complete the proof of
    completeness for erasing steps, we use our inductive definition
    of strong normalization.

\begin{lem}[Typing normal forms]
  \label{l:strong-normal-forms-typable} \mbox{}
  \begin{enumerate}
    \item For all $t\in{\nf\djn}$, there exists $\Gamma$, $\sigma$ such that $\derp
      \Gamma {t}{\sigma}{\sysFullJ}$.
    \item For all $t\in{\nef\djn}$, for all $\sigma$, there exists $\Gamma$ such
      that $\derp \Gamma {t}{\sigma}{\sysFullJ}$.
  \end{enumerate}
\end{lem}
\begin{proof}
  By simultaneous induction on $t \in {\nf\djn}$ and $t\in {\nef\djn}$.

  First, the cases relative to statement~(1).
  \begin{cas}
    \case{$t = x$} Pick an arbitrary $\sigma$. We have $\der{x:\mult{\sigma}}{x}{\sigma}$ by
    rule~$\ruleAxJ$.

    \case{$t = \lx.s$ where $s \in {\nf\djn}$}
    By \ih\ on $s$ there exists $\Gamma'$ and $\tau$ such that
    $\der{\Gamma'}{s}{\tau}$. Let $\Gamma$ and $\MN$ be such that $\Gamma'=\Gamma;x:\MN$ ($\MN$ is possibly empty).
    We get $\der{\Gamma}{\lx.s}{\MN \to \tau}$ by rule~$\ruleAbsJ$.
    We conclude by taking $\sigma = \MN \to \tau$.

    \case{$t = \gap {s} u y r$ where $u, r \in {\nf\djn}$ and
    $s \in {\nef\djn}$} By the \ih\ on $r$ there is a derivation of 
    $\der{\Lambda'}{r}{\sigma}$. Let $\Lambda$ and $\mult{\tau_i}_{\iI}$ be such that $\Lambda'=\Lambda;y:\mult{\tau_i}_{\iI}$. Now we
    construct a derivation
    $\der{\Pi}{s}{\choice{\mult{\emult \to {\tau_i}}_{\iI}}}$ as
    follows.

    \begin{itemize}
      \item If $I = \emptyset$, then the \ih\ on $s$ gives a
        derivation $\der{\Pi}{s}{\tau}$ and we use rule~$\many$ to get
        $\der{\Pi}{s}{\mult\tau}$. We conclude by setting
        $\choice{\mult{\emult \to {\tau_i}}_{\iI}} = \mult\tau$.
      \item If $I \neq \emptyset$, then 
        the \ih\ on $s$ gives a derivation of $\der{\Pi_i}{s}{\emult \to {\tau_i}}$ for each $\iI$. We take
        $\Pi = \inter_{\iI} \Pi_i$ and we conclude with rule~$\many$ since
        $\choice{\mult{\emult \to {\tau_i}}_{\iI}} = \mult{\emult \to {\tau_i}}_{\iI}$.
    \end{itemize}
    Finally, the \ih\ on $u$ gives a
    derivation $\der\Delta{u}{\rho}$ from which we get
    $\der\Delta{u}{\mult\rho}$, by choosing
    $\choice{\multunion_{\iI} \emult} = \mult\rho$. We conclude
    with rule~$\ruleAppJ$ as follows:
    \[ \inferrule*{
        \der{\Pi}{s}{\choice{\mult{{\emult} \to {\tau_i}}_{\iI}}} \and
        \der{\Delta}{u}{\choice{\multunion_{\iI} \emult}} \and
      \der{\Lambda; y:\mult{\tau_i}_{\iI}}{r}{\sigma}}{
  \seq{\Pi \inter \Delta \inter \Lambda}{\gap{s}{u}{y}{r}}{\sigma }} \]
\end{cas}

Next, the cases relative to statement~(2).
\begin{cas}
  \case{$t=x$} As seen above, given an arbitrary type $\sigma$, we can take $\Gamma=\mult{\sigma}$.
  \case{$t = \gap {s} u y r$ where $u \in {\nf\djn}$ and $s,r \in {\nef\djn}$}
  Pick an arbitrary $\sigma$. The proof proceeds \textit{ipsis verbis} as in the case $t = \gap {s} u y r$ above.
  \qedhere
\end{cas}
\end{proof}

\begin{lem}[Anti-substitution]
  \label{l:anti-substitution-lemma-right-strong}%
  If $\der{\Gamma}{\rsub{u}{x}t}{\sigma}$ where $x \in \fv{t}$, then
  there exist $\Gamma_t$, $\Gamma_u$ and $\MM \neq \emult$
  such that $\der{\Gamma_t; x:\MM}{t}{\sigma}$,
  $\der{\Gamma_u}{u}{\MM}$ and
  $\Gamma = \Gamma_t \inter \Gamma_u$.
\end{lem}
\begin{proof}
  By induction on the derivation $\der{\Gamma}{\rsub{u}{x}t}{\sigma}$.
  We extend the statement to derivations ending with $\many$, for which the
  property is straightforward by the \ih\ We
  reason by cases on $t$.
  \begin{cas}
    \case{$t = x$} Then $\rsub u x t = u$.
    We take $\Gamma_t = \emptyset$, $\Gamma_u = \Gamma$, $\MM = \mult{\sigma}$, and we have
    $\der{x:\mult{\sigma}}{x}{\sigma}$ by rule~$\ruleAxJ$
    and $\der{\Gamma}{u}{\MM}$ by rule $\many$ on the derivation of the hypothesis.

    \case{$t = \ly.s$ where $y \neq x$ and $y \notin \fv u$ and $x \in
    \fv{s}$}\!\!
    Then $\rsub u x t = \ly.\rsub u x s$.
    We have $\sigma = {\MN} \to {\tau}$
    and $\der{\Gamma;y:\MN}{\rsub u x s}{\tau}$. 

    By the \ih\ there exists $\Gamma', \Gamma_u, \MM \neq \emult$ such that
    $\der{\Gamma';y:\MN;x:\MM}{s}{\tau}$,
    $\der{\Gamma_u}{u}{\MM}$,
    and $\Gamma;y:\MN= (\Gamma';y:\MN) \inter \Gamma_u$.
    Moreover, by $\alpha$-conversion and \autoref{l:relevance-sysFullJ}
    we know that $y \notin \dom{\Gamma_u}$ so that
    $\Gamma = \Gamma'\inter \Gamma_u$.
    We conclude by deriving $\der{\Gamma';y:\MN}{\lx. s}{ {\MN} \to {\tau}}$
    with rule~$\ruleAbsJ$. Indeed, by letting $\Gamma_t = \Gamma'$
    we have $\Gamma = \Gamma_t \inter \Gamma_u$ as required.
    \case{$t = \gap {t_1} {t_2} y r$, where $y \neq x$, $y \notin \fv u$
    and $x \in \fv{t_1} \cup \fv{t_2} \cup (\fv{r} \setminus y)$}\,
    We detail the case where $x \in \fv {t_1} \cap \fv {t_2} \cap \fv{r}$,
    the other cases are similar.
    By construction, we have derivations
    $\der{\Gamma_1}{\rsub u x {t_1}}{ \choice{\mult{{\MN_i} \to {\tau_i}}_{\iI}}}$,
    $\der{\Gamma_2}{\rsub u x {t_2}}{ \choice{\multunion_{\iI}\MN_i}}$ and
    $\der{\Gamma_3;y:\mult{\tau_i}_{\iI}}{\rsub u x r}{\sigma}$,
    with $\Gamma = \Gamma_1 \inter \Gamma_2 \inter \Gamma_3$.

    By the \ih\ there are environments
    $\Gamma_{t_1}, \Gamma_{t_2}, \Gamma_r$,
    $\Gamma_u^1, \Gamma_u^2, \Gamma_u^3$ and multitypes
    $\MM_1, \MM_2, \MM_3$ all different from $\emult$ such that
    $\der{\Gamma_{t_1};x:\MM_1}{t_1}{ \choice{\mult{{\MN_i} \to {\tau_i}}_{\iI}}}$,
    $\der{\Gamma_{t_2};x:\MM_2}{t_2}{ \choice{\multunion_{\iI}\MN_i}}$,
    $\der{\Gamma_r;x:\MM_3}{r}{ \sigma}$,
    $\der{\Gamma_u^1}{u}{ \MM_1}$,
    $\der{\Gamma_u^2}{u}{ \MM_2}$,
    $\der{\Gamma_u^3}{u}{ \MM_3}$ and
    $\Gamma_1 = \Gamma_{t_1} \inter \Gamma_u^1$,
    $\Gamma_2 = \Gamma_{t_2}\inter \Gamma_u^2$,
    $\Gamma_3 = \Gamma_r \inter \Gamma_u^3$.
    Let $\Gamma_t = \Gamma_{t_1} \inter \Gamma_{t_2} \inter \Gamma_r$,
    $\Gamma_u = \Gamma_u^1 \inter \Gamma_u^2 \inter \Gamma_u^3$ and
    $\MM = \MM_1 \multunion \MM_2 \multunion \MM_3$.
    We can build a derivation
    $\der{\Gamma_t;x:\MM}{\gap {t_1} {t_2} y r}{\sigma}$
    with rule~$\ruleAppJ$ and a derivation
    $\der{\Gamma_u}{u}{\MM}$ with \autoref{l:split-many-strong}.
    We conclude since $\Gamma = \Gamma_1 \inter \Gamma_2 \inter \Gamma_3 = \Gamma_{t_1} \inter \Gamma_u^1 \inter
    \Gamma_{t_2} \inter \Gamma_u^2 \inter
    \Gamma_r \inter \Gamma_u^3 = \Gamma_t \inter \Gamma_u$. 
    \qedhere
  \end{cas}
\end{proof}

\begin{lem}[Non-erasing subject expansion]
  \label{l:qse-strong}%
  If $\derp{\Gamma}{t_2}{\sigma}{\sysFullJ}$ and $t_1 \rew{\djn} t_2$ is a non-erasing step, then
  $\derp{\Gamma}{t_1}{\sigma}{\sysFullJ}$.
\end{lem}
\begin{proof}
  By induction on $t_1 \rew{\djn} t_2$.
  \begin{cas}
    \case{$t_1 = \gap{\ctx\dc{\lx.t}} u y r \rrule{\beta}
    \rsub{\ctx\dc {\rsub u x t}}{y}r=t_2$} Since the reduction is
    non-erasing, we have $y \in \fv r$ and $x \in \fv{t}$. By
    \autoref{l:anti-substitution-lemma-right-strong}, there exists
    $\Gamma_r$, $\Gamma'$ and $\MN$ such that
    $\der{\Gamma_r;y:\MN}{r}{\sigma}$,
    $\der{\Gamma'}{\ctx\dc{\rsub u x t}}{\MN}$ and
    $\Gamma = \Gamma' \inter \Gamma_r$. Let
    $\MN = \mult{\tau_i}_{\iI} \neq \emult$ since $y \in \fv r$. By
    rule~$\many$, we have a decomposition
    $(\der{\Gamma'_i}{\ctx\dc{\rsub u x t}}{\tau_i})_{\iI}$ with
    $\Gamma' = \inter_{\iI} \Gamma'_i$. Since
    $\ctx\dc{\rsub u x t} = \rsub u x {\ctx\dc t}$, by
    \autoref{l:anti-substitution-lemma-right-strong} again, for each
    $\iI$ there are $\Gamma_t^{i}$, $\Gamma_u^{i}$ and
    $\MM_i \neq \emult$ such that
    $\der{\Gamma_t^{i};x:\MM_{i}}{\ctx\dc t}{\tau_i}$,
    $\der{\Gamma_u^{i}}{u}{\MM_{i}}$ and
    $\Gamma'_{i} = \Gamma_t^{i} \inter \Gamma_u^{i}$. By
    rule~$\ruleAbsJ$ followed by $\many$, there are derivations
    $\der{\Gamma_t}{\lx.\ctx\dc
    t}{\mult{{\MM_i} \to {\tau_i}}_{\iI}}$ with
    $\Gamma_t = \inter_{\iI} \Gamma_t^i$. By \autoref{l:perml-typ},
    there is a derivation
    $\der{\Gamma_t}{\ctx\dc{\lx.t}}{\mult{{\MM_i} \to {\tau_i}}_{\iI}}$.
    Finally, by \autoref{l:split-many-strong}, there is a derivation
    $\der{\Gamma_u}{u}{\multunion_{\iI} \MM_i}$ with
    $\Gamma_u = \inter_{\iI} \Gamma_u^i$. Since neither $I$ nor the
    $\MM_i$’s are empty, the choice operator is in both cases
    the identity and
    we can build the following derivation using
    rule~$\ruleAppJ$:
    \[ \inferrule*{
        \der{\Gamma_t}{\ctx\dc{\lx.t}}{\mult{{\MM_i} \to {\tau_i}}_{\iI}}
        \and \der{\Gamma_u}{u}{\multunion_{\iI} \MM_i}
      \and \der{\Gamma_r;y:\mult{\tau_i}_{\iI}}{r}{\sigma}}{
  \seq{\Gamma}{\gap{\ctx\dc{\lx.t}} u y r}{\sigma}} \]
  We verify $\Gamma = \Gamma' \inter \Gamma_r
  = \inter_{\iI} \Gamma'_i \inter \Gamma_r
  = \inter_{\iI} (\Gamma_t^i \inter \Gamma_u^i) \inter \Gamma_r
  = \Gamma_t \inter \Gamma_u \inter \Gamma_r$.

  \case{$t_1 =\lx.t$ and $t_1 = \gap t u x r$ and the reduction is internal}
  These cases are direct by the \ih
  \qedhere
\end{cas}
\end{proof}

We cannot conclude completeness straightaway, given that subject expansion
was only shown for non-erasing cases.
Instead, we prove that from any term on the right of a reduction, we can build a
derivation for the term on the left.
We rely on the previous lemma for the non-erasing steps, and construct
derivations for erasing ones, in which the typing environment grows with
anti-reduction.
We use the inductive characterization of strong normalization $\isn\djn$ to
recognize the left terms that are indeed strongly normalizing, which are the
only ones for which we can build a typing derivation.
\begin{lem}[Completeness for $\djncalc$]
  \label{l:completeness}%
  If $t \in \sn\djn$, then $t$ is $\sysFullJ$-typable.
\end{lem}
\begin{proof}
  In the statement, we replace $\sn\djn$ by $\isn\djn$, using
  \autoref{l:cbn-sn-isn}.
  We use induction on $\isn\djn$ to show the
  following stronger property $\mathcal P$: If $t \in \isn\djn$ then
  there are $\Gamma, \sigma$ such that $\der\Gamma{t}{\sigma}$, and if
  $t \in \nl$, then the property holds for any $\sigma$.
  \begin{cas}
    \case{$t = x$} We get $\seq{x:\mult{\sigma}}{x}{\sigma}$ by rule~$\ruleAxJ$, for any $\sigma$.

    \case{$t = \lx.s$, where $s \in \isn{\djn}$} By the \ih\ we
    have 
    $\der{\Delta}{s}{\tau}$. Let us write $\Delta$ as $\Gamma; x:\MM$,
    where $\MM$ is possibly empty. Then we get 
    $\der{\Gamma}{\lx.s}{\sigma}$, where $\sigma =\MM \to \tau$,
    by using rule~$\ruleAbsJ$ on the previous derivation. 

    \case{$t = \gap {\nl} u x r$,
    where $\nl, u, r \in \isn\djn$ and $r \in {\nf\whdj}$} By the \ih\ there
    are derivations $\der\Delta{u}{\rho}$ and
    $\der{\Lambda;x:\mult{\tau_i}_{\iI}}{r}{\sigma}$ with $I$
    possibly empty. Moreover, $\der\Delta{u}{\mult\rho}$
    holds by rule $\many$. If $r \in \nl$, we have a derivation
    for any type $\sigma$ by the stronger \ih

    We now construct a derivation
    $\der{\Pi}{\nl}{\choice{\mult{\emult \to {\tau_i}}_{\iI}}}$
    as follows:
    \begin{itemize}
      \item If $I = \emptyset$, then the \ih\ gives
        $\der{\Pi}{\nl}{\tau}$ for an arbitrary $\tau$, and
        then we obtain $\der{\Pi}{\nl}{\mult\tau}$ by rule~$\many$.
        We conclude by setting $\choice{\mult{\emult \to {\tau_i}}_{\iI}} = \mult \tau$. 
      \item If $I \neq \emptyset$, then by the stronger \ih\ we can derive
        $\der{\Pi_i}{\nl}{\emult \to {\tau_i}}$ for each $\iI$.
        We take $\Pi = \inter_{\iI} \Pi_i$ and we conclude
        with rule $\many$
        since $\choice{\mult{\emult \to {\tau_i}}_{\iI}} =
        \mult{\emult \to {\tau_i}}_{\iI}$. 
    \end{itemize}
    We conclude with rule
    $\ruleAppJ$ as follows, by setting in particular
    $\choice {\multunion_{\iI} \emult} = \mult \rho$.
    \[ \inferrule*{
        \der{\Pi}{\nl}{\choice{\mult{{\emult} \to {\tau_i}}_{\iI}}} \and
        \der{\Delta}{u}{\choice{\multunion_{\iI} \emult}} \and
      \der{\Lambda; y:\mult{\tau_i}_{\iI}}{r}{\sigma}}{
  \seq{ \Pi \inter\Delta \inter \Lambda}{\gap{s}{u}{y}{r}}{\sigma }} \]

  \case{$t \notin {\nf\whdj}$} That is, $t = \ctx\lrc{\gap{\ctx\dcnl{\lx.s}} u y r}$,
  where $t' = \ctx\lrc{\rsub {\rsub u x {\ctx\dcnl s}} y r}\in \isn\djn$, $ \ctx\dcnl s\in \isn\djn$, and $u \in \isn\djn$.
  Notice that $t \notin \nl$ by \autoref{l:nfwh}.
  By the \ih\ $t', \ctx\dcnl s$ and $u$ are typable.
  We show by a second induction on $\lrc$ that
  $\der{\Sigma}{t'}{\sigma}$ implies $\der{\Gamma}{t}{\sigma}$,
  for some $\Gamma$.
  For the base case $\lrc = \ec$, there are three cases.
  \begin{cas}
    \subcase{$x \in \fv s$ and $y \in \fv r$}
    Since $t'=\rsub {\rsub u x {\ctx\dcnl s}} y r$ is typable
    and $t \rew{\beta} t'$, then $t$ is also typable
    with $\Sigma$ and $\sigma$ by the non-erasing subject expansion \autoref{l:qse-strong}. We conclude with $\Gamma = \Sigma$.
    \subcase{$x \notin \fv s$ and $y \in \fv r$} Then
    $t' = \rsub {\ctx\dcnl s} y r$ and by \ih\ there is
    a derivation
    $\der{\Sigma}{\rsub {\ctx\dcnl s} y r}{\sigma}$. The
    anti-substitution
    \autoref{l:anti-substitution-lemma-right-strong}, gives
    $\der{\Lambda;y:\MN}{r}{\sigma}$,
    $\der{\Pi}{\ctx\dcnl s}{\MN}$ with
    $\Sigma = \Lambda \inter \Pi$. Let
    $\MN = \mult{\sigma_i}_{\iI}$. We have $I \neq \emptyset$ by
    \autoref{l:relevance-sysFullJ} since $y \in \fv r$. By the Split
    \autoref{l:split-many-strong} there are derivations
    $\der{\Pi_i}{\ctx\dcnl s}{\sigma_i}$ such that
    $\Pi = \inter_{\iI} \Pi_i$. Since $u \in \isn\djn$, the
    \ih\ gives a derivation $\der{\Delta}{u}{\rho}$ and by rule
    $\many$ we get $\der{\Delta}{u}{\mult{\rho}}$. Moreover,
    \autoref{l:relevance-sysFullJ} implies that
    $x \notin \dom{\Pi_i}$ for each $\iI$ because
    $x \notin \fv{\ctx\dcnl s}$, then we can construct
    derivations
    $\left(\der{\Pi_i}{\lx.\ctx\dcnl s}{\emult \to
    {\sigma_i}}\right)_{\iI}$. By \autoref{l:perml-typ} applied
    for each $\iI$,
    we retrieve
    $\left(\der{\Pi_i}{\ctx\dcnl{\lx.s}}{
    {\emult} \to {\sigma_i}}\right)_{\iI}$.
    And by rule~$\many$ we get
    $\der{\Pi}{\ctx\dcnl{\lx.s}}{\mult{\emult \to {\sigma_i}}_{\iI}}$.
    Finally, since $\choice{\mult{ {\emult} \to {\sigma_i}}_{\iI}}
    = \mult{\emult \to {\sigma_i}}_{\iI}$,
    it is sufficient to set $\choice{\multunion_{\iI} \emult} = \mult \rho$ and we obtain the following derivation:
    \[ \inferrule*{
        \der{\Pi}{\ctx\dcnl{\lx.s}}{\choice{\mult{
        {\emult} \to {\sigma_i}}_{\iI}}} \and
        \der{\Delta}{u}{\choice{\multunion \emult} } \and
      \der{\Lambda; y:\emult }{r}{\sigma}}{
  \seq{ \Gamma}{\gap{\ctx\dcnl{\lx.s}} u y r}{\sigma}} \]
  where
  $\Gamma = \Pi
  \inter \Delta\inter \Lambda$.
  We then conclude.
  \subcase{$y \notin \fv r$} Since
  $t'=\rsub {\rsub u x {\ctx\dcnl s}} y r$ is typable and
  $t' = r$, then
  there is a derivation $\der{\Lambda}{r}{\sigma}$ where
  $y \notin \dom \Lambda$ holds by relevance (so that $\Sigma = \Lambda$). We can then write
  $\der{\Lambda; y:\emult}{r}{\sigma}$. We construct a
  derivation of $t$ ending with rule~$\ruleAppJ$. For this we
  need two witness derivations for $u$ and
  $\ctx\dcnl{\lx.s}$.
  Since $u \in \isn\djn$, the \ih\ gives a derivation
  $\der{\Delta}{u}{\rho}$, and then we get $\der{\Delta}{u}{\mult\rho}$ by application of rule $\many$. Similarly, since
  $\ctx\dcnl s \in \isn\djn$, the \ih\ gives a derivation
  $\der{\Pi;x:\MM}{\ctx\dcnl s}{\tau}$ where $\MM$ can be
  empty. Thus
  $\der{\Pi}{\lx. \ctx\dcnl s}{ {\MM} \to \tau}$.
  By \autoref{l:perml-typ}, we get
  $\der{\Pi}{\ctx\dcnl{\lx.s}}{ {\MM} \to \tau}$, and then we get
  $\der{\Pi}{\ctx\dcnl{\lx.s}}{ \mult{{\MM} \to \tau}}$ by application of rule $\many$.
  Finally, by setting $\choice \emult = \mult{{\MM} \to \tau}$ and
  $\choice {\multunion \emult} = \mult \rho$ we construct the following derivation:
  \[ \inferrule*{
      \der{\Pi}{\ctx\dcnl{\lx.s}}{\choice{\emult}} \and
      \der{\Delta}{u}{\choice{\multunion \emult} } \and
    \der{\Lambda; y:\emult }{r}{\sigma}}{
\seq{ \Gamma}{\gap{\ctx\dcnl{\lx.s}} u y r}{\sigma}} \]
where
$\Gamma =\Pi \inter\Delta\inter
\Lambda$. We then conclude.
\end{cas}

Then, there are two inductive cases.
We extend the second \ih\ to multi-types trivially.
\begin{cas}
  \subcase{$\lrc = \gap{\lrc'}{u'} z {r'}$} Let consider the terms $t_0 =
  \ctx{\lrc'}{\gap{\ctx\dcnl{\lx.s}} u y r}$ and
  $t_1 = \ctx{\lrc'}{\rsub {\rsub u x {\ctx\dcnl
  s}} y r}$ so that $t = \gap {t_0}{u'}{z}{r'}$ and $t'
  = \gap{t_1}{u'}{z}{r'}$. The type derivation of $t'$ ends
  with a rule~$\ruleAppJ$ with the premises:
  $\der{\Sigma_1}{t_1}{\choice{\mult{{\MM_i} \to {\tau_i}}_{\iI}}}$,
  $\der{\Delta}{u'}{\choice{\multunion_{\iI} \MM_i}}$ and
  $\der{z:\mult{\tau_i}_{\iI}; \Lambda}{r'}{\sigma}$, where
  $\Sigma = \Sigma_1 \inter \Delta \inter \Lambda$. By the
  second \ih\ we get a derivation
  $\der{\Gamma_0}{t_0}{\choice{\mult{{\MM_i} \to {\tau_i}}_{\iI}}}$ for some $\Gamma_0$. We build a derivation for $t$
  with type $\sigma$ ending with rule $\ruleAppJ$ and using
  the derivations for $t_0$ and the ones for $u'$ and $r'$, so that the corresponding typing environment is $\Gamma = \Gamma_0 \inter \Delta \inter \Lambda$. We then conclude.

  \subcase{$\lrc = \gap{\nl}{u'} z {\lrc'}$}
  Let $t_0, t_1$ be the same as before so that $t = \gap {\nl}{u'}{z}{t_0}$ and $t' = \gap{\nl}{u'}{z}{t_1}$.
  We detail the case where $z \in \fv{t_0}$ and $z \notin \fv{t_1}$, the other ones being similar to case 1.
  The type derivation of $t'$ is as follows,
  with $\Sigma = \Gamma_{\nl} \inter \Delta \inter \Sigma'$.
  \[ \inferrule*[right=\ruleAppJ]{
      \der{\Gamma_{\nl}}{\nl}{\mult\tau} \and
      \der{\Delta}{u'}{\mult\rho} \and
      \der{\Sigma'}{t_1}{\sigma}}{ \der{\Gamma_{\nl} \inter \Delta \inter
  \Sigma'}{\gap{s'}{u'}{z}{t_1}}{\sigma}} \] By the second
  \ih\ we have a derivation $\der{z:\mult{\tau_i}_{\iI};
  \Gamma'}{t_0}{\sigma}$ for some $\Gamma'$. Also by relevance
  \autoref{l:relevance-sysFullJ}) we have $I \neq
  \emptyset$. By the \ih\ on property
  $\mathcal P$, we can build derivations
  $\der{\Pi_i}{\nl}{{\emult} \to {\tau_i}}$
  for each $\iI$ and thus a derivation
  $\der{\Pi}{\nl}{\mult{\emult \to {\tau_i}}_{\iI}}$
  by rule~$\many$ with $\Pi=
  \inter_{\iI}\Pi_i$. Setting
  $\choice{\multunion_{\iI} \emult} = \mult \rho$, we
  then build the following derivation:
  \[ \inferrule*[right=\ruleAppJ]{
      \der{\Pi}{\nl}{\mult{{\emult} \to {\tau_i}}_{\iI}}
      \and \der{\Delta}{u'}{\choice{\multunion_{\iI} \emult}}
    \and \der{z:\mult{\tau_i}_{\iI};\Gamma'}{t_0}{\sigma}}{
\seq{\Gamma}{\gap{\nl}{u'}{z}{t_0}}{\sigma}} \]
where $\Gamma = \Pi \inter \Delta \inter \Gamma'$.
We thus conclude.
\qedhere
\end{cas}
\end{cas}
\end{proof}

Soundness and completeness together entail:
\begin{thm}[Characterization]
  \label{l:sysFullJ-dbn}%
  System $\sysFullJ$ characterizes strong normalization, i.e.
  $t$ is $\sysFullJ$-typable if and only if
  $t \in \sn\dbeta $.
  Moreover, if $\dern{\Gamma}{t}{\sigma}{n}$
  then the number of reduction steps in any reduction sequence from $t$ to
  normal form is bounded by $n$.
\end{thm}

\begin{proof}
  Soundness holds by
  \autoref{l:bound}, while completeness holds by \autoref{l:completeness}.
  The bound is given by \autoref{l:bound}.
\end{proof}

\subsection{Quantitative Behavior of \texorpdfstring{$\pi$}{pi}}
\label{s:pi-not-quantitative}

We have mentioned already that $\pi$ is rejected by the quantitative type
systems $\sysFullJ$.
Concretely, this happens in the critical case when
$x \notin \fv r$ and $y \in \fv{r'}$ in
\[
  t_0 = \gap{\gap t u x r}{u'} y {r'}
  \rew\pi \gap t u x {\gap r {u'} y {r'}} = t_1
\]

\begin{exa}
  Consider $t_1 \rew\pi t_2$ with
  $t_1 = \gap{\gap{x_1}{y_1}{z_1}{x_2}}{y_2}{z_2}{\gap{z_2}{z_2}{z_3}{z_3}}$
  and $t_2 = \gap{x_1}{y_1}{z_1}{\gap{x_2}{y_2}{z_2}{\gap{z_2}{z_2}{z_3}{z_3}}}$.
  Let $\rho_1 = {\mult\sigma} \to {\tau}$ and
  $\rho_2 = {\mult\sigma} \to {{\mult\tau} \to \tau}$.
  For each $i\in\{1,2\}$ let $\Delta_i = x_1:\mult{\sigma_1};
  y_1:\mult{\sigma_2};x_2:\mult{\rho_i}$. Consider
  \[ \Psi = \inferrule*[vcenter]{
      \inferrule*{
        \inferrule*{ }
        {\seq{z_2:\mult{{\mult\tau} \to \tau}}{z_2}{\mult\tau \to \tau}}
      }{
        \seq{z_2:\mult{{\mult\tau} \to \tau}}{z_2}{\mult{{\mult\tau} \to \tau}}
      }
      \and \inferrule*{
      \inferrule*{ }{\seq{z_2:\mult\tau}{z_2}{\tau}}}
      {\seq{z_2:\mult\tau}{z_2}{\mult\tau}}
      \and \inferrule*{ }{\seq{z_3:\mult\tau}{z_3}{\tau}}
    }{
      \seq{z_2:\mult{{\mult\tau} \to \tau,\tau}}{\gap{z_2}{z_2}{z_3}{z_3}}{\tau}
  } \]
  and the derivation $\Phi_i$ for $i \in \{1,2\}$:
  \[ \Phi_i =
    \inferrule*[vcenter]{
      \inferrule*{
      \inferrule*{ }{\seq{x_1:\mult{\sigma_1}}{x_1}{\sigma_1}}}
      {\seq{x_1:\mult{\sigma_1}}{x_1}{\mult{\sigma_1}}}
      \and \inferrule*{
      \inferrule*{ }{\seq{y_1:\mult{\sigma_2}}{y_1}{\sigma_2}}}
      {\seq{y_1:\mult{\sigma_2}}{y_1}{\mult{\sigma_2}}}
    \and \inferrule*{ }{\seq{x_2:\mult{\rho_i}}{x_2}{\rho_i}}}{
\seq{\Delta_i}{\gap{x_1}{y_1}{z_1}{x_2}}{\rho_i}} \]
Then, for the term $t_1$, we have the following derivation:
\[ \inferrule*{
    \inferrule*{\Phi_1 \and \and \Phi_2}{
    \seq{\Delta_1 \inter \Delta_2}{\gap{x_1}{y_1}{z_1}{x_2}}{\mult{\rho_1,\rho_2}}}
    \and \inferrule*{
      \inferrule*{ }{\seq{y_2:\mult{\sigma}}{y_2}{\sigma}}
    \and \inferrule*{ }{\seq{y_2:\mult{\sigma}}{y_2}{\sigma}}}
    {\seq{y_2:\mult{\sigma,\sigma}}{y_2}{\mult{\sigma,\sigma}}}
    \and \Psi
  }{
    \seq{\Gamma_1}{\gap{\gap{x_1}{y_1}{z_1}{x_2}}{y_2}{z_2}{\gap{z_2}{z_2}{z_3}{z_3}}}{\tau}
} \]
where $\Gamma_1 = x_2:\mult{\rho_1,\rho_2}; y_2:\mult{\sigma,\sigma};
x_1:\mult{\sigma_1,\sigma_1}; y_1:\mult{\sigma_2,\sigma_2}$. 

While for the term $t_2$, we have:
\[ \inferrule*{
    \inferrule*{
    \inferrule*{ }{\seq{x_1:\mult{\sigma_1}}{x_1}{\sigma_1}}}
    {\seq{x_1:\mult{\sigma_1}}{x_1}{\mult{\sigma_1}}}
    \and \inferrule*{
    \inferrule*{ }{\seq{y_1:\mult{\sigma_2}}{y_1}{\sigma_2}}}
    {\seq{y_1:\mult{\sigma_2}}{y_1}{\mult{\sigma_2}}}
    \and \Phi
  }{
    \seq{\Gamma_2}{\gap{x_1}{y_1}{z_1}{
    \gap{x_2}{y_2}{z_2}{\gap{z_2}{z_2}{z_3}{z_3}}}}{\tau}
} \]
where
\[ \Phi = \inferrule*[vcenter]{
    \inferrule*{
      \left(\inferrule*{ }{\seq{x_2:\mult{\rho_i}}{x_2}{\rho_i}}\right)_{i \in
    \{1,2\}}}
    {\seq{x_2:\mult{\rho_1,\rho_2}}{x_2}{\mult{\rho_1,\rho_2}}}
    \and \inferrule*{
      \inferrule*{ }{\seq{y_2:\mult{\sigma}}{y_2}{\sigma}}
    \and \inferrule*{ }{\seq{y_2:\mult{\sigma}}{y_2}{\sigma}}}
    {\seq{y_2:\mult{\sigma,\sigma}}{y_2}{\mult{\sigma,\sigma}}}
    \and \Psi
  }{
\seq{\Gamma_2}{\gap{x_2}{y_2}{z_2}{\gap{z_2}{z_2}{z_3}{z_3}}}{\tau}} \]
and $\Gamma_2 = x_2:\mult{\rho_1,\rho_2}; y_2:\mult{\sigma,\sigma};
x_1:\mult{\sigma_1}; y_1:\mult{\sigma_2}$.

Thus, the multiset types of $x_1$ and $y_1$ are not the same in $\Gamma_1$ and $\Gamma_2$.
Despite the fact that the step $t_1 \rew{\pi} t_2$ does not erase any subterm,
the typing environment is losing quantitative information.
If we were to use sets of types instead of multisets, then  $\Gamma_1$ and $\Gamma_2$
  would be identical.
  This is exactly what happens in the idempotent framework~\cite{matthes00}
where subject reduction and expansion hold for $\pi$.
\end{exa}

Despite the fact that quantitative subject reduction fails for
some $\pi$-steps, the following weaker property is sufficient
to recover (qualitative) soundness of our typing system $\sysFullJ$
w.r.t. the reduction relation $\rew{\jn}$.
We will use soundness in \autoref{sec:equivalences} to show equivalence
between $\sn{\djn}$ and $\sn{\jn}$.

\begin{lem}[Typing behavior of $\pi$]
  \label{l:non-sr-pi}%
  Let $\dernp{\Gamma}{t_1}{\sigma}{n_1}{\sysFullJ}$.
  If $t_1 = \gap{\gap t u x r}{u'}{y}{r'} \rrule{\pi}
  t_2 = \gap t u x {\gap r {u'}{y}{r'}}$,
  then there are $n_2$ and $\Sigma \sqsubseteq \Gamma$ such that
  $\dernp{\Sigma}{t_2}{\sigma}{n_2}{\sysFullJ}$ with $n_1 \geq n_2$.
\end{lem}
\begin{proof}
  The derivation of $t_1$ ends with~$\ruleAppJ$, with $\Gamma = \Gamma' \inter \Delta_{u'} \inter \Lambda_{r'}$
  and $n_1 = 1 + n' + n_{u'} + n_{r'}$.
  \[ \inferrule*{
      \dern{\Gamma'}{\gap t u x r}{ \choice{\mult{{\MM_i} \to {\tau_i}}_{\iI}}}{n'}
      \and\mkern-18mu \dern{\Delta_{u'}}{u'}{\choice{\multunion_{\iI} \MM_i}}{n_{u'}}
    \and\mkern-18mu \dern{\Lambda_{r'};y:\mult{\tau_i}_{\iI}}{r'}{\sigma}{n_{r'}}}{
\seq{\Gamma}{\gap{\gap t u x r}{u'} y {r'}}{\sigma}} \]
There are two possibilities.
\begin{cas}
  \case{$I \neq \emptyset$}
  Then $\choice{\mult{{\MM_i} \to {\tau_i}}_{\iI}} = \mult{{\MM_i} \to {\tau_i}}_{\iI}$ and for each $\iI$ there is one derivation of $\gap t u x r$ having the following form:
  {\small \[ \inferrule*[right=\ruleAppJ]{
        \dern{\Gamma_t^i}{t}{ \choice{\mult{{\MN_j} \to {\rho_j}}_{\jJ_i}}}{n_t^i}
        \and\mkern-18mu \dern{\Delta_u^i}{u}{\choice{\multunion_{\jJ_i} \MN_j}}{n_u^i}
      \and\mkern-18mu \dern{\Lambda_r^i; x:\mult{\rho_j}_{\jJ_i}}{r}{{\MM_i} \to {\tau_i}}{n_r^i}}{
      \seq{\Gamma_t^i \inter \Delta_u^i \inter \Lambda_r^i}{\gap t u x
r}{{\MM_i} \to {\tau_i}}}\]}
where $\Gamma' = \inter_{\iI} (\Gamma_t^i \inter \Delta_u^i \inter \Lambda_r^i)$
and $n' = \sum_{\iI} n_t^i + n_u^i + n_r^i$.
From $(\dern{\Lambda_r^i; x:\mult{\rho_j}_{\jJ_i}}{r}{{\MM_i} \to {\tau_i}}{n_r^i})_{\iI}$
we can construct a derivation
$\Phi_r = \dern{\interii \Lambda_r^i; x:\multjj{\rho_j}}{r}{\multii{{\MM_i}
\to {\tau_i}}}{+_{\iI} n_r^i}$
using rule $\many$, where $J = \uplus_{\iI} J_i$.
We then construct the following derivation:
\[ \Psi =
  \inferrule*[vcenter]{
    \Phi_r
    \and \dern{\Delta_u'}{u'}{\chmultii \MM}{n_{u'}}
  \and \dern{\Lambda_{r'}; y:\multii{\tau_i}}{r'}{\sigma}{n_{r'}}}{
  \seq{\interii \Lambda_r^i \inter \Delta_{u'} \inter
\Lambda_{r'};x:\multjj{\rho_j} }{\gap r {u'} y {r'}}{\sigma}} \]
We then build two derivations $\dern{\Gamma_t}{t}{\choice{\multjj{{\MN_j}
\to {\rho_j}}}}{n_t}$
with $\Gamma_t \sqsubseteq \interii \Gamma_t^i$ and $n_t \leq +_{\iI} n_t^i$
and $\dern{\Delta_u}{u}{\chmultjj \MN}{n_u}$
with $\Gamma_u \sqsubseteq \interii \Gamma_u^i$ and $n_u \leq +_{\iI} n_u^i$
as follows:
\begin{itemize}
  \item If $x \in \fv r$, then all the $J_i$'s, and thus also $J$, are non-empty by relevance
    so that $\choice{\mult{{\MN_j} \to {\rho_j}}_{\jJ_i}} = \mult{{\MN_j}
    \to {\rho_j}}_{\jJ_i}$.
    Also, $\choice{\mult{{\MN_j} \to {\rho_j}}_{\jJ}} = \mult{{\MN_j}
    \to {\rho_j}}_{\jJ}$.
    We obtain the expected derivation for $t$ by \autoref{l:split-many-strong}, with
    $\Gamma_t = \interii \Gamma_t^i$, $n_t = +_{\iI} n_t^i$.
    Now for $u$, notice that for each $\iI$ we can have either
    $\choice{\multunion_{\jJ_i} \MN_j} = \multunion_{\jJ_i} \MN_j$
    or, if all the $\MN_j$’s are empty, $\choice{\multunion_{\jJ_i} \MN_j} =
    \mult{\sigma_i}$ for some $\sigma_i$
    derived by $\dern{\Delta_u^k}{u}{\mult{\sigma_k}}{n_u^k}$.
    Then, there are two possibilities.
    \begin{enumerate}
      \item If $\multunion_{\jJ} \MN_j = \emult$, we take an arbitrary $k \in I$
        and let $\chmultjj \MN = \mult{\sigma_k}$
        so that we can give a derivation $\dern{\Delta_u}{u}{\mult{\sigma_k}}{n_u}$
        with $\Delta_u = \Delta_u^k \sqsubseteq \interii \Delta_u^i$
        and $n_u = n_u^k \leq +_{\iI} n_u^i$.
      \item Otherwise, we have $\chmultjj \MN = \multunion_{\jJ} \MN_j$.
        Let $I'$ be the subset of $I$ such that for each $\iI'$ we have $\multunion_{\jJ_i} \MN_j \neq \emult$
        and $J' = \uplus_{\iI'} J_i$.
        By Lem.~30 we build a derivation $\dern{\Delta_u}{u}{\multunion_{\jJ'} \MN_j}{n_u}$
        such that $\Delta_u = \inter_{\iI'} \Delta_u^i \sqsubseteq \interii \Delta_u^i$
        and $n_u = +_{\iI'} n_u^i \leq \interii n_u^i$.
    \end{enumerate}
  \item If $x \notin \fv r$, then all the $J_i$'s are empty by relevance.
    Therefore, for each $\iI$ there are a $\sigma_i, \sigma_i'$ such that
    $\choice{\mult{{\MN_j} \to {\rho_j}}_{\jJ_i}} = \mult{\sigma_i}$
    is derived by $\dern{\Gamma_t^i}{t}{\mult{\sigma_i}}{n_t^i}$
    and $\choice{\multunion_{\jJ_i} \MN_j} = \mult{\sigma_i'}$
    is derived by $\dern{\Gamma_u^i}{u}{\mult{\sigma_i'}}{n_u^i}$.
    We take an arbitrary $k \in I$
    and we take $\choice{\multjj{{\MN_j} \to {\tau_j}}} = \mult{\sigma_k}$ and
    $\chmultjj\MN = \mult{\sigma_k'}$.
    We obtain the expected derivation by taking
    $\Gamma_t = \Gamma_t^k \sqsubseteq \interii \Gamma_t^i$,
    $n_t = n_t^k \leq +_{\iI} n_t^i$,
    $\Gamma_u = \Gamma_u^k \sqsubseteq \interii \Gamma_u^i$
    and $n_u = n_u^k \leq +_{\iI} n_u^i$.
\end{itemize}
Finally, we build the following derivation of size $n_2$.
\[ \inferrule*{
    \dern{\Gamma_t}{t}{\choice{\mult{{\MN_j} \to {\tau_j}}_{\jJ}}}{n_t}
    \and \dern{\Delta_u}{u}{\chmultjj \MN}{n_u}
  \and \Psi}{
\seq \Sigma {\gap t u x {\gap r {u'} y {r'}}}{\sigma}} \]
We have $\Sigma = \Gamma_t \inter \Delta_u \interii \Lambda_r^i \inter \Delta_{u'} \inter \Lambda_{r'} \sqsubseteq \Gamma$
and $n_2 = n_t + n_u +_{\iI} n_r^i + n_{u'} + n_{r'} \leq n_1$.

\case{$I = \emptyset$}
Then there is some $\tau$ such that $\choice{\mult{{\MM_i} \to {\tau_i}}_{\iI}} = \mult{\tau}$
and the derivation of $\gap t u x r$ ends as follows:
\[ \inferrule*[right=\ruleAppJ]{
    \dern{\Gamma_t}{t}{ \choice{\mult{{\MN_j} \to {\rho_j}}_{\jJ}}}{n_t}
    \and\mkern-18mu \dern{\Delta_u}{u}{\choice{\multunion_{\jJ} \MN_j}}{n_u}
  \and\mkern-18mu \dern{\Lambda_r; x:\mult{\rho_j}_{\jJ}}{r}{\tau}{n_r}}{
  \inferrule*[right=\many]{
  \seq{\Gamma_t \inter \Delta_u \inter \Lambda_r}{\gap t u x r}{\tau}}{
\seq{\Gamma_t \inter \Delta_u \inter \Lambda_r}{\gap t u x r}{\mult\tau}}} \]
with $\Gamma' = \Gamma_t \inter \Delta_u \inter \Lambda_r$
and $n' = n_t + n_u + n_r$.

We construct the following derivation of size $n_2$:
\[ \inferrule*{
    \dern{\Gamma_t}{t}{\choice{\multjj{{\MN_j} \to {\rho_j}}}}{n_t}
    \and \dern{\Delta_u}{u}{\chmultjj \MN}{n_u}
  \and \Psi}{
\seq \Sigma {\gap t u x {\gap r {u'} y {r'}}}{ \sigma}} \]
where
\[ \Psi =
  \inferrule*[vcenter]{
    \dern{\Lambda_r;x:\multjj{\rho_j}}{r}{\mult \tau}{n_r}
    \and \dern{\Delta_u'}{u'}{\chmultii \MM}{n_{u'}}
  \and \dern{\Lambda_{r'}}{r'}{\sigma}{n_{r'}}}{
  \seq{\Lambda_r \inter \Delta_{u'} \inter
\Lambda_{r'};x:\multjj{\rho_j}}{\gap r {u'} y {r'}}{\sigma}} \]
We have $\Sigma = \Gamma_t \inter \Delta_u \inter \Lambda_r \inter \Delta_{u'} \inter \Lambda_{r'} = \Gamma$
and $n_2 = n_t + n_u + n_r + n_{u'} + n_{r'} = n_1$.
\qedhere
\end{cas}
\end{proof}

\subsection{Soundness for \texorpdfstring{$\jcalc$}{Lambda J}}
The previous lemma states that reducts of typed terms are also typed.
To show that reduction of typed terms terminates, we show that the maximal length of
reduction to normal form is bounded 
by the size of the type derivation,
and so is finite.
This is similar to what we have done for $\rew\djn$.

We recall that for each $t \in \sn{\jn}$, $\maxred[\jn]{t}$ represents
the maximal length of a $(\jn)$-reduction sequence to the $(\jn)$-normal form
starting at $t$.  We also define $\rbmaxredbp{t}$ as the maximal
number of $\beta$-steps in $(\jn)$-reduction sequences from $t$ to
its $(\jn)$-normal form.
Like $\maxred[\jn]{t}$, $\rbmaxredbp t$ is bounded by the size of any type
derivation for $t$.
Notice that, in general, $\rbmaxredbp{t}
\geq \maxred[\beta] t$, simply because $\pi$ creates
$\beta$-redexes, as already discussed.
Statements \ref{l:commutation-beta-pi} to \ref{l:pi-preserves-length-beta} are
needed to define $\maxred[\jn] t$ inductively.  We will
write $\pi(t)$ for the (unique) $\pi$-normal form of $t$.

\begin{lem}
  \label{l:commutation-beta-pi}
  If $t_1 \rew{\beta} t_2$ and $ t_1 \rew{\pi} t_3$, then
  there is $t_4$ such that $t_3 \rew{\beta} t_4$ and
  $t_2 \rewn{\pi} t_4$.
\end{lem}
\begin{proof}
  By case analysis of the possible overlaps of the two contracted redexes.
\end{proof}

\begin{lem}
  \label{l:commutation-beta-pi-to-nf}
  If $t_1 \rew{\beta} t_2$, 
  then there is $t_3$ such that
  $\pi(t_1) \rew{\beta} t_3$
  and $t_2 \rewn{\pi} t_3$.
\end{lem}
\begin{proof}
  By induction on the reduction sequence from $t_1$ to $\pi(t_1)$
  using \autoref{l:commutation-beta-pi} for the base case.
\end{proof}

\begin{lem}
  \label{l:from-t-to-pit}
  If there is a $(\jn)$-reduction sequence $\rho$ starting at $t$ and containing $k$ $\beta$-steps, then there is a $(\jn)$-reduction sequence $\rho'$ 
  starting at $\pi(t)$ and also containing $k$ $\beta$-steps.
\end{lem}
\begin{proof}
  By induction on the (necessarily finite) reduction sequence $\rho$.
  If the length of $\rho$ is $0$, then $k=0$ and the property is trivial.
  If the length of $\rho$ is $1+n$, we analyze the two possible cases:
  \begin{enumerate}
    \item If $\rho$ is $t \rew{\beta} t'$ followed by $\rho_0$ of length $n$ and containing $k_0 = k-1$ $\beta$-steps, then
      the property holds for $t'$ w.r.t. $\pi(t')$.
      But \autoref{l:commutation-beta-pi-to-nf}
      gives a term $t''$ such that $\pi(t) \rew{\beta} t''$ and $t' \rewn{\pi} t''$.
      Then we construct the $(\jn)$-reduction sequence
      $\pi(t) \rew{\beta} t'' \rewn{\pi} \pi(t'') = \pi(t')$ followed by the one obtained by the \ih\ This new sequence has $1+k_0=k$ $\beta$-steps.
    \item If $\rho$ is $t \rew{\pi} t'$ followed by $\rho_0$ of length $n$ and containing $k_0=k$ $\beta$-steps, then
      the property holds for $t'$ w.r.t. $\pi(t')$. Since $\pi(t)=\pi(t')$, we are done by the \ih
      \qedhere
  \end{enumerate}
\end{proof}

\begin{lem}
  \label{l:pi-nf-preserves-length-beta}
  $\rbmaxredbp{t} = \rbmaxredbp{\pi(t)}$.
\end{lem}
\begin{proof}
  First we prove $\rbmaxredbp{t} \leq \rbmaxredbp{\pi(t)}$. If there is a $(\jn)$-reduction sequence starting at
  $t$ and containing $k$ $\beta$-steps,
  then the same happens for $\pi(t)$ by \autoref{l:from-t-to-pit}.
  Next we prove $\rbmaxredbp{t} \geq \rbmaxredbp{\pi(t)}$. If there is a $(\jn)$-reduction sequence starting at
  $\pi(t)$ and containing $k$ $\beta$-steps,
  then the same happens for $t$ because
  it is sufficient to prefix this sequence with the steps $t \rewn{\pi} \pi(t)$.
  We conclude $\rbmaxredbp{t} = \rbmaxredbp{\pi(t)}$.
\end{proof}

\begin{cor}
  \label{l:pi-preserves-length-beta}
  If $t \rew{\pi} t'$, then $\rbmaxredbp{t} = \rbmaxredbp{t'}$.
\end{cor}
\begin{proof}
  Suppose $t \rew{\pi} t'$, hence $\pi(t)=\pi(t')$. Then we have $\rbmaxredbp{t} =\rbmaxredbp{\pi(t)}
	= \rbmaxredbp{\pi(t')}=\rbmaxredbp{t'}$. The first and last equalities are justified by \autoref{l:pi-nf-preserves-length-beta}.
\end{proof} 

Following \cite{matthes00}, $\sn{\jn}$ admits an inductive characterization $\isn {\jn}$, given in \autoref{fig:SN}, which uses the following inductive generation for $\jterms$-terms:
\begin{equation} \label{eq:alternative-generation-of-terms}
	t,u,r \, \Coloneqq \, x\vec{S} \mid \lx.t \mid (\lx.t)S\vec{S} \qquad  S \,
	\Coloneqq \, \gap{}{u}{y}{r}
\end{equation}
Hence $S$ stands for a \emph{generalized} argument, while $\vec{S}$ denotes a possibly empty list of $S$'s. Notice that at most one rule applies to a given term, so the rules are deterministic (and thus invertible).

As argued before for the $\lambda$-calculus, the use of vectors $\vec{S}$ of generalized arguments could be avoided by employing \emph{generalized weak-head contexts} $\ctxName S \Coloneqq \ec \mid \gap {\ctxName S} u y r$. There is a one-to-one correspondence between such contexts and the vectors $\vec{S}$, and the formal $\ctx{\ctxName S}r$ corresponds to the informal notation $r\vec{S}$. Contexts $\ctxName S$ are particular cases of left-right contexts $\ctxName R$. Hence, rule $\isnbpbetarule$ in Fig.~\ref{fig:SN} is the same rule as the particular case $\dcnl=\ec$ of rule $\isnbetarule$ in \autoref{d:isn}. Notice also that rules $\isnbpvarrule$ and $\isnbpabsrule$ in Fig.~\ref{fig:SN} are like rules $\isnvarrule$ and $\isnabsrule$ in that definition.
	
	\begin{figure}[t]
		\begin{mathpar}
			\inferrule*[right=\isnbpvarrule]{ }{x\in \isn {\jn}}
			\and \inferrule*[right=\isnbphvarrule]{u,r\in \isn {\jn}}{
				\gap{x}{u}{z}{r}\in \isn{\jn}}
			\and \inferrule*[right=\isnbpabsrule]{t\in \isn {\jn}}{
				\lx.t\in \isn {\jn}}
			\and \inferrule*[right=\isnbppirule]{\gap{x}{u}{y}{rS}\vec{S}\in \isn {\jn}}{
				\gap{x}{u}{y}{r}S\vec{S}\in \isn {\jn}}
			\and \inferrule*[right=\isnbpbetarule]{\rsub{\rsub{u}{x}t}{y}r \vec{S}\in \isn {\jn} \quad t,u \in \isn {\jn}}{
				\gap{(\lx.t)}{u}{y}{r} \vec{S}\in \isn {\jn}}
		\end{mathpar}
		\caption{Inductive characterization of the strong $(\jn)$-normalizing $\jcalc$-terms}
		\label{fig:SN}
	\end{figure}

\begin{lem} 	The function
    $\rbmaxredbp{\cdot}:\sn{\jn}\to\mathbb{N}_{0}$
    verifies the following equalities\footnote{These equalities can be seen as giving an alternative, recursive definition of function $\rbmaxredbp{\cdot}$, based on the inductive definition of $\sn{\jn}$ given in Fig.~\ref{fig:SN}.}:
  \label{l:equalities}
  \[ \begin{array}{llll}
    \rbmaxredbp{x} & = & 0 \\
    \rbmaxredbp{\lx. t} & = & \rbmaxredbp{t}\\
    \rbmaxredbp{\gap x u y r} & = & \rbmaxredbp{u} + \rbmaxredbp{r} \\
    \rbmaxredbp{\gap {\gap {x} u y r}{u'} z {r'} \vec{S}} & = & \rbmaxredbp{\gap {x} u y {\gap r {u'} z {r'}} \vec{S}} \\
    \rbmaxredbp{\gap{(\lx.t)} u y r \vec{S}} & = &
    \begin{cases}
      1 + \rbmaxredbp{\rsub {\rsub u x t} y r \vec{S}} &\!\! \mbox{if } x \in \fv{t} \mbox{ and } y \in \fv{r} \\
      1 + \rbmaxredbp{\rsub t y r \vec{S}} + \rbmaxredbp{u} &\!\! \mbox{if } x \notin \fv{t} \mbox{ and } y \in \fv{r} \\
      1 + \rbmaxredbp{r \vec{S}} + \rbmaxredbp{t} + \rbmaxredbp{u} &\!\! \mbox{if } y \notin \fv{r} \\
    \end{cases}
  \end{array} \]
\end{lem}

\begin{proof}
  The first three equalities are straightforward. The fourth is justified by \autoref{l:pi-preserves-length-beta}.   As to the fifth, notice $\rbmaxredbp{\gap{(\lx.t)} u y r \vec{S}}=\rbmaxredbp{\gap{(\lx.t)} u y {r \vec{S}}}$, again due to \autoref{l:pi-preserves-length-beta}; and notice that, in the
    \rhs\  of the equation, the scope of substitutions $\rsub {\_}y{\_}$ can be understood as encompassing $\vec{S}$. So, the fifth equation can be rewritten as
\[
 \begin{array}{llll}
	\rbmaxredbp{\gap{(\lx.t)} u y s} & = &
	\begin{cases}
		1 + \rbmaxredbp{\rsub {\rsub u x t} y s} & \mbox{ if } x \in \fv{t} \mbox{ and } y \in \fv{s} \\
		1 + \rbmaxredbp{\rsub t y s} + \rbmaxredbp{u} & \mbox{ if } x \notin \fv{t} \mbox{ and } y \in \fv{s} \\
		1 + \rbmaxredbp{s} + \rbmaxredbp{t} + \rbmaxredbp{u} & \mbox{ if } y \notin \fv{s} \\
	\end{cases}
\end{array} 
\]


The inequality \lhs\ $\geq$ \rhs\ is obvious since, in each case, there is a $(\beta,\pi)$-reduction sequence from $\gap{(\lambda x.t)}uys$ containing the number of $\beta$-steps specified by \rhs

As to \lhs\ $\leq$ \rhs, consider an arbitrary $(\beta,\pi)$-reduction sequence from $\gap{(\lambda x.t)}uys$ to the $\beta\pi$-normal form $r'$: it has necessarily the form
\[
\gap{(\lambda x.t)}uys \rewn{\beta\pi} \gap{(\lambda x.t')}{u'}y{s'}\rew{\beta}\rsub{\rsub{u'}{x}{t'}}{y}{s'}=:r\rewn{\beta\pi}r' \qquad(*)
\]
where, for each $E=t,u,s,r$, one has the reduction $E\rewn{\beta\pi}E'$ witnessed by a reduction sequence containing $l_E$ $\beta$-reduction steps. Hence, the reduction sequence $(*)$ contains $l:=1+l_t+l_u+l_s+l_r$ $\beta$-reduction steps. Moreover, for each $E=t,u,s$, one has 
\[
\rbmaxredbp{E}\geq\rbmaxredbp{E'}+l_E \qquad(**)
\]

  \begin{cas}
    \case{$x\in\fv t$ and $y\in\fv s$} We want $1 + \rbmaxredbp{\rsub {\rsub u x t} y s} \geq l$. Now
\[
\begin{array}{cll}
&1 + \rbmaxredbp{\rsub {\rsub u x t} y s} \\
= &  1 + \rbmaxredbp{\rsub uxt} \times  |s|_y + \rbmaxredbp{s}\\
= &  1 + (\rbmaxredbp{u} \times |t|_x + \rbmaxredbp{t})\times |s|_y + \rbmaxredbp{s}\\
\geq &  1 + ((\rbmaxredbp{u'}+l_u)\times |t|_x + \rbmaxredbp{t'}+l_t)\times |s|_y + \rbmaxredbp{s'}+l_s & \text{(by $(**)$)}\\
\geq &  1 +  (\rbmaxredbp{u'}\times |t'|_x + \rbmaxredbp{t'})\times |s'|_y + \rbmaxredbp{s'} + (l_u\times |t|_x +l_t)\times |s|_y + l_s& \text{(a)}\\
= &  1 +  \rbmaxredbp{r} + (l_u\times |t|_x +l_t)\times |s|_y + l_s& \text{(b)}\\
\geq &  1 +  \rbmaxredbp{r} + l_u +l_t + l_s& \text{(c)}\\
\geq & l &\text{(d)}
\end{array}
\]
that are justified:
(a) by $|t|_x\geq|t'|_x$ and $|s|_y\geq|s'|_y$;
(b) by $r=\rsub{\rsub{u'}{x}{t'}}{y}{s'}$;
(c) by $|t|_x,|s|_y\geq1$;
and (d) by $\rbmaxredbp{r}\geq l_r$.

\case{$x\notin\fv t$ and $y\in\fv s$ } Similar but simpler.

\case{$y\notin\fv s$} Then $s\rewn{\beta\pi}s'=r\rewn{\beta\pi}r'$, hence $l_s+l_r\leq\rbmaxredbp{s}$. Then $l\leq 1 + l_t + l_u + \rbmaxredbp{s}  \leq 1 + \rbmaxredbp{s} + \rbmaxredbp{t} + \rbmaxredbp{u}$, as required.
\qedhere
\end{cas}
\end{proof}

\begin{lem}
  \label{l:bound-pi}
  If $\dernp{\Gamma}{t}{\sigma}{k}{\sysFullJ}$, then $t\in\sn{\jn}$ and $\rbmaxredbp t \leq k$.
\end{lem}

\begin{proof} 
%
Let $\rbmaxp{t}$ be the length of the longest $\pi$-reduction sequence of $t$.
We proceed by induction on the pair
$\pair k {\rbmaxp{t}}$ with respect to the lexicographic order and we reason by case analysis on $t$, according to the inductive definition (\ref{eq:alternative-generation-of-terms}) of terms.

The proofs for cases $t = x$, $t = \lx.u$, $t = \gap x u y r$ and $t = \gap{(\lx.s)} u y r \vec{S}$ are similar to the ones in \autoref{l:bound}, only replacing  $\maxred[\djn]{t}$ by $\rbmaxredbp t$, and using the inductive characterization in Fig.~\ref{fig:SN} instead of that in \autoref{d:isn}. We only show here the most interesting case, which is $t = \gap {\gap {x} u y r}{u'} z {r'} \vec{S}$.

Let $t' = \gap {x} u y {\gap r {u'} z {r'}} \vec{S}$. Since $t\to_{\pi}t'$, $\rbmaxredbp{t}=\rbmaxredbp{t'}$, due to \autoref{l:pi-preserves-length-beta}. By \autoref{l:non-sr-pi} there is a type derivation $\dernp{\Delta}{t'}{\sigma}{k'}{\sysFullJ}$ with $k' \leq k$ and $\Delta \sqsubseteq \Gamma$.
Since $k'\leq k$ and $\rbmaxp{t}>\rbmaxp{t'}$, we can use the \ih\ and we get $t'\in\sn{\jn}$ and $\rbmaxredbp{t'} \leq k'$. By rule $\isnbppirule$ in Fig.~\ref{fig:SN}, we obtain $t\in\sn{\jn}$. Given that $\rbmaxredbp{t}=\rbmaxredbp{t'}$, we obtain $\rbmaxredbp{t} \leq k' \leq k$.
\end{proof}

As a corollary we obtain:
\begin{lem}[Soundness for $\jcalc$]
  \label{l:soundness-pi}
  If $t$ is $\sysFullJ$-typable, then $t \in \sn{\jn}$.
\end{lem}
\begin{proof}
  By \autoref{l:bound-pi}, the number of $\beta$-reduction steps in any
  $(\jn)$-reduction sequence starting at $t$ is finite. So
  in any infinite $(\jn)$-reduction sequence starting at
  $t$, there is necessarily a term $u$ from which there is an infinite
  amount of $\pi$-steps only. But this is impossible since $\pi$
  terminates, so we conclude by contradiction. 
\end{proof}

%% file: faithfulness.tex
\section{Faithfulness of the Translation}
\label{sec:faithfulness}

The original translation of generalized applications into ES
(see~\cite{espiritosanto07}), based on $\mapjes{\gap t u x r} = \esub
{\mapjes t\mapjes u} x \mapjes r$ (full details are given below), is not conservative with respect to
strong normalization; this is also true for the original translation
to $\lambda$-terms given by~\cite{joachimski03},
which is based on $\mapjes{\gap
    t u x r} = \rsub {\mapjes t\mapjes u} x \mapjes r$: it preserves
  strong normalization but normalizes too much. Indeed, in a
$\beta$-redex $s\eqdef\gap {(\lx.t_0)} u y r$, the
interaction of $\lx.t_0$ with the argument $u$ is
materialized by the internal substitution in the contractum term
$\rsub{\rsub u x t_0} y r$.
Such interaction may be elusive: if the
external substitution is vacuous (that is, if $y$ is not free in $r$),
$\beta$-reduction will simply throw away the $\lambda$-abstraction
$\lx.t_0$ and its argument $u$. In the
  translated term $\mapjes s$, the $\beta$-redex
$\mapjes{(\lx.t_0)}\mapjes u= (\lx. \mapjes {t_0})\mapjes u$ is also
thrown away in the case of translation to $\lambda$-terms, whereas it
may reduce in the context of the explicit substitution $\esub
{(\lx.\mapjes{t_0}) \mapjes u} y \mapjes r$.

The different interactions between the abstraction and its argument in the two
mentioned models of computation has important consequences. Here is an example.
\begin{exa}
  \label{e:not-faithful}
Let $\delta \eqdef \lx.\gap x x z z$.
Let $r$ be a $\jterms$-term with no free occurrences of $y$, \eg\ $r = \ly.y$.
The only possible reduction from the $\jterms$-term $\gap \delta \delta y r$ is to
$r = \lx.x$, which is a normal form in $\jcalc$  or $\djncalc$, whereas
the subterm $\mapjes\delta \mapjes\delta =
(\lx. \esub{xx}{z}z)(\lx. \esub{xx}{z}z)$ may reduce forever in the context of a vacuous explicit substitution, \ie\  $\esub
{\mapjes\delta\mapjes\delta} y \mapjes r \rewp{} \esub
{\mapjes\delta\mapjes\delta} y \mapjes r $ holds in the ES calculus.
\end{exa}

In this section we define an alternative encoding to the original one and prove
it faithful: a term in $\jterms$ is $\dbeta$-strongly normalizing \textit{iff}
its alternative encoding is strongly normalizing in the ES framework.
In a later section, we use this connection with ES to establish the equivalence
between strong normalization of $\djncalc$ and $\jcalc$.

\subsection{A New Translation}%
\label{l:new_translation}

We define the syntax and semantics of an ES calculus borrowed
from~\cite{accattoli12b} to which we relate $\djncalc$.
It is a simple calculus where $\beta$ is implemented in two independent steps:
one creating a let-binding, and another one substituting the term bound.
It has a notion of distance which allows to reduce redexes such as $(\esub N x
{(\ly.M)}) P \rew\db \esub N x \esub P y M$, where the ES $\esub N x$ lies
between the abstraction and its argument.
Terms and list contexts are given by:
\[\begin{array}{rrcl}
  (\esterms) & M,N,P,Q & \Coloneqq & x \mid \lx.M \mid MN \mid \esub N x M\\
  \gramTitle{List contexts} & \lc & \Coloneqq & \ec \mid \esub N x \lc
\end{array} \]
The calculus $\escalc$ is defined by $\calcwithrules{\esterms}{\db,\sub}$, meaning that $\esterms$ is the set of terms and that this set is equipped with $\to_{\db}$ and $\to_{\sub}$, the reduction relations obtained by closing $\db,\sub$ under all contexts, where:
\begin{align*}
  \ctx\lc{\lx.M} N &\rrule\db \ctx\lc{\esub N x M}\\
  \esub N x M &\rrule\sub \rsub N x M
\end{align*}

Now, consider the (original) translation from $\jterms$ to $\esterms$~\cite{espiritosanto07}:
\begin{mathpar}
  \mapjes x \eqdef x \and \mapjes{(\lx.t)} \eqdef \lx.\mapjes t
  \and \mapjes{\gap t u y r} \eqdef \esub {\mapjes t \mapjes u} y \mapjes r
\end{mathpar}
According to it, the notion of distance in $\escalc$ corresponds to our notion
of distance for $\djncalc$.
For instance, the application $\gap t u x \_$ in the term $\gap{\gap t u x {\ly.r}}{u'}{z}{r'}$ can be seen as
a substitution $\esub {\mapjes t\mapjes u} x$ inserted between the abstraction $\ly.r$ and the argument $u'$.
But how can we now (informally) relate $\pi$ to the notions of existing permutations for $\escalc$?
Using the previous translation, we can see that
$t_0
= \gap{\gap{t}{u}{x}{r}}{u'}{y}{r'}\rrule{\pi}\gap{t}{u}{x}{\gap{r}{u'}{y}{r'}}
= t_1$
simulates as
\[
  \mapjes{t_0} = \esub {(\esub {\mapjes t \mapjes u} x \mapjes r) \mapjes{u'}} y
  \mapjes{r'}
  \rew{} \esub {\esub {\mapjes t \mapjes u} x ({\mapjes r} \mapjes{u'})} y
  \mapjes{r'}
  \rew{} \esub {\mapjes t \mapjes u} x \esub {\mapjes r \mapjes{u'}} y \mapjes{r'}
  = \mapjes{t_1} \enspace.
\]

The first step is an instance of a rule in ES known
as $\sigma_1$: $(\esub u x t)v \rrule{} \esub u x (tv)$,
and the second one of a rule we call $\sigma_4$: $\esub{\esub u x t} y v
\rrule{} \esub u x \esub y t v$.
Quantitative types for ES tell us that only rule $\sigma_1$,
but not rule $\sigma_4$, is valid for a call-by-name calculus.
This is why it is not surprising that $\pi$ is rejected by our type system, as detailed in \autoref{s:pi-not-quantitative}.

The alternative encoding we propose is as follows (noted
$\mapjesn{(\cdot)}$ instead
of $\mapjes{(\cdot)}$): 
\begin{defi}[Translation from $\jterms$ to $\esterms$]
  \[ \begin{array}{lll}
    \mapjesn x \eqdef x &\quad
    \mapjesn{(\lx.t)} \eqdef \lx.\mapjesn t &\quad
    \mapjesn{\gap t u x r} \eqdef \mapjesapp t u x r
     \end{array} \]
   where $\lvar{x}$ and $\rvar{x}$ are fresh variables. 
\end{defi}
Notice the above $\pi$-reduction $t_0\to t_1$ is still simulated: $\mapjesn{t_0} \rew{\sigma_4}^2 \mapjesn{t_1}$. Moreover, consider again the counterexample $t = \gap \delta \delta y r$ to faithfulness (\autoref{e:not-faithful}).
The alternative encoding of $t$ is now given by
$\esub {\mapjesn\delta}{\lvar y} \esub{\mapjesn\delta}{\rvar y} \rsub {\lvar y
\rvar y} y {\mapjesn r}$,
which is just
$\esub{\mapjesn \delta}{\lvar y} \esub {\mapjesn \delta}{\rvar y} {\mapjesn r}$,
because $y \notin \fv{\mapjesn r}$.
The only hope to have an interaction between the two copies of $\mapjesn\delta$
in the previous term is to execute the ES,
but such executions will just throw away those two copies,
because $\lvar y, \rvar y \notin \fv{\mapjesn r}$.
This hopefully gives an intuitive idea of the faithfulness of our encoding.

\subsection{Proof of Faithfulness}

We need to prove the equivalence between two notions of strong normalization:
the one of a term in $\djncalc$ and the one of its encoding in $\escalc$.
While this proof can be a bit involved using traditional methods,
quantitative types will make it very straightforward.
Indeed, since quantitative types correspond exactly to strong normalization,
we only have to show that a term $t$ is typable exactly when its encoding is typable,
for two appropriate quantitative type systems.
For $\escalc$, we will use the following system~\cite{kesner20a}:
\begin{defi}[The Type System $\sysFullES$]
  \begin{mathpar}
    \inferrule*[right=\ruleAxES]{ }{
    \seq{x:\mult\sigma} x \sigma}
    \and \inferrule*[right=\many]{
    \left(\seq {\Gamma_i} M {\sigma_i}\right) \and I \neq \emptyset}{
  \seq {\interii \Gamma_i} M {\multii{\sigma_i}}}
  \and \inferrule*[right=\ruleAbsES]{
  \seq {\Gamma;x:\MM} M \sigma}{
\seq \Gamma {\lx.M} {\MM \to \sigma}}
\and \inferrule*[right=\ruleAppES]{
  \seq \Gamma M {\MM \to \sigma}
\and \seq \Delta N {\choice\MM}}{
\seq {\Gamma \inter \Delta} {MN} \sigma}
\and \inferrule*[right=\ruleES]{
  \seq{\Gamma;x:\MM} M \sigma
\and \seq \Delta N {\choice\MM}}{
\seq{\Gamma \inter \Delta}{\esub N x M} \sigma}
\end{mathpar}
\end{defi}

\begin{lem}
  \label{l:sysFullES-characterizes-sn}%
  Let $M \in \esterms$.
  Then $M$ is typable in $\sysFullES$ \textit{iff} $M \in \sn{\db,\sub}$.
\end{lem}

A simple induction on the type derivation shows that the encoding
$(\mapjesn\cdot)$ is sound.
\begin{lem} 
  \label{l:typJ-implies-typES}%
  Let $t \in \jterms$. Then $\derp \Gamma {t}{\sigma} \sysFullJ \implies
  \derp \Gamma {\mapjesn t}{\sigma} \sysFullES$.
\end{lem}
\begin{proof}
  By induction on the type derivation.
  Notice that the statement also applies by straightforward \ih\ to rule~$\many$.
  \begin{cas}
    \case{$\ruleAxJ$}
      Then $t = x$ and we type $\mapjesn t = x$ with rule $\ruleAxJ$.

      \case{$\ruleAbsJ$}
      Then $t = \lx.s$ and $\mapjesn t = \lx.\mapjesn s$.
      We conclude by \ih\ using $\ruleAbsES$.

      \case{$\ruleAppJ$}
      Then $t = \gap s u x r$ and
      $\mapjesn t = \mapjesapp s u x r$.
      By the \ih\ we have derivations
      $\derp{\Pi}{\mapjesn s}{ \choice{\mult{{\MM_i} \to {\tau_i}}_{\iI}}}\sysFullES$,
      $\derp{\Delta}{\mapjesn u}{\choice{\multunion_{\iI}\MM_i}}\sysFullES$
      and $\derp{\Lambda;x:\mult{\tau_i}_{\iI}}{\mapjesn r}{\sigma}\sysFullES$
      with $\Gamma = \Pi  \inter \Delta \inter \Lambda$.

      If $I \neq \emptyset$, it is easy to construct a derivation
      $\derp{\lvar x:\mult{{\MM_i} \to {\tau_i}}_{\iI};\rvar
      x:\multunion_{\iI} \MM_i}{\lvar x \rvar x}{\mult{\tau_i}_{\iI}}\sysFullES$.
      By \autoref{l:substitution-lemma-right-strong},
      we get 
      $\Phi = \derp{\Lambda;\lvar x:\mult{{\MM_i} \to {\tau_i}}_{\iI};\rvar x:\multunion_{\iI} \MM_i}{
      \rsub{\lvar x \rvar x} x {\mapjesn r}}{\sigma}\sysFullES$.
      We conclude by building the following derivation.
      \[ \inferrule*{
          \inferrule*{\Phi
          \and \seq{\Delta}{\mapjesn u}{ \choice{\multunion_{\iI} \MM_i}} }{
          \seq{\Lambda \inter \Delta; \lvar x}{\mult{{\MM_i} \to {\tau_i}}_{\iI}}{
            {\rsub{\lvar x \rvar x} x {\mapjesn r}} \esub {\rvar
          x}{\mapjesn u}}{\sigma}
        }
      \and \der{\Pi}{\mapjesn s}{\mult{{\MM_i} \to {\tau_i}}_{\iI}}}{
      \seq{\Pi  \inter \Delta \inter \Lambda}{\mapjesapp s u x r}{\sigma}
  } \]

  If  $I = \emptyset$, then $x \notin \fv{r}=\fv{\mapjesn r}$ by relevance,
  so that $\mapjesn t = \mapjesappempty s u x r$.
  By the \ih\ we have derivations
  $\derp\Pi{\mapjesn s}{\mult \tau}\sysFullES$,
  $\derp\Delta{\mapjesn u}{\mult \rho}\sysFullES$
  and $\derp\Lambda{\mapjesn r}{\sigma}\sysFullES$
  with $\Gamma = \Pi \inter \Delta \inter  \Lambda $.
  We conclude by building the following derivation.
  \[ \inferrule*{
      \inferrule*{
        \der{\Lambda;\lvar x:\emult;\rvar x:\emult}{\mapjesn r}{\sigma}
      \and \seq{\Delta}{\mapjesn u}{\mult\rho}}{
      \seq{\Lambda \inter \Delta; \lvar x:\emult}{
      \mapjesn r \esub{\rvar x}{\mapjesn u}}{\sigma}
    }
  \and \der\Pi{\mapjesn s}{\mult\tau}}{
\seq{\Lambda \inter \Pi \inter \Delta}{\mapjesapp s u x r}{\sigma}}
\qedhere \]
\end{cas}
\end{proof}

This last result, together with the two characterization
\autoref{l:sysFullJ-dbn} and \autoref{l:sysFullES-characterizes-sn},
gives:
\begin{cor}
  \label{l:snJ-implies-snES}%
  Let $t \in \jterms$. If $t \in \sn\dbeta$ then $\mapjesn t \in \sn{\db,\sub}$.
\end{cor}

We show the converse by a detour through the encoding of $\esterms$ to $\jterms$.

\begin{defi}[Translation from  $\esterms$ to $\jterms$]
  \label{def:es-j}
  \[ \begin{array}{lll}
    \mapesj x \eqdef x & \mapesj{(M N)} \eqdef \gap{\mapesj{M}}{\mapesj{N}}{x}{ x} & \\
    \mapesj{(\lx.M)} \eqdef \lx.\mapesj M & 
    \mapesj{(M \esub N x)} \eqdef \gap {\id}{ \mapesj{N}}{ x}{ \mapesj{M}}
  \end{array} \]
\end{defi}

The two following lemmas, shown by induction on the type derivations,
give in particular that $\mapjesn t$ typable implies $t$ typable.
\begin{lem}
  \label{l:typES-implies-typJ}%
  Let $M \in \esterms$. Then $\derp \Gamma {M}{\sigma} \sysFullES \implies
  \derp \Gamma {\mapesj M}{\sigma} \sysFullJ$.
\end{lem}
\begin{proof}
  By induction on the derivation.
  The cases where the derivation ends with $\ruleAxJ$, $\ruleAbsJ$ or $\many$ (generalizing the statement) are straightforward.
  \begin{cas}
    \case{$\ruleAppJ$}
      Then $M = PN$ and $\mapesj M = \gap{\mapesj P}{\mapesj N} z z$.
      By the \ih\ we have derivations
      $\derp\Lambda{\mapesj P}{\MM \to \sigma}\sysFullJ$ and
      $\derp\Delta{\mapesj N}{\choice{\MM}}\sysFullJ$
      with $\Gamma = \Lambda \inter \Delta$.
      By application of rule $\many$ we obtain
      $\derp\Lambda{\mapesj P}{\mult{\MM \to \sigma}}\sysFullJ$. 
      We conclude by building the following derivation.
      \[ \inferrule*{
          \der\Lambda{\mapesj P}{\mult{\MM \to \sigma}}
          \and \der\Delta{\mapesj N}{\choice{\MM}}
        \and \inferrule*{ }{\seq{x:\mult\sigma}{x}{\sigma}}}{
    \seq{\Lambda \inter \Delta}{\gap{\mapesj P}{\mapesj N} x x}{\sigma}} \]
    \case{$\ruleES$}\!\!
    Then $M = P \esub x N$ and we have a translation of the form
    $\mapesj M = \gap {(\lz.z)}{\mapesj N} x {\mapesj P}$.
    By the \ih\ we have derivations
    $\derp{\Lambda;x:\MM}{\mapesj P}{\sigma}\sysFullJ$ and
    $\derp\Delta{\mapesj N}{\choice\MM}\sysFullJ$
    with $\Gamma = \Lambda \inter \Delta$.
    Let $\MM = \mult{\tau_i}_{\iI}$.

    If $I \neq \emptyset$, We conclude by building the following derivation.
    \[ \inferrule*{
        \inferrule*{
          \left( \inferrule{
              \inferrule*{ }{
            \seq{z:\mult {\tau_i}}{z}{\tau_i}}}{
      \seq{\emptyset}{\lz. z}{\mult {\tau_i}\to\tau_i}} \right)_{\iI}}{
    \seq{\emptyset}{\lz. z}{\mult{\mult {\tau_i}\to\tau_i}}_{\iI}}
    \and \der\Delta{\mapesj N}{\choice{\MM}}
  \and \der{\Lambda;x:\MM}{\mapesj P}{\sigma}}{
\seq{\Delta \inter \Lambda}{\gap{(\lz.z)}{\mapesj N} x {\mapesj P}}{\sigma}} \]
If $I = \emptyset$, We conclude by building the following derivation (where $\tau$ is arbitrary).
\[ \inferrule*{
    \inferrule*{
      \inferrule*{
      \inferrule*{ }{\seq{z:\mult {\tau}}{z}{\tau}}}{
  \seq{\emptyset}{\lz. z}{\mult {\tau}\to\tau}}}{
\seq{\emptyset}{\lz. z}{\mult{\mult {\tau}\to\tau}}}
\and \der\Delta{\mapesj N}{\choice{\MM}}
\and \der{\Lambda;x:\MM}{\mapesj P}{\sigma}}{
\seq{\Delta \inter \Lambda}{\gap{(\lz.z)}{\mapesj N} x {\mapesj P}}{\sigma}}
\qedhere \]
\end{cas}
\end{proof}

\begin{lem}
  \label{l:typJ-from-typJESJ}%
  Let $t \in \jterms$. Then $\derp\Gamma{\mapjesj t}{\sigma}\sysFullJ \implies
  \derp\Gamma{t}{\sigma}\sysFullJ$.
\end{lem}
\begin{proof}
  By induction on $t$.
  The cases where $t = x$ or $t = \lx.s$ are straightforward by the \ih\ We reason by cases for the generalized application. 
  \begin{cas}
    \case{$t = \gap s u x r$ where $x \in \fv r$}
      We have
      \begin{align*}
        \mapjesj t &= \mapesj{(\mapjesapp s u x r)}
        = \gap \id {\mapjesj{s}}{\lvar x}{
          \gap \id {\mapjesj{u}}{\rvar x}{
      \rsub{\gap{\lvar x}{\rvar x} z z} x {\mapjesj r}}}\end{align*}
      By construction and also by the anti-substitution \autoref{l:anti-substitution-lemma-right-strong} it is not difficult to see that $\Gamma = \Gamma_s \inter \Gamma_u \inter \Gamma_r$ and there exist derivations having the following conclusions, where $I \neq \emptyset$:
      \begin{enumerate}
        \item \label{la} $\derp{\Gamma_r;x:\mult{\tau_i}_{\iI}}{\mapjesj r}{\sigma}\sysFullJ$
        \item \label{lb} $\derp{\lvar
            x:\mult{{\mult{\tau_i}} \to \tau_i}_{\iI}}{\lvar x}{\mult{{\mult{\tau_i}}
          \to \tau_i}_{\iI}}\sysFullJ$
        \item \label{lc} $\derp{\rvar x:\mult{\tau_i}_{\iI}}{\rvar x}{\mult{\tau_i}_{\iI}}\sysFullJ$
        \item \label{ld} $\derp{\emptyset}{\id}{\mult{{\mult{\tau_i}} \to \tau_i}_{\iI}}\sysFullJ$
        \item \label{le} $\derp{\Gamma_u}{\mapjesj u}{\mult{\tau_i}_{\iI}}\sysFullJ$
        \item \label{lf}
          $\derp{\emptyset}{\id}{\mult{{\mult{{\mult{\tau_i}} \to \tau_i}} \to
          {{\mult{\tau_i}} \to \tau_i}}_{\iI}}\sysFullJ$
        \item \label{lg} $\derp{\Gamma_s}{\mapjesj s}{\mult{{\mult{\tau_i}} \to \tau_i}_{\iI}}\sysFullJ$
      \end{enumerate}

      The \ih\ on points~\ref{la}, \ref{le} and~\ref{lg} give
      $\derp{\Gamma_r;x:\mult{\tau_i}_{\iI}}{r}{\sigma}\sysFullJ$,
      $\derp{\Gamma_u}{u}{\mult{\tau_i}_{\iI}}\sysFullJ$ and $\derp{\Gamma_s}{
      s}{\mult{{\mult{\tau_i}} \to \tau_i}_{\iI}}\sysFullJ$ resp., so that we conclude with the following derivation:

      \[ \inferrule*{
          \der{\Gamma_s}{ s}{\mult{{\mult{\tau_{i}}} \to {\tau_{i}}}_{\iI}}
          \and \der{\Gamma_u}{u}{\mult{\tau_i}_{\iI}}
        \and \der{\Gamma_r;x:\mult{\tau_i}_{\iI}}{r}{\sigma}}{
    \seq{\Gamma}{\gap{s}{u}{x}{r}}{\sigma}} \]

  \case{$t = \gap s u x r$ where $x \notin \fv r$} Then
    we have
    \[ \mapjesj t
      = \mapesj{(\mapjesappempty s u x r)}
      = \gap \id {\mapjesj{s}}{\lvar x}{
        \gap \id {\mapjesj{u}}{\rvar x}{
    \mapjesj r}}\] 
    We have the following derivation, where
    $\Gamma = \Gamma_s \inter \Gamma_r \inter \Gamma_r$, ${\mult{\tau_1}} \to
    \tau_1, {\mult{\tau_2}} \to \tau_2, \rho$ and $\rho'$ are witness types.
    \[ \inferrule*{
        \inferrule*{\vdots}{
        \seq{\emptyset}{\id}{\mult{{\mult{\tau_1}} \to \tau_1}}}
        \and \der{\Gamma_{s}}{\mapjesj{s}}{\mult{\rho}}
        \and \Phi}{
    \seq{\Gamma_{s} \inter \Gamma_{u} \inter \Gamma_r}{
      \gap \id {\mapjesj {s}}{\lvar x}{
\gap \id {\mapjesj{u}}{\rvar x}{\mapjesj r}}}{\sigma}} \]
Where
\[ \Phi =
  \inferrule*[vcenter]{
    \inferrule*{
    \vdots}{
    \seq{\emptyset}{\id}{\mult{{\mult{\tau_2}} \to \tau_2}}}
    \and \der{\Gamma_{u}}{\mapjesj{u}}{\mult{\rho'}}
  \and \der{\Gamma_r}{\mapjesj r}{\sigma}}{
  \seq{\Gamma_{u} \inter \Gamma_r}{\gap \id {\mapjesj{u}}{\rvar
x}{\mapjesj r}}{\sigma}} \]
By the \ih\ we have derivations
$\derp{\Gamma_r}{r}{\sigma}\sysFullJ$,
$\derp{\Gamma_s}{s}{\mult \rho}\sysFullJ$ and
$\derp{\Gamma_u}{u}{\mult{\rho'}}\sysFullJ$.
We then derive $\derp\Gamma{\gap s u x r}{\sigma}\sysFullJ$ by
rule~$\ruleAppJ$.\qedhere
\end{cas}
\end{proof}


Putting everything together, we get this equivalence:
\begin{cor}
  \label{l:typJ-iff-typES}%
  Let $t \in \jterms$. Then $\derp\Gamma{t}{\sigma}{\sysFullJ} \iff
  \derp\Gamma{\mapjesn t}{\sigma}{\sysFullES}$.
\end{cor}

This corollary, together with the two characterization
\autoref{l:sysFullJ-dbn} and \autoref{l:sysFullES-characterizes-sn}, provides
the main result of this section:
\begin{thm}[Faithfulness]
  \label{l:snJ-iff-snES}%
  Let $t \in \jterms$. Then $t \in \sn\dbeta \iff \mapjesn t \in \sn{\db,\sub}$.
\end{thm}

%% file: equivalences.tex
\section{Equivalent Notions of Strong Normalization}
\label{sec:equivalences}

In the previous section, we related strong $\dbeta$-normalization with strong
normalization of ES.
In this section we compare the various concepts of strong normalization
that are induced on $\jterms$ by $\beta$, $\dbeta$, $(\beta,\ptwo)$ and $(\jn)$.
This comparison makes use of several results obtained in the previous
sections.
From it, we also obtain new results about the original calculus $\jcalc$.

\paragraph{\texorpdfstring{$\beta$}{beta}-normalization is not enough}
Obviously, $\sn\dbeta\subseteq\sn\beta$, since $\beta\subseteq\dbeta$.
Similarly, $\sn\jn\subseteq\sn\beta$ and
$\sn{\beta,\ptwo}\subseteq\sn\beta$.  These inclusions are
  strict, as shown in \autoref{sec:semantics-djn} using the term $t_\Omega
  \eqdef \gap{\gap {x_1} y {x_2} \delta} \delta z z$.
  Indeed, this term is a premature normal form in 
  $\sn\beta$, but is not $(\jn)$- strongly normalizable. In all the three
  cases, $\beta$-strong normalization is not preserved by
permutation, as there is a term $t_\Omega \in
\sn{\beta}$ such that $t_\Omega \notin \sn{\jn}$, $t_\Omega \notin
\sn{\beta, \ptwo}$ and $t_\Omega \notin \sn{\dbeta}$.

\subsection{Comparing \texorpdfstring{$\dbeta$}{dbeta}  with \texorpdfstring{$\beta+\ptwo$}{beta+p2}}
\label{sec:comp_bptwo}

We now formalize the fact that our calculus
$\calcwithrules\jterms\dbeta$ is a version with distance of
$\calcwithrules\jterms{\beta,\ptwo}$, so that they are equivalent from
a normalization point of view.  To achieve this, we will
establish the equivalence of strong normalization
with $\dbeta$ and with $(\beta,\ptwo)$
by providing a long chain of equivalences.  One of them is
\autoref{l:snJ-iff-snES}, proved in the previous
section; the other is a result about $\sigma$-rules in the
\textlambda-calculus -- which is why we have to go through the
\textlambda-calculus again.
\Autoref{l:typES-implies-typL}, \autoref{l:snL-implies-snJ} and the two
  upcoming translations $\mapesl{(.)}$ and $\mapjesl{(.)}$ prepare the
equivalence result by relating strong normalization in different calculi.

\begin{defi}[Translation $\mapesl\cdot$ from $\esterms$ to $\lterms$]
  \[
    \mapesl x \eqdef x \quad
    \mapesl{(\lx.M)} \eqdef \lx.\mapesl M\quad
    \mapesl{(MN)} \eqdef \mapesl M \mapesl N\quad
    \mapesl{\esub N x M} \eqdef (\lx.\mapesl M) \mapesl N
  \]
\end{defi}

\begin{lem}
  \label{l:typES-implies-typL}%
  Let $M \in \esterms$. Then 
  $M \in \sn{\db,\sub} \implies \mapesl M \in \sn\beta$.
\end{lem}

\begin{proof}
  For typability in the \textlambda-calculus, we use the type system
  $\mathcal{S'}_{\lambda}$ with choice operators of \cite{kesner20a}.
  It can be seen as a restriction of the system $\sysFullES$ to $\lambda$-terms.
  Suppose $M \in \sn{\db,\sub}$.
  By \autoref{l:sysFullES-characterizes-sn} $M$ is typable in $\sysFullES$, and it
  is straightforward to show that $\mapesl M$ is typable in $\mathcal{S'}_\lambda$.
  Moreover, $\mapesl M$ typable implies that $\mapesl M \in \sn\beta$
  (\cite{kesner20a}), which is what we want.
\end{proof}

For $t\in\jterms$, let $\mapjesl t \eqdef \mapesl{(\mapjesn t)}$.
So, we are just composing the alternative encoding of generalized application
into ES with the map into \textlambda-calculus just introduced.
The translation $\mapjesl{(\cdot)}$ may be given directly by recursion as follows:
\[
  \mapjesl{x}=x \qquad \mapjesl{(\lx.t)}=\lx.\mapjesl t \qquad \mapjesl{\gap t u
  y r}=(\lambda\rvar{y}.(\lambda\lvar{y}.\rsub{\lvar{y}\rvar{y}}y{\mapjesl{r}})\mapjesl t)\mapjesl u
\]


\begin{lem}
  \label{l:snL-implies-snJ}%
  $\mapjesl t \in \sn{\beta,\sigma_2}
  \implies t \in \sn{\beta,\ptwo}$.
\end{lem}

\begin{proof}
  Because $\mapjesl{(\cdot)}$ produces a strict simulation from $\jterms$ to $\lterms$. More precisely: (i) if $t_1\to_{\beta} t_2$ then $\mapjesl{t_1}\to_{\beta}^+\mapjesl{t_2}$; (ii) if $t_1\to_{\ptwo} t_2$ then $\mapjesl{t_1}\to_{\sigma_2}^2\mapjesl{t_2}$.
\end{proof}

\begin{thm}\label{l:equivalence-dbn-betap2}
  Let $t \in \jterms$. Then $t \in \sn{\beta,\ptwo}$ iff $t \in \sn{\dbeta}$.
\end{thm}
\begin{proof}
  We prove that the following conditions are equivalent:
  1) $t \in \sn{\beta,\ptwo}$.
  2) $t \in \sn{\dbeta}$.
  3) $\mapjesn t \in \sn{\db,\sub}$.
  4) $\mapjesl t \in \sn{\beta}$.
  5) $\mapjesl t \in \sn{\beta,\sigma_2}$.
  Now, $1) \implies 2)$ is because $\rew\dbeta \subset \rewp{\beta,\ptwo}$.
  $2) \implies 3)$ is by \autoref{l:snJ-implies-snES}.
  $3) \implies 4)$ is by \autoref{l:typES-implies-typL}.
  $4) \implies 5)$ is shown by \cite{regnier94}.
  $5) \implies 1)$ is by \autoref{l:snL-implies-snJ}.
\end{proof}

Incidentally, the previous proof also contains a new proof of \autoref{l:snJ-iff-snES}.

\subsection{Comparing \texorpdfstring{$\dbeta$}{dbeta} with
\texorpdfstring{$(\beta,\pi)$}{beta+pi}}
\label{sec:comp_bpi}
We now  prove the equivalence between strong normalization  for $\dbeta$ and for $(\jn)$. One of the implications already follows from the properties of the typing system.
\begin{lem}\label{lem:sn-dbn-implies-sn-betapi}
  Let $t\in \jterms$. If $t\in\sn{\dbeta}$ then $t\in\sn{\jn}$.
\end{lem}
\begin{proof}
  Follows from the completeness of the typing system (\autoref{l:completeness})
  and soundness of $\sysFullJ$ for $(\jn)$ (\autoref{l:soundness-pi}).
\end{proof}

The proof of the other implication requires more work, organized in 4 parts:
1) a remark about ES;
2) a new translation of ES into the $\jcalc$-calculus with strict simulation;
3) the admissibility of two logical implications for Matthes' inductive definition
of $\sn{\beta,\pi}$ (\autoref{lem:ruleI-and-ruleII}); and
4) preservation of strong $(\jn)$-normalization by a certain map from the set
$\jterms$ into itself (\autoref{lem:mapjj-preserves-SN}).

The remark about explicit substitutions is the following one, where B-reduction means dB-reduction when the list
context is empty:

\begin{lem}\label{lem:remark-es}
  For all $M\in\esterms$, $M\in\sn{\db,\sub}$ iff $M\in\sn{\brule,\sub}$.
\end{lem}

The translation $\mapesj{(\cdot)}$ in \autoref{def:es-j} induces a simulation of
each reduction step $\rew{\sub}$ on $\esterms$ into a reduction step $\rew{\beta}$ on
$\jterms$, but cannot simulate the creation of an ES effected by rule $\db$.
A solution is to refine the translation $\mapesj{(\cdot)}$  for applications,
yielding the following alternative translation:
\[ \begin{array}{rclcrcl}
  \mapesjj x &:=& x &\qquad&
  \mapesjj{(\lx.M)} &:=& \lx.\mapesjj M \\
  \mapesjj{(MN)} &:=& \gap \id {\mapesjj N} y {\gap{\mapesjj M} y z z} &&
  \mapesjj{\esub{N}{x}M} &:=& \gap \id {\mapesjj N} x {\mapesjj M}
\end{array} \]

Since the clause for ES is not changed, simulation of each reduction step $\rew{\sub}$
by a reduction step $\rew{\beta}$ holds as before.
The improvement lies in the simulation of each $\db$-reduction step:
\[
  \mapesjj{((\lx.M)N)}
  = \gap \id {\mapesjj N} y {\gap{(\lx.\mapesjj M)} y z z}
  \rew\beta \gap \id {\mapesjj N} y {\rsub y x \mapesjj M}
  =_{\alpha} \mapesjj{(\esub N x M)}
\]

This strict simulation gives immediately:

\begin{lem}\label{lem:mapesjj-sn}
  For all $M\in\esterms$, if $\mapesjj M\in\sn{\beta}$ then $M\in\sn{\brule,\sub}$.
\end{lem}

We now prove two properties of strong normalization  for $(\jn)$ in $\jcalc$. 
%
A preliminary fact is the following:

\begin{lem}\label{lem:preliminary}
  The set
  $\sn {\jn}$ is closed under prefixing of arbitrary $\pi$-reduction steps:
  \[ 
  \inferrule{t\rew{\pi}t' \mbox{ and } t'\in \sn {\jn}}{t\in \sn {\jn}} \]
\end{lem}
\begin{proof} We first consider the following three facts:
  \begin{enumerate}
    \item Every $t\in\jterms$ has a unique $\pi$-normal form $\pi(t)$.
    \item The map $\pi(\cdot)$ preserves $\beta$-reduction steps, that is,
      $t_1\rew{\beta}t_2$ implies $\pi(t_1) \rew{\beta} \pi(t_2)$
      (\autoref{l:commutation-beta-pi-to-nf}).
    \item $\rew{\pi}$ is terminating.
  \end{enumerate}
  Now, suppose $t \notin \sn{\jn}$, so that there is an infinite
  $(\jn)$-reduction sequence starting at $t$.
  Then by the previous facts it is possible to construct an infinite
  $\beta$-reduction sequence starting at $\pi(t)$.
  But $\pi(t) = \pi(t')$ and $t' \rewn{\pi} \pi(t')$, so there is an infinite
  $(\jn)$-reduction sequence starting at $t'$, which leads to a
  contradiction.
\end{proof}

Recall the inductive characterization $\isn {\jn}$, given in \autoref{fig:SN}.
Given that $\sn{\jn}=\isn{\jn}$, the ``rule'' in \autoref{lem:preliminary}, when
written with $\isn{\jn}$, is admissible for the predicate $\isn{\jn}$.
Now, consider:
\begin{mathpar}
  \inferrule*[right=\isnbpone]{u,r\in \isn {\jn}}{
  \rsub{\gap{y}{u}{z}{z}}x r\in \isn {\jn}}
  \and \inferrule*[right=\isnbptwo]{\rsub{\rsub{\rsub{u}{y}t}{z}r}{x}s \in \isn {\jn} \qquad t,u\in \isn {\jn}\qquad x\notin \fv{t,u,r}}{
  \rsub{\gap{(\ly.t)}{u}{z}{r}}{x}s \in \isn {\jn}}
\end{mathpar}
Notice rule~$\isnbptwo$ generalizes rule~$\isnbpbetarule$: just take $s=x\vec{S}$, with $x\notin\vec{S}$.

\begin{lem}
  \label{lem:ruleI-and-ruleII}
  Rules~$\isnbpone$ and~$\isnbptwo$ are admissible rules for the predicate $\isn {\jn}$.
\end{lem}
\begin{proof}
  Proof of~$\isnbpone$.
  By induction on $t \in \isn {\jn}$, we prove that
  $\rsub{\gap y u z z} x t \in \isn{\jn}$.
  The most interesting case is~$\isnbppirule$, which we spell out in detail.
  We will use a device to shorten the writing: if $E$ is $t$, or $S$, or
  $\vec{S}$, then $\underline{E}$ denotes $\rsub{\gap y u z z} x E$.
  Suppose $t = \gap{y'}{u'}{z'}{t'}S\vec{S} \in \isn{\jn}$ with
  $\gap{y'}{u'}{z'}{t'S}\vec{S} \in \isn{\jn}$.
  We want $\underline t \in \isn{\jn}$.
  If $y'\neq y$, then the thesis follows by the \ih\ and one application
  of~$\isnbppirule$.
  Otherwise,
  $\underline t =
  \gap{\gap y u z z}{\underline{u'}}{z'}{\underline{t'}}
  \underline{S} \underline{\vec S}$.
  By the \ih,
  \[ \gap {\gap y u z z} {\underline{u'}} {z'}
    {\underline{t'}\underline S}\underline{\vec S}
  \in \isn{\jn}.  \]
  By inversion of~$\isnbppirule$, we get
  \[ \gap y u z {
    \gap z {\underline{u'}}{z'}{\underline{t'}\underline S}}
  \underline{\vec S} \in \isn{\jn}. \]
  From this, \autoref{lem:preliminary} gives
  \[ \gap y u z
    {\gap z {\underline{u'}}{z'}{\underline{t'}} \underline S}
  \underline{\vec S} \in \isn{\jn}. \]
  Finally, two applications of~$\isnbppirule$ yield
  $\underline t \in \isn{\jn}$.

  Proof of~$\isnbptwo$.
  We prove the following: for all $t_1\in \isn {\jn}$,
  for all $n\geq 0$,
  if $t_1$ has at least $n$ occurrences of the sub-term $\rsub{\rsub{u}{y}t}{z}r$, where $t,u\in \isn {\jn}$, then,
  for any choice of $n$ such occurrences, $t_2 \in \isn{\jn}$,
  where $t_2$ is the term that results from $t_1$ by replacing each of those
  $n$ occurrences by $\gap{(\ly.t)}{u}{z}{r}$.

  Notice the statement we are going to prove entails the admissibility of (II).
  Indeed, given $s$, let $n$ be the number of free occurrences of $x$ in $s$.
  The term $t_1 = \rsub{\rsub{\rsub u y t} z r} x s$
  has well determined $n$ occurrences of the sub-term
  $\rsub{\rsub u y t} z r$ (those resulting from substituting for $x$; it may have others), and
  $\rsub{\gap{(\ly.t)} u z r} x s$ is the term that results from $t_1$ by
  replacing each of those $n$ occurrences by $\gap{(\ly.t)} u z r$.

  Suppose $t_1 \in \isn{\jn}$ and consider $n$ occurrences of the
  sub-term $\rsub{\rsub u y t} z r$ in $t_1$.
  The proof is by induction on $t_1 \in \isn{\jn}$ and sub-induction on $n$.
  A term $s$ is determined, with $n$ free occurrences of $x$, such that $x
  \notin t, u, r$ and $t_1 = \rsub{\rsub{\rsub u y t} z r} x s$.
  We want to prove that $\rsub{\gap{(\ly.t)} u z r} x s \in \isn{\jn}$.
  We will use a device to shorten the writing: if $E$ is $t$, or $S$, or $\vec S$,
  then $\underline E$ denotes $\rsub{\rsub{\rsub u y t} z r} x E$ and
  $\underline{\underline E}$ denotes $\rsub{\gap{(\ly.t)} u z r} x E$.
  The proof proceeds by case analysis on $s$.

  We show the critical case $s = x \vec S$, where use is made of the
  sub-induction hypothesis.
  We are given $\underline s = \rsub{\rsub u y t} z r \underline{\vec S} \in
  \isn{\jn}$.
  We want to show $\underline{\underline s} = \gap{(\ly.t)} u z r
  \underline{\underline{\vec S}} \in \isn{\jn}$.
  Given that $t, u \in \isn{\jn}$, it suffices to prove
  \begin{equation}
    \label{eq:ruleII-proof}
    \rsub{\rsub u y t} z r \underline{\underline{\vec S}} \in \isn{\jn}
  \end{equation}
  due to rule~$\isnbpbetarule$.
  Let $s'\eqdef \rsub{\rsub u y t} z r \vec S$.
  Since $x \notin t, u, r$, we have $\underline{s'} = \underline s$
  (whence $\underline{s'} \in \isn{\jn}$),
  and the number of free occurrences of $x$ in $s'$ is $n-1$.
  By sub-induction hypothesis, $\underline{\underline{s'}} \in \isn{\jn}$.
  But $\underline{\underline{s'}}
  = \rsub{\rsub u y t} z r \underline{\underline{\vec S}}$,
  again due to $x \notin t, u, r$.
  Therefore \autoref{eq:ruleII-proof} holds.
\end{proof}


We now move to the fourth part of the ongoing reasoning.
Consider the map from $\jterms$ to itself obtained by composing
$\mapjesn{(\cdot)}: \jterms \to \esterms$ with
$\mapesjj{(\cdot)}: \esterms \to \jterms$.
Let us write $\mapjj\cdot$ this composition.
A recursive definition is also possible, as follows:
\[ \mapjj x = x \qquad \mapjj{(\lx.t)} = \lx.\mapjj t \qquad
  \mapjj{\gap t u y r} =
  \gap \id {\mapjj t} {y_1} {
\gap \id {\mapjj u}{y_2}{\rsub{\gap{y_1}{y_2} z z} y {\mapjj r}}} \]

\begin{lem}
  \label{lem:mapjj-preserves-SN}%
  If $t \in \sn{\jn}$ then $\mapjj t \in \sn{\jn}$.
\end{lem}
\begin{proof}
  For $t \in \sn {\jn}$, $\maxred[\jn] t$ denotes the length of
  the longest $(\jn)$-reduction sequence starting at $t$.
  We prove $\mapjj t \in \isn{\jn}$ by induction on
  the longest $(\jn)$-reduction sequence starting at $t$ ($\maxred[\jn]{t}$), with
  sub-induction on the size of $t$.
  We proceed by case analysis of $t$.
  \begin{cas}
    \case{$t=x$} We have $\mapjj x = x\in \isn{\jn}$.
    \case{$t=\lx.s$}
      We have $\mapjj t = \lx.\mapjj s$.
      The sub-inductive hypothesis gives $\mapjj{s}\in \isn {\jn}$.
      By rule~$\isnbpabsrule$, $\lx.\mapjj s \in \isn{\jn}$.
      \case{$t = \gap y u x r$}
      We have $\mapjj t = \gap \id y {x_1} {
      \gap \id {\mapjj u}{x_2}{\rsub{\gap{x_1}{x_2} z z} x {\mapjj r}}}$.
      By the (sub)-\ih, $\mapjj u, \mapjj r \in \isn{\jn}$.
      Rule~$\isnbpone$ yields
      $\rsub{\gap y {\mapjj u} z z} x {\mapjj r} \in \isn{\jn}$.
      Applying rule~$\isnbpbetarule$ twice, we obtain $\mapjj t \in \isn{\jn}$.
      \case{$t = \gap{(\ly.s)} u x r$}
      We have $\mapjj t = \gap \id {\ly.\mapjj s}{x_1}{
      \gap \id {\mapjj u}{x_2}{\rsub{\gap{x_1}{x_2} z z} x {\mapjj r}}}$.
      Notice that $\maxred[\jn] t$ is greater than $\maxred[\jn]{s}$
      and $\maxred[\jn] u$.
      By the induction hypothesis, $\mapjj s, \mapjj u \in \isn{\jn}$.
      Also $\maxred[\jn] t > \maxred[\jn]{\rsub{\rsub u y s} x r}$.
      Hence $\mapjj{(\rsub{\rsub u y s} x r)} \in \isn{\jn}$, again by the
      \ih\ Since map $\mapjj{(\cdot)}$ commutes with substitution,
      $\rsub{\rsub{\mapjj u} y {\mapjj s}} x {\mapjj r} \in \isn{\jn}$.
      This, together with $\mapjj s, \mapjj u \in \isn{\jn}$, gives
      $\rsub{\gap{(\ly.\mapjj s)}{\mapjj u} z z} x {\mapjj r}
      \in \isn{\jn}$, due to rule~$\isnbptwo$.
      Applying rule~$\isnbpbetarule$ twice, we obtain $\mapjj t \in \isn{\jn}$.
      \case{$t = \gap{\gap{t_0}{u_1} x {r_1}}{u_2} y {r_2}$}
      Let $s \eqdef \gap{t_0}{u_1}{x_1}{\gap{r_1}{u_2}{y}{r_2}}$.
      Since $t \rew{\pi} s$, the \ih\ gives $\mapjj s \in \isn{\jn}$.
      The induction hypothesis also gives $\mapjj{t_0}, \mapjj{u_1} \in \isn{\jn}$.
      The term $\mapjj s$ is
      \[ \gap \id {\mapjj{t_0}} {x_1}
        {\gap \id {\mapjj{u_1}} {x_2}
          {\rsub{\gap{x_1}{x_2} z z} x
            {\gap \id {\mapjj{r_1}} {y_1}
              {\gap \id {\mapjj{u_2}} {y_2}
      {\rsub{\gap{y_1}{y_2} z z} y {\mapjj{r_2}}}}}}} \]
      From $\mapjj s \in \isn{\jn}$, by four applications of
      $\isnbpbetarule$ we obtain
      \begin{equation}
        \label{eq:mapjj-preserves-sn}%
        \rsub
        {\gap{(\rsub{\gap{\mapjj{t_0}}{\mapjj{u_1}}{z}{z}}{x}{\mapjj{r_1}})}{\mapjj{u_2}}{z}{z}}
        y {\mapjj{r_2}} \in \isn{\jn}
      \end{equation}
      We want $\mapjj t \in \isn{\jn}$, where $\mapjj t$ is
      \[ \gap \id
        {\gap \id {\mapjj{t_0}} {x_1}
          {\gap \id {\mapjj{u_1}} {x_2} {\rsub{\gap{x_1}{x_2} z z} x {\mapjj{r_1}}}}
        } {y_1}
      {\gap \id {\mapjj{u_2}} {y_2} {\rsub{\gap{y_1}{y_2} z z} y {\mapjj{r_2}}}} \]
      From \autoref{eq:mapjj-preserves-sn} and $\mapjj{u_1} \in \isn{\jn}$,
      rule (II) obtains
      \[ \rsub {\gap \id {\mapjj{u_1}} {x_2}
          {\gap {\rsub{\gap{\mapjj{t_0}}{x_2}{z}{z}}{x}{\mapjj{r_1}}}
      {\mapjj{u_2}} z z}} y {\mapjj{r_2}} \in \isn{\jn} \]
      From this, \autoref{lem:preliminary} (prefixing of $\pi$-reduction steps) obtains
      \[ \rsub {\gap {\gap \id {\mapjj{u_1}} {x_2}
          {\rsub {\gap{\mapjj{t_0}}{x_2} z z} x {\mapjj{r_1}}}}
      {\mapjj{u_2}} z z} y {\mapjj{r_2}} \in \isn{\jn} \]
      From this and $\mapjj{t_0} \in \isn{\jn}$, rule (II) obtains
      \[ \rsub {\gap \id {\mapjj{t_0}} {x_1}
          {\gap {\gap \id {\mapjj{u_1}} {x_2}
            {\rsub{\gap{x_1}{x_2} z z} x {\mapjj{r_1}}}}
      {\mapjj{u_2}} z z}} y {\mapjj{r_2}} \in \isn{\jn} \]
      From this, \autoref{lem:preliminary} (prefixing of $\pi$-reduction steps) obtains
      \[ \rsub {\gap {\gap \id {\mapjj{t_0}} {x_1}
            {\gap \id {\mapjj{u_1}} {x_2}
          {\rsub{\gap{x_1}{x_2} z z} x \mapjj{r_1}}}}
      {\mapjj{u_2}} z z} y {\mapjj{r_2}} \in \isn{\jn} \]
      Finally, two applications of $\isnbpbetarule$ obtain
      $\mapjj t \in \isn{\jn} = \sn{\jn}$.
      \qedhere
  \end{cas}
\end{proof}

All is in place to obtain the desired result:
\begin{thm}\label{l:equivalence-dbn-betapi}
  Let $t\in \jterms$. Then $t\in\sn{\dbeta}$ iff $t\in\sn{\jn}$.
\end{thm}
\begin{proof}
  The implication from left to right is \autoref{lem:sn-dbn-implies-sn-betapi}. For
  the converse, suppose $t\in\sn{\jn}$. By \autoref{lem:mapjj-preserves-SN},
  $\mapjj t\in\sn{\jn}$. Trivially, $\mapjj t\in\sn{\beta}$.
  Since $\mapjj t=\mapesjj{(\mapjesn t)}$, \autoref{lem:mapesjj-sn} gives $\mapjesn
  t\in\sn{\brule,\sub}$.
  By \autoref{lem:remark-es}, $\mapjesn t \in\sn{\db,\sub}$. By \autoref{l:snJ-iff-snES}, $t\in\sn{\dbeta}$. 
\end{proof}

\subsection{Consequences for \texorpdfstring{$\jcalc$}{Lambda J}}
\label{s:consequences-lj}
Our previous results for $\djncalc$ provide new ones for the original $\jcalc$
as immediate consequences of Theorems~\ref{l:sysFullJ-dbn}, \ref{l:snJ-iff-snES}
and \ref{l:equivalence-dbn-betapi}: a quantitative type system characterizing
strong normalization, and a faithful translation into ES.
\begin{thm}
  Let $t\in\jterms$. Then:
  \begin{itemize}
    \item[1.] \textbf{(Characterization)} $t \in \sn{\jn}$ \textit{iff} $t$ is $\sysFullJ$-typable.
    \item[2.] \textbf{(Faithfulness)} $t \in \sn{\jn}$ \textit{iff} $\mapjesn t \in \sn{\db,\sub}$.
  \end{itemize}
\end{thm}

Beyond strong normalization, $\jcalc$ gains a new normalizing strategy, which
reuses the notion of left-right normal form introduced in
\autoref{subsec:isn-dbeta}.
We take the definitions of neutral terms, answer and left-right context $\lrc$
given there for $\djncalc$, in order to define a new left-right strategy and a
new predicate $\isnj$ for $\jcalc$.
The strategy is defined as the closure under $\lrc$ of rule $\beta$ and of the
particular case of rule $\pi$ where the redex has the form
$\gap{\nl}{u}{x}{\nans}S$.
\begin{defi}
  Predicate $\isnj$ is defined by the rules $\isnvarrule$, $\isnapprule$,
  $\isnabsrule$ in \autoref{d:isn}, together with the following two rules (which
  replace rule $\isnbetarule$)\footnote{Notice how a redex has the two possible forms
    $(\lx.t)S$ or $\gap{\nl}{u}{x}{\nans}S$, that can be written as $\nans S$, that
    is, the form $\ctx\dcnl{\lx.t} S$ of a left-right redex in $\djncalc$.
  Notice that left-right redexes are the same in $\djncalc$ and $\jcalc$.}:
  \begin{mathpar}
    \inferrule*[right=\isnredexa]{\ctx\lrc{\gap{\nl}{u}{y}{\nans S}} \in \isnj}{
    \ctx\lrc{\gap \nl u y \nans S} \in \isnj}
    \and \inferrule*[right=\isnredexb]{\ctx\lrc{\rsub{\rsub{u}{x}t}{y}r}, t, u
    \in \isnj}{
    \ctx\lrc{\gap {(\lx.t)} u y r}\in \isnj}
  \end{mathpar}
\end{defi}
The corresponding normalization strategy is organized as usual: an initial phase
obtains a left-right normal form, whose components are then reduced by internal
reduction.
Is this new strategy any good? \autoref{l:final} answers positively with the
equivalence between $\isnj$ and $\isn{\jn}$.
Before proving it, we need an intermediate lemma.

\begin{lem}
  \label{lem:case-hvar-and-pi}%
  The following rules are admissible for the predicate $\isnj$:
  \begin{mathpar}
  \inferrule{u, r \in \isnj}{\gap x u y r \in \isnj}
  \and \inferrule {\gap \nl u y {sS} \vec S \in \isnj}
  {\gap \nl u y s S \vec S \in \isnj}
\end{mathpar}
\end{lem}
\begin{proof}
  We start with the first rule, by induction on $r \in \isnj$.
  If $r$ is generated by rules $\isnvarrule$, $\isnapprule$ or $\isnabsrule$,
  then $r$ is a weak-head normal form and rule $\isnapprule$ applies.
  Otherwise $r = \ctx\lrc{redex}$.
  By inversion of rules $\isnredexa$ and $\isnredexb$, one obtains
  $\ctx{\lrc}{contractum}\in \isnj$, plus two other subterms of the redex also in
  $\isnj$ in case of $\isnredexa$.
  Let $\lrc'\eqdef\gap{x}{u}{y}{\lrc}$.
  By the \ih\ $\ctx{\lrc'}{contractum}\in \isnj$.
  By one of the rules $\isnredexa$/$\isnredexb$, $\ctx{\lrc'}{redex} \in \isnj$,
  that is $\gap x u y r \in \isnj$.

 For the second rule, we prove by induction on $r \in \isnj$, that, if
  $r=\gap{\nl}{u}{y}{sS}\vec{S}$, then $\gap {\nl}uys S\vec{S} \in \isnj$.
  We do a case analysis of $s$.
  \begin{cas}
    \case{$s = \nans$} Follows by rule $\isnredexa$ by taking $\lrc = \ec \vec S$.
    \case{$s=\ctx\lrc{redex}$}
      Let $\lrc_1 \eqdef \gap \nl u y {\lrc S} \vec S$
      and $\lrc_2 \eqdef \gap \nl u y \lrc S \vec S$.
      Since $r = \ctx{\lrc_1}{redex}$, inversion of rule
      $\isnredexa$/$\isnredexb$ gives $\ctx{\lrc_1}{contractum}\in \isnj$,
      plus two other subterms of the redex also in $\isnj$ in case of
      $\isnredexb$.
      By \ih\ $\ctx{\lrc_2}{contractum} \in \isnj$.
      A final application of $\isnredexa$/$\isnredexb$ gives
      $\ctx{\lrc_2}{redex}\in \isnj$, as required.
      \case{$s = \nl'$}
      First, notice there are exactly four sub-cases:
      \begin{cas}
        \subcase{$\nl'S$ is a weak-head normal form and $\vec S$ empty}
          By inversion of~$\isnapprule$, we take $sS$ apart, obtain its
          components in $\isnj$ and, using $\isnapprule$, we reconstruct the
          term $\gap \nl u y {\nl'} S$ in $\isnj$.
        \subcase{$S$ has the form $\ap{u'}{y'}{\ctx\lrc{redex}}$ and $\vec S$ is
        arbitrary}
          By inversion of the rule $\isnredexa$/$\isnredexb$, we have
          $\gap{\nl}{u}{y}{\nl' \ap{u'}{y'}{\ctx\lrc{contractum}}}\vec{S}\in \isnj$,
          plus two other subterms of the redex also in $\isnj$ in case of
          $\isnredexb$.
          By the \ih, we have that
          $\gap{\nl}{u}{y}{\nl'}\ap{u'}{y'}{\ctx\lrc{contractum}}\vec{S}\in \isnj$.
          As required, we obtain
          $\gap{\nl}{u}{y}{\nl'}\ap{u'}{y'}{\ctx\lrc{redex}}\vec{S}\in \isnj$
          by rule $\isnredexa$/$\isnredexb$.
        \subcase{$S$ has the form $\ap{u'}{y'}{\nans}$ and $\vec S$ is
        non-empty}
          Let $\vec S = R \vec R$.
          By applying inversion of $\isnredexa$ twice, we obtain
          $\gap \nl u y {\gap{\nl'}{u'}{y'}{\nans R}} \vec R \in \isnj$.
          By the \ih, $\gap{\gap \nl u y {\nl'}}{u'}{y'}{\nans R} \vec R \in \isnj$.
          By $\isnredexa$, $\gap{\gap \nl u y {\nl'}}{u'}{y'} \nans R \vec R\in
          \isnj$, as required.
        \subcase{$S$ has the form $\ap{u'}{y'}{\nl''}$ and $\vec S$ is
        non-empty}
          We have to analyze $\vec{S}$.
          For that, we introduce some notation.
          $R^{nl}$ (respectively $R^{ans}$, $R^{whnf}$, $R^{rdx}$) will denote a
          generalized argument of the form $\ap t z \nl$ (resp. $\ap t z \nans$,
          $\ap t z whnf$, $\ap t z {\ctx\lrc{redex}}$).

          Let $\nl_0=\gap{\nl}{u}{y}{\gap{\nl'}{u'}{y'}{\nl''}}$
          and $\nl_1=\gap{\gap{\nl}{u}{y}{\nl'}}{u'}{y'}{\nl''}$.
          The non-empty $\vec{S}$ has exactly 3 possible forms (in all cases
          $m \geq 0$).
          \begin{cas}
            \subsubcase{$R^{nl}_1\cdots R^{nl}_m R^{whnf}_{m+1}$}
              We apply the same kind of reasoning as in subcase~1.
              \subsubcase{$R^{nl}_1\cdots R^{nl}_m R^{rdx}\vec{R}$}
              Let $R^{rdx} = \ap{u''}{y''}{\ctx{\lrc''}{redex}}$ and let
              \[ \begin{array}{rcl}
                \lrc_0 & = & \nl_0R^{nl}_1\cdots R^{nl}_m \ap{u''}{y''}{\lrc''}\vec{R}\\
                \lrc_1 & = & \nl_1R^{nl}_1\cdots R^{nl}_m \ap{u''}{y''}{\lrc''}\vec{R}
              \end{array} \]
              Inversion of rule $\isnredexa$/$\isnredexb$ gives
              $\ctx{\lrc_0}{contractum}\in \isnj$, plus two other subterms of
              the redex also in $\isnj$ in case of $\isnredexb$.
              By the \ih, we have that $\ctx{\lrc_1}{contractum}\in \isnj$.
              We obtain $\ctx{\lrc_1}{redex}\in \isnj$ by rule
              $\isnredexa$/$\isnredexb$, as required.
              \subsubcase{$R^{nl}_1\cdots R^{nl}_m R^{ans}_{m+1}R_{m+2}\vec{R}$}
              Let $R^{ans}_{m+1}=\ap{u''}{y''}{\nans}$ and let
              \[ \begin{array}{rcl}
                \nl_2&=&\nl_0 R^{nl}_1\cdots R^{nl}_m\\
                \nl_3&=&\nl_1 R^{nl}_1\cdots R^{nl}_m\\
              \end{array} \]
              By inversion of $\isnredexa$, we obtain
              $\gap{\nl_2}{u''}{y''}{\nans R_{m+2}}\vec{R}\in \isnj$.
              Next \ih\ gives $\gap{\nl_3}{u''}{y''}{\nans R_{m+2}}\vec{R}\in \isnj$.
              By $\isnredexa$, $\gap{\nl_3}{u''}{y''}{\nans}R_{m+2}\vec{R}\in
              \isnj$ as required.
              \qedhere
          \end{cas}
      \end{cas}
  \end{cas}
\end{proof}

\begin{thm}
  \label{l:final}
  Let $t \in \jterms$. Then $t \in \isnj$ iff $t \in \isn{\jn}$.
\end{thm}
\begin{proof}
  $\Rightarrow$) We show that each rule defining $\isnj$ is admissible for the
  predicate $\isn{\jn}$ defined in \autoref{fig:SN}.
  Cases $\isnvarrule$ and $\isnabsrule$ are straightforward.
  Case $\isnredexa$ is by the \ih\ and \autoref{lem:preliminary}.
  Case $\isnredexb$ is by the \ih\ and rule~$\isnbptwo$.
  Case $\isnapprule$ is proved by a straightforward induction on $\nl$.

  $\Leftarrow$) We show that each rule in \autoref{fig:SN}
  defining the predicate $\isn{\jn}$ is admissible for the
  predicate $\isnj$.
  Cases~$\isnbpvarrule$ and~$\isnbpabsrule$ are straightforward.
  Case~$\isnbpbetarule$ is by rule $\isnredexb$ and the \ih, by just taking
  $\lrc=\ec\vec{S}$.
  Case~$\isnbphvarrule$ follows by the first rule of
  \autoref{lem:case-hvar-and-pi} and the
  \ih\ Case~$\isnbppirule$ is by the second and the \ih
\end{proof}

\subsection{Alternative Proof of Equivalence} 

The last theorem can also be shown as a corollary of $\isnj = \sn{\jn}$
and the fact that $\sn{\jn} = \isn{\jn}$ proved by
\cite{joachimski03}.
We will show the first equality $\isnj = \sn{\jn}$ in a similar way as for
$\dbeta$ (\autoref{l:cbn-sn-isn}).

\begin{lem}
  \label{l:stability-betapi}%
  If $t_0 \rew{\jn} t_1$, then
  \begin{itemize}
    \item $\rsub u x {t_0} \rew{\jn} \rsub u x {t_1}$, and
    \item $\rsub {t_0} x u \rewn{\jn} \rsub {t_1} x u$.
  \end{itemize}
\end{lem}
\begin{proof}
  The first statement is proved by induction on $t_0 \rew{\jn}
  t_1$ using \autoref{l:perm-rsub}.
  The second is proved by induction on $u$.
\end{proof}

\begin{lem}
  \label{l:whpi-deterministic}
  The strategy introduced in \autoref{s:consequences-lj} is deterministic.
\end{lem}
\begin{proof}
  For every term there is a unique decomposition in terms of a $\lrc$ context
  and a redex.
  Besides that, $\beta$ and $\pi$ redexes do not overlap.
\end{proof}

\begin{lem}
  \label{l:snbeta}%
  Let $t_0 = \ctx\lrc{\rsub{\rsub u x t} y r} \in \sn{\jn}$, $t \in \sn{\jn}$
  and $u \in \sn{\jn}$.
  Then $t'_0 = \ctx\lrc{\gap{(\lx.t)} u y r} \in \sn{\jn}$.
\end{lem}
\begin{proof}
  By hypothesis we also have $r \in \sn{\jn}$.  We use
  the lexicographic order to reason by induction on
  $\langle \maxred[\jn]{t_0}, \maxred[\jn]{t}, \maxred[\jn]{u}, \lrc \rangle$.
  To show $t'_0 \in \sn{\jn}$ it is sufficient to show that all its
  reducts are in $\sn{\jn}$. We analyze all possible cases.
  \begin{description}
    \item[Case] $t'_0 \rew{\beta} t_0$. We conclude by the hypothesis.
    \item[Case] $t'_0 \rew{\jn} \ctx\lrc{\gap{(\lx.t')}{u}{y}{r}} =t'_1$,
      where $t \rew{\jn} t'$.
      We have $t', u \in \sn{\jn}$ and by item~(2)
      $t_0 = \ctx\lrc{\rsub{\rsub u x t} y r} \rewn{\jn} \ctx\lrc{\rsub{\rsub{u}{x}{t'}}{y} r} = t_1$, so that also $t_1 \in \sn{\jn}$.
      We conclude $t'_1\in \sn{\jn}$ by the \ih\ since
      $\maxred[\jn]{t_1}  \leq \maxred[\jn]{t_0}$ and $\maxred[\jn]{t'} < \maxred[\jn] t$.
    \item[Case] $t'_0 \rew{\jn} \ctx\lrc{\gap{(\lx.t)}{u'}{y}{r}} = t'_1$,
      where $u \rew{{\jn}} u'$.
      We have $t, u' \in \sn{\jn}$ and by item~(2)
      $t_0 = \ctx\lrc{\rsub{\rsub{u}{x}{t}}{y} r} \rewn{{\jn}} \ctx\lrc{\rsub{\rsub{u'}{x}{t}}{y} r}= t_1$,
      so that also $t_1\in \sn{\jn}$.
      We conclude $t'_1\in \sn{\jn}$ by the \ih\ since
      $\maxred[\jn]{t_1}\leq  \maxred[\jn]{t_0}$
      and $\maxred[\jn]{u'} < \maxred[\jn] u$.
    \item[Case] $t'_0 \rew{{\jn}} \ctx\lrc{\gap{(\lx.t)}{u}{y}{r'}} = t'_1$,
      where $r \rew{{\jn}} r'$.
      We have $t, u \in \sn{\jn}$ and by item~(1)
      $t_0 = \ctx\lrc{\rsub{\rsub{u}{x}{t}}{y} r} \rew{{\jn}} \ctx\lrc{\rsub{\rsub{u}{x}{t}}{y}{r'}} = t_1$,
      so that also $t_1 \in \sn{\jn}$.
      We conclude $t'_1\in \sn{\jn}$ by the \ih\ since
      $\maxred[\jn]{t_1} < \maxred[\jn]{t_0}$.
    \item[Case] $t'_0 \rew{{\jn}} \ctx{\lrc'}{\gap{(\lx.t)}{u}{y}{r}} = t'_1$,
      where $\lrc \rew{\jn} \lrc'$.
      Thus we also have that $t_0 = \ctx\lrc{\rsub{\rsub u x t} y r} \rew{{\jn}} \ctx{\lrc'}{\rsub{\rsub{u}{x}{t}}{y} r} = t_1$.
      We have $t_1, t, u \in \sn{\jn}$.
      We conclude that $t'_1\in \sn{\jn}$ by the \ih\  since 
      $\maxred[\jn]{t_1} < \maxred[\jn]{t_0}$.
    \item[Case] $\lrc = \ctx{\lrc'}{\ec S}$
      and $t_0' = \ctx{\lrc'}{\gap{(\lx.t)}{u}{y}{r} S} \rew\pi
      \ctx{\lrc'}{\gap{(\lx.t)} u y {r S}} = t_1'$.
      This is the only case left.
      We have $t_0 = \ctx{\lrc'}{\rsub{\rsub u x t} y r S}
      = \ctx{\lrc'}{\rsub{\rsub u x t} y {(r S)}} = t_1$.
      We also have $t, u \in \sn{\jn}$.
      We conclude $t_1' \in \sn{\jn}$ by the \ih\ on $\lrc$ since $\langle
      \maxred[\jn]{t_1}, \maxred[\jn] t, \maxred[\jn] u
      \rangle = \langle \maxred[\jn]{t_0}, \maxred[\jn] t,
      \maxred[\jn] u \rangle$. Notice that when $\lrc =\ec$, then $\pi$-reduction can only take place in some subterm of $t'_0$, already considered in the previous cases.
      \qedhere
  \end{description}
\end{proof}

\begin{lem}
  \label{l:snpi}%
  If $t_0 = \ctx\lrc{\gap \nl u y {\nans S}} \in \sn{\jn}$,
  then $t'_0 = \ctx\lrc{\gap \nl u y \nans S} \in \sn{\jn}$.
\end{lem}
\begin{proof}
  We use the lexicographic order to reason by induction on
  $\pair{\maxred[\jn]{t_0}} \nl$.
  To show $t'_0 \in \sn{\jn}$ it is sufficient to show that all its
  reducts are in $\sn{\jn}$. We analyze all possible cases.
  \begin{cas}
    \case{$t'_0 \rew{\pi} t_0$} We conclude by the hypothesis.
    \case{$t'_0 \rew{\jn} \ctx\lrc{\gap{\nl'} u y \nans S} = t'_1$,
    where $\nl \rew{\jn} \nl'$}
      We have $t_0 \rew{\jn} \ctx\lrc{\gap{\nl'} u y {\nans S}} = t_1$, so that also $t_1 \in \sn{\jn}$.
      We conclude $t'_1\in \sn{\jn}$ by the \ih\ since
      $\maxred[\jn]{t_1}  < \maxred[\jn]{t_0}$.
    \case{$t'_0 \rew{\jn} \ctx\lrc{\gap \nl {u'} y \nans S} = t'_1$,
    where $u \rew{\jn} u'$}
      We have $t_0 \rew{\jn} \ctx\lrc{\gap \nl {u'} y {\nans S}} = t_1$, so that also $t_1 \in \sn{\jn}$.
      We conclude $t'_1\in \sn{\jn}$ by the \ih\ since
      $\maxred[\jn]{t_1}  < \maxred[\jn]{t_0}$.
    \case{$t'_0 \rew{\jn} \ctx\lrc{\gap \nl u y {a'} S} = t'_1$,
    where $\nans \rew{\jn} \nans'$}
      We have $t_0 \rew{\jn} \ctx\lrc{\gap \nl u y {\nans' S}} = t_1$, so that also $t_1 \in \sn{\jn}$.
      We conclude $t'_1\in \sn{\jn}$ by the \ih\ since
      $\maxred[\jn]{t_1}  < \maxred[\jn]{t_0}$.
    \case{$t'_0 \rew{\jn} \ctx\lrc{\gap \nl u y \nans {S'}} = t'_1$,
    where $S \rew{\jn} S'$}
      We have $t_0 \rew{\jn} \ctx\lrc{\gap \nl u y {\nans {S'}}} = t_1$, so that also $t_1 \in \sn{\jn}$.
      We conclude $t'_1 \in \sn{\jn}$ by the \ih\ since
      $\maxred[\jn]{t_1}  < \maxred[\jn]{t_0}$.
    \case{$\lrc = \ctx{\lrc'}{\ec S'}$} Thus,
      $t_0' = \ctx{\lrc'}{\gap{\gap \nl u y \nans} {u'} z r S'}
      \rew\pi \ctx{\lrc'}{\gap{\gap \nl u y \nans} {u'} z {r S'}} = t_1'$,
      where $S = \ap{u'}{z}{r}$.
      Then, $t_0 = \ctx{\lrc'}{\gap \nl u y {\gap \nans {u'} z r} S'} \rew\pi^2 \ctx{\lrc'}{\gap{\nl} u y {\gap \nans {u'} z {r S'}}} = t_1$, so that also $t_1 \in \sn{\jn}$.
      We conclude $t_1' \in \sn{\jn}$ by the \ih\ since $\maxred[\jn]{t_1} < \maxred[\jn]{t_0}$.
      \case{$\nl = \gap{\nl''}{u'}{z}{\nl'}$}
      Thus $t_0' = \ctx\lrc{\gap{\gap{\nl''}{u'}{z}{\nl'}} u y \nans S}
      \rew\pi^2 \ctx{\lrc}{\gap {\nl''}{u'}{z}{\gap{\nl'} u y {\nans }S}} = t_1'$.
      We do a case analysis on all the one-step reducts of $t_0'$ so we need to consider $t_1'$ with $S$ outside.
      We have $t_0 \rew\pi \ctx{\lrc}{\gap {\nl''}{u'}{z}{\gap{\nl'} u y {\nans S}}} =  t_1$,
      so that also $t_1 \in \sn{\jn}$.
      Let $\lrc' = \ctx\lrc{\gap{\nl''}{u'}{z}{\ec}}$.
      We have $\maxred[\jn]{t_1} < \maxred[\jn]{t_0}$ so by the \ih
      $\ctx{\lrc'}{\gap{\nl'} u y \nans S} \in \sn{\jn}$.
      Because $\gap{\nl'} u y \nans$ is an answer we can apply
      the \ih\ on $\nl''$ and we conclude $t_1' \in \sn{\jn}$.
      \qedhere
  \end{cas}
\end{proof}

\begin{lem}
  $\isnj = \sn{\jn}$.
\end{lem}
\begin{proof}
  First, we show $\isnj \subseteq \sn{\jn}$.
  We proceed  by induction on $t \in \isnj$.
  \begin{cas}
    \case{$t = x$} Straightforward.
    \case{$t = \lx.s$, where $s \in \isnj$}
      By the \ih\ $s \in \sn{\jn}$, so that $t \in \sn{\jn}$
      trivially holds.
    \case{$t = \gap \nl u x r$ where $\nl, u, r \in \isnj$ and $r \in
    {\nf\whdj}$}
      Since $\nl$ is stable by reduction, $\nl$ cannot in particular reduce to an answer. 
      Therefore any kind of reduction starting at $t$ only occurs in the subterms $\nl$, $u$ and $r$.
      We conclude since  $\nl, u, r \in \sn{\jn}$ hold by the \ih
    \case{$t = \ctx\lrc{\gap \nl u y \nans S}$, where $\ctx\lrc{\gap \nl u y
    {\nans S}} \in \isnj$}
      The \ih\ gives $\ctx\lrc{\gap \nl u y {\nans S}} \in \sn{\jn}$,
      so that $t \in \sn{\jn}$ holds by \autoref{l:snpi}.
    \case{$t = \ctx\lrc{\gap{(\lx.s)} u y r}$, where $\ctx\lrc{\rsub{\rsub u x s} y r},
    s, u \in \isnj$}
      Induction hypothesis gives $\ctx\lrc{\rsub{\rsub u x s} y r} \in \sn{\jn}$,
      $s \in \sn{\jn}$ and $u \in \sn{\jn}$ so that
      $t \in \sn{\jn}$ holds by \autoref{l:snbeta}.
  \end{cas}

  Next, we show $\sn{\jn} \subseteq \isnj$.
  Let $t \in \sn{\jn}$. We reason by induction on $\tuple{\maxred[\jn] t, \tmsz t}$
  w.r.t. the lexicographic order.
  If $\tuple{\maxred[\jn] t, \tmsz t}$ is minimal, \ie\ $\tuple{0, 1}$,
  then $t$ is a variable and thus in $\isnj$ by rule~$\isnvarrule$. Otherwise we proceed by case analysis.
  \begin{cas}
    \case{$t = \lx.s$} Since $\maxred[\jn] s = \maxred[\jn] t$
      and $\tmsz s < \tmsz t$, we conclude by the \ih\ and rule~$\isnabsrule$.
      \case{$t$ is an application}
      There are three cases.
      \begin{cas}
        \subcase{$t \in {\nf\whdj}$}
          Then $t = \gap \nl u x r$ with $\nl, u, r \in \sn{\jn}$
          and $r \in {\nf\whdj}$.
          We have $\maxred[\beta] \nl \leq \maxred[\jn] t$,
          $\maxred[\jn] u \leq \maxred[\jn] t$,
          $\maxred[\jn] r \leq \maxred[\jn]  t$,
          $\tmsz{\nl} < \tmsz{t}$, $\tmsz{u} < \tmsz{t}$ and $\tmsz{r} < \tmsz{t}$.
          By the \ih\ $\nl, u,r \in \isnj$
          and thus we conclude by rule~$\isnapprule$.
          \subcase{$t = \ctx\lrc{\gap{(\lx.s)} u y r}$}
          $t \in \sn{\jn}$ implies
          $\ctx\lrc{\rsub{\rsub u x s} y r},
          s, u \in \sn{\jn}$,
          so that they are in $\isnj$ by the \ih\
          We conclude that $t \in \isnj$ by rule~$\isnredexb$.
          \subcase{$t = \ctx\lrc{\gap \nl u y \nans S}$}
          $t \in \sn{\jn}$ implies $\ctx\lrc{\gap \nl u y {\nans S}} \in \sn{\jn}$,
          so that this term is in $\isnj$ by the \ih\ We conclude $t \in \isnj$
          by rule $\isnredexa$.
          \qedhere
      \end{cas}
  \end{cas}
\end{proof}

%% file: conclusion.tex
\section{Conclusion}\label{sec:conclusion}

\paragraph{Contributions.} This paper presents and studies
several properties of the call-by-name $\djncalc$-calculus, a
formalism implementing  an appropriate notion of distant reduction  to
unblock the $\beta$-redexes arising in generalized application
notation.

Strong normalization of simply typed terms was shown by translating
the $\djncalc$-calculus into the \textlambda-calculus.  A full
characterization of strong normalization was developed by means of a
quantitative type system, where the length of reduction to normal form
is bounded by the size of the type derivation of the starting term.  
An inductive definition of strong normalization was defined and proved
correct in order to achieve this characterization.  It was also shown
how the traditional permutative $\pi$-rule is rejected by the
quantitative system, thus emphasizing the choice of distant reduction
for a quantitative generalized application framework.

We have also defined a faithful translation from the $\djncalc$-calculus into ES.
The translation preserves strong normalization, in contrast to the traditional
translation from generalized applications to ES \eg\ in~\cite{espiritosanto07}.
Last but not least, we related strong normalization of $\djncalc$ with that of
other calculi, including in particular the original $\jcalc$
of Joachimski and Matthes~\cite{joachimski03,joachimski00}.
New results for the latter were found by means of the techniques developed for
$\djncalc$.
In particular, a quantitative characterization of strong normalization was
developed for $\jcalc$, where the bound on reduction given by the size of type
derivations
only holds for $\beta$-steps (and not for $\pi$-steps).

This paper is an extended version of \cite{espiritosanto22}.
In this version we provide full proofs, and improve the presentation and
discussion.
The proof of confluence for $\djncalc$ given in \autoref{sec:some-properties}
comes from the third author's thesis~\cite{Peyrot22}.

\paragraph{Related works.}

Generalizing elimination rules of natural
deduction is an old idea, occurring several times in the literature,
most notably by Schroeder-Heister~\cite{schroederheister84,schroederheister84a} or
Tennant~\cite{tennant92,tennant02}, before being coined in the version at the
origin of $\jcalc$ by von Plato~\cite{plato01}.  The generalization of
implication elimination itself has come up independently along the
years, as pointed out in \cite{schroederheister14}.

Concerning $\jcalc$, several results motivated by
a proof-theoretical approach are found in the literature. In parallel to his works with Joachimski
\cite{joachimski00,joachimski03} introducing the calculus,
\cite{matthes01} proves an interpolation theorem (with information on
terms) for $\jcalc$ extended with pairs and sum datatypes. In his PhD
thesis, Barral~\cite{barral08} defines a set of conversions for $\jcalc$
beyond $\beta$ and $\pi$. Some of these conversions where already
given by Matthes~\cite{matthes01}, another one is an undirected version of
$\ptwo$. Espírito Santo and his coauthors have used $\jcalc$, and his
multiary extension $\jmcalc$~\cite{espiritosanto11} to compare the
computational content of natural deduction and the sequent
calculus~\cite{espiritosanto09,espiritosanto18,espiritosanto23}.

The first non-idempotent type system for generalized applications was
proposed in our conference paper \cite{espiritosanto22}. Intersection
type systems for $\jcalc$ have been given before in \cite{matthes00}
and \cite{espiritosanto12}, but these systems are
based on idempotent types, so that they are not able to
characterize quantitative properties.
In \cite{kesner22a}, the solvability property is characterized for
  $\djncalc$ and  $\jcalc$, both operationally and logically, by means
  of non-idempotent types. It is also shown that solvability in $\djncalc$,
$\jcalc$ and $\lambda$-calculus are equivalent.
Other calculi based on different logical systems have been adapted to enable
quantitative analyzes: this is for instance the case of $\lmcalc$ based on
classical logic~\cite{kesner20a}, or the Curry-Howard interpretation of the
intuitionistic sequent calculus $\bar\lambda$~\cite{kesner15}.

\paragraph{Future work.}  Quantitative type systems, introduced here for
the call-by-name system $\djncalc$, have been successfuly adapted to
the call-by-value setting in \cite{kesner22a}.  A version with
  distance is also introduced, relying on the  $\pi$-rule, which is shown
  sound in a quantitative type system for CbV.  Solvability and
    strong normalization in the call-by-value setting are characterized with appropriate reduction
  relations and through quantitative type systems.  Further
unification between call-by-name and call-by-value with the help of
generalized applications could be considered in the setting of
call-by-push-value~\cite{levy06} or the polarized
lambda-calculus~\cite{espiritosanto17}.

The size of the typing derivations in the typing system we introduced for $\djncalc$ provides an upper bound on the length of reduction sequences, as spelled out in the proof of soundness.
The topic of estimating the exact length of the longest reduction sequence of strongly normalizing terms has been investigated for the $\lambda$-calculus at least since the work of de Vrijer~\cite{DeVrijer87}. Recently, typing systems based on non-indempotent intersection types were proposed which provide tight bounds for the length of evaluation sequences and for the size of results, in the context of several evaluation strategies for the $\lambda$-calculus~\cite{accattoli20,kesner22b}. It would be interesting to see if these techniques can be adapted to the setting of generalized applications.  The precise measures on reduction length obtained would
enable us to precisely measure the quantitative relationship between the call-by-name \textlambda-calculus and $\djncalc$. Such techniques
could also be adopted for call-by-value, to sharpen the relation
between generalized applications and call-by-value calculi.

An interesting line of work involving generalized applications is
  currently being developed, starting with Geuvers and Hurkens~\cite{geuvers16}.
  In these works, inference systems are derived from a truth table, with
  elimination rules following a generalized shape, akin to von Plato’s system.
  Proof terms are then used to anotate proofs~\cite{geuvers18}, and strong
  normalization is proved~\cite{geuvers19,abel21}.
  Interestingly, the standard implication introduction rule is replaced by two
  different rules.
  It would be interesting to understand the peculiarities of the corresponding
  λ-calculus with generalized applications designed in the spirit of these two
derived forms of abstractions.